\documentclass[10pt,journal,compsoc]{IEEEtran}

\makeatletter
	\usepackage{amsmath}
	\usepackage{amssymb}
	\usepackage{amsthm}
	\usepackage{amsfonts}
	\usepackage{mathtools}
		\long\def\Comment#1{}

		\newcommand{\Debug@Out}[3]{%
			\typeout{*MCLabPaper_Minimal.#1%
				\ifempty{#2}{: #3}{. #2: #3}}%
		}%

		\newif\ifDraft

		\long\def\DraftNote@Off#1{}%
		\long\def\DraftWarning@Off#1{}%
		\long\def\DraftError@Off#1{}%

		\newcommand{\Draft@Box}[2]{%
			\par\noindent%
			\framebox{%
				\begin{minipage}[t]{\linewidth}%
				\noindent\textbf{#1:} #2%
				\end{minipage}%
			}%
		}
		\long\def\DraftNote@On#1{%
			\Draft@Box{Note}{#1}%
		}%
		\long\def\DraftWarning@On#1{%
			\Draft@Box{Warning}{#1}%
		}%
		\long\def\DraftError@On#1{%
			\Draft@Box{Error}{#1}%
		}%

		\newcommand{\DraftOn}[1][]{% #1 ignored
			\Drafttrue%	
			\let\DraftNote\DraftNote@On%
			\let\DraftWarning\DraftWarning@On%
			\let\DraftError\DraftError@On%
		}
		\newcommand{\DraftOff}{%
			\Draftfalse%	
			\let\DraftNote\DraftNote@Off%
			\let\DraftWarning\DraftWarning@Off%
			\let\DraftError\DraftError@Off%
		}

		\DraftOff

		\def\ThisFolder{./}
		\newcommand{\Input}[2][.]{%
			\begingroup%
				\filename@parse{#1/#2}%
				\edef\ThisFolder{\ThisFolder\filename@area}%
				\input{\ThisFolder\filename@base.\filename@ext}%
			\endgroup%
		}

		\newcommand{\eg}{\textit{e.g.}, }%
		\newcommand{\ie}{\textit{i.e.}, }%
		\newcommand{\wrt}{with respect to\space}
		
		\newcommand{\respectively}{respectively\xspace}

		\def\compactthickmuskip{1mu plus 1mu minus 2mu}
		\def\compactmedmuskip{1mu plus 1mu minus 2mu}
		\newenvironment{CompactMath}
			{%	
			 	\thickmuskip=\compactthickmuskip%
			 	\medmuskip=\compactmedmuskip%
			}
			{}

		\newcommand{\DataPut}[2]{%
			\expandafter%
				\xdef\csname Data:#1\endcsname{#2}%
		}

		\newcommand{\DataIfKeyExistsTF}[3]{%
			\ifcsname Data:#1\endcsname%
				#2%
			\else%
				#3%
			\fi%
		}

		\newcommand{\DataGet}[2][]{%
			\DataIfKeyExistsTF{#2}{%
				\ifempty{#1}{%
					\csname Data:#2\endcsname%
				}{%
					\def#1{\csname Data:#2\endcsname}%
				}%
			}{%
				\PackageError{MCLabPaper_Minimal}%
		 		{Data-key '#2' undefined}{Before using \string\DataGet{#2}, you need to use \string\DataPut{#2}{value}.}%
			}%
		}

		\ifdefined\mathbb\else%
			\Debug@Out{Math}{Redefining mathbb as mathbf}%
			\let\mathbb\mathbf%
		\fi

		\newcommand{\Reals}{% Reals
			\ensuremath{
				\mathbb{R}%
			}\xspace%
		} 
		\newcommand{\RealsGEZ}{% 
			\ensuremath{
				\Reals_{0+}
			}\xspace%
		} 
		\newcommand{\RealsGZ}{% 
			\ensuremath{
				\Reals_{+}
			}\xspace%
		}
		\newcommand{\IntegersGZ}{% Naturals excluding 0
			\ensuremath{
				\mathbb{N}_{+}
			}\xspace%
		} 
		\newcommand{\IntegersGEZ}{% Naturals (including 0)
			\ensuremath{
				\mathbb{N}
			}\xspace%
		} 
		\newcommand{\Fun}[1]{\ensuremath{\textup{#1}}}
		\newcommand{\MathComma}{\ensuremath{, \allowbreak}}
		\newcommand{\Set}[1]{\ensuremath{\left\{#1\right\}}}
		\newcommand{\SetCardinality}[1]{\ensuremath{\left|#1\right|}}%
		\newcommand{\Floor}[1]{%
		    \ensuremath{\left\lfloor {#1} \right\rfloor}%
		}

		\newcommand{\Fail}{\textsl{FAIL}\xspace} % FAIL
		\newcommand{\Pass}{\textsl{PASS}\xspace} % PASS

		\def\Acronym@List{}%
		\newcommand{\MCLab@expandonce}[1]{%
		  \unexpanded\expandafter{#1}}

		\newcommand{\NewAcronym}[5]{%
			\def\New@Acro@id{#1}%
			\ifempty{#2}{%
				\def\New@Acro@s{\MCLab@expandonce\New@Acro@id}% 
			}{%
				\def\New@Acro@s{#2}%
			}%
			\ifempty{#3}{%
				\def\New@Acro@sp{\New@Acro@s s}%
			}{%
				\def\New@Acro@sp{#3}%
			}%
			\def\New@Acro@l{#4}%
			\ifempty{#5}{%
				\def\New@Acro@lp{\New@Acro@l s}%
			}{%
				\def\New@Acro@lp{#5}%
			}%
			\DataPut{Acronym/#1/s}{\New@Acro@s}%
			\DataPut{Acronym/#1/sp}{\New@Acro@sp}%
			\DataPut{Acronym/#1/l}{\New@Acro@l}%
			\DataPut{Acronym/#1/lp}{\New@Acro@lp}%
			\DataPut{Acronym/#1/used}{0}%
			\edef\Acronym@List{\Acronym@List{#1}}%
		}

		\newcommand{\Acs}[1]{%
			\DataGet{Acronym/#1/s}%
			\xspace%
		}
		\newcommand{\Acsp}[1]{%
			\DataGet{Acronym/#1/sp}%
			\xspace%
		}
		\newcommand{\Acl}[1]{%
			\DataGet{Acronym/#1/l}%
			\xspace%
		}
		\newcommand{\Aclp}[1]{%
			\DataGet{Acronym/#1/lp}%
			\xspace%
		}

		\newcommand{\Acf}[2][ (]{% [#1]: left-sep, {#2}: acro id
			\def\Acf@left{{#1}}\Acf@o{#2}%
		}
		\newcommand{\Acf@o}[1]{% {#1}: acro id
			\def\Acf@acroid{#1}\Acf@om%
		}
		\newcommand{\Acf@om}[1][)]{% [#1]: right-sep
			\DataGet{Acronym/\Acf@acroid/l}%
			\Acf@left%
			\DataGet{Acronym/\Acf@acroid/s}%
			#1% right sep
			\AcronymUsed{\Acf@acroid}%
			\xspace%
		}

		\newcommand{\Acfp}[2][ (]{% [#1]: left-sep, {#2}: acro id
			\def\Acfp@left{{#1}}\Acfp@o{#2}%
		}
		\newcommand{\Acfp@o}[1]{% {#1}: acro id
			\def\Acfp@acroid{#1}\Acfp@om%
		}
		\newcommand{\Acfp@om}[1][)]{% [#1]: right-sep
			\DataGet{Acronym/\Acfp@acroid/lp}%
			\Acfp@left%
			\DataGet{Acronym/\Acfp@acroid/sp}%
			#1% right sep
			\AcronymUsed{\Acfp@acroid}%
			\xspace%
		}

		\newcommand{\Ac}[2][ (]{% [#1]: left-sep, {#2}: acro id
			\def\Ac@left{{#1}}\Ac@o{#2}%
		}
		\newcommand{\Ac@o}[1]{% {#1}: acro id
			\def\Ac@acroid{#1}\Ac@om%
		}
		\newcommand{\Ac@om}[1][)]{% [#1]: right-sep
		 	\DataGet[\Ac@used]{Acronym/\Ac@acroid/used}%
			\ifnum\Ac@used=0%
				\Acf[\Ac@left]{\Ac@acroid}[#1]%
			\else%
				\Acs{\Ac@acroid}%
			\fi%
		} 	

		\newcommand{\Acp}[2][ (]{% [#1]: left-sep, {#2}: acro id
			\def\Acp@left{{#1}}\Acp@o{#2}%
		}
		\newcommand{\Acp@o}[1]{% {#1}: acro id
			\def\Acp@acroid{#1}\Acp@om%
		}
		\newcommand{\Acp@om}[1][)]{% [#1]: right-sep
		 	\DataGet[\Acp@used]{Acronym/\Acp@acroid/used}%
			\ifnum\Acp@used=0%
				\Acfp[\Acp@left]{\Acp@acroid}[#1]%
			\else%
				\Acsp{\Acp@acroid}%
			\fi%
		}

		\newcommand{\AcronymReset}[1]{%
			\DataPut{Acronym/#1/used}{0}%
		}
		\newcommand{\AcronymUsed}[1]{%
			\DataPut{Acronym/#1/used}{1}%
		}

		\newcommand{\AcronymResetAll}{%
			\expandafter\Util@For@Each\expandafter{\Acronym@List}{%
				\Debug@Out{Acronym}{AcronymListResetAll}{ - Processing '##1'}%
					\AcronymReset{##1}%
			}%
		}

		\newcommand{\AcronymListIsEmpty}{No acronyms defined.}% could be redefined by user
		\newcommand{\AcronymList}[1][description]{%
			\Debug@Out{Acronym}{AcronymList}{Start. List: '\Acronym@List'}%
			\ifempty{\Acronym@List}{%
				\AcronymListIsEmpty%
			}{%		
				\csname AcronymList@#1\endcsname%
			}%
		}

		\newcommand{\AcronymList@description}{%	
			\begin{description}
				\expandafter\Util@For@Each\expandafter{\Acronym@List}{%
					\Debug@Out{Acronym}{AcronymList@description}{ - Processing '##1'}%
					\item[\Acs{##1}] \Acl{##1}%
				}
			\end{description}%
		}

		\newtoks\AcronymList@tabtoks
		\newcommand\AcronymList@addtoks[1]{\AcronymList@tabtoks\expandafter{\the\AcronymList@tabtoks#1}}
		\newcommand{\AcronymList@tabular}{%
			\Debug@Out{Acronym}{AcronymList@tabular}{Start. List: '\Acronym@List'}%
			\ifempty{\Acronym@List}{%
				No acronyms defined.
			}{%		
				\expandafter\Util@For@Each\expandafter{\Acronym@List}{%
					\Debug@Out{Acronym}{AcronymList@tabular}{ - Processing '##1'}%
					\expandafter\AcronymList@addtoks\expandafter{%
						\textbf{\Acs{##1}} & \quad \Acl{##1}%
						\\
					}
				}%
				\begin{tabular}{ll}
					\the\AcronymList@tabtoks
				\end{tabular}
			}%
		}

		\newcommand{\Pretty}[2][]{#2\ifempty{#1}{}{\text{ #1}}}

		\newcommand{\Algo@Kw}[1]{\textbf{#1}}%	

		\newlength{\AlgoSingleIndentWidth}
		\newlength{\AlgoHangingWidth}
		\newlength{\AlgoLineIndent}
		\newlength{\AlgoWidth}
		\newsavebox{\AlgoBlockBodyBox}
		\newlength{\AlgoBlockDepth}
		\newlength{\AlgoBlockHeight}
		\newenvironment{Algo@Indent@Minipage} %% UNUSED, TO BE DELETED (creates left bar, but does not support page breaks)
			{%		
				\addtolength
					{\AlgoWidth}
					{-\AlgoSingleIndentWidth}%		
				\par
				\begin{lrbox}{\AlgoBlockBodyBox}%
				\begin{minipage}[t]{\AlgoWidth}
				\setlength{\leftskip}{\AlgoHangingWidth}%
				\setlength{\parindent}{-\AlgoHangingWidth}%
			}
			{%
				\end{minipage}\end{lrbox}%
		  		\settodepth{\AlgoBlockDepth}{\usebox{\AlgoBlockBodyBox}}%
		  		\settoheight{\AlgoBlockHeight}{\usebox{\AlgoBlockBodyBox}}%
		  		\addtolength{\AlgoBlockHeight}{\AlgoBlockDepth}%
		  		\noindent%
		  		\makebox[0pt]{%
		  			\hspace{\AlgoLineIndent}%
		  			\rule[-\AlgoBlockDepth]{0.5pt}{\AlgoBlockHeight}%
		  		}%
		  		\hspace{\AlgoSingleIndentWidth}%
		  			\usebox{\AlgoBlockBodyBox}%  			
		  		\Algo@EndIndent% % Shows end or not depending on style
			}

		\newenvironment{Algo@Indent}
			{%	
				\par%
				\addtolength{\leftskip}{\AlgoSingleIndentWidth}
			}
			{%
				\addtolength{\leftskip}{-\AlgoSingleIndentWidth}
		  		\par\Algo@EndIndent% % Shows end or not depending on style
			}

		\long\def\Algo@While#1#2{%
				\par%\noindent
				\Algo@Kw{while} #1 
				\begin{Algo@Indent}
					#2%
				\end{Algo@Indent}
		}

		\long\def\Algo@ForEach#1#2{%
				\par%\noindent
				\Algo@Kw{foreach} #1 \Algo@Kw{do}%
				\begin{Algo@Indent}%
					#2%
				\end{Algo@Indent}%
		}

		\long\def\Algo@If#1#2{%
				\par\Algo@Kw{if} #1 \Algo@Kw{then}
					\begin{Algo@Indent}
						#2%
					\end{Algo@Indent}
		}
		\long\def\Algo@lIf#1#2{%
				\par\Algo@Kw{if} #1
					\Algo@Kw{then}
							#2%
		}

		\long\def\Algo@Else#1{%
				\Algo@Kw{else}
				\begin{Algo@Indent}
					#1%
				\end{Algo@Indent}
		}
		\long\def\Algo@lElse#1{%
				\Algo@Kw{else}
					#1%
		}

		\newcommand{\Algo@BlankLine}{\par\vspace{\baselineskip}}

		\long\def\Algo@EndNoBegin#1{%
			\par
			\begin{Algo@Indent}
				#1%	
			\end{Algo@Indent}
			\par	
			\Algo@Kw{end}
			\par
		}

		\long\def\Algo@Comment#1{%
			/* #1 */
			\par
		}

		\long\def\Algo@Function#1#2{%
			\Algo@Kw{function} \AlgoFun{#1}%
			\begin{Algo@Indent}#2\end{Algo@Indent}%
		}

		\long\def\AlgoTypeDef#1#2{%
			\begingroup%
			\AlgoHideEnd% Change style: hide end
			\def\Field##1##2##3{%
				\AlgoVar{##1}%
				\ifequals{##2}{}{}{\space ##2}%
				\ifequals{##3}{}{\;}{:%
					\begin{Algo@Indent}%
						##3%
					\end{Algo@Indent}%
				}%
			}%
			\let\AlgoField\Field%
			\AlgoType{#1}%
			\par
			\Algo@EndNoBegin{#2}%
			\endgroup%
		}
		\newcommand{\AlgoType}[1]{\textbf{#1}}

		\newcommand{\AlgoFun}[1]{\ensuremath{\Fun{#1}}}
		
		\newcommand{\AlgoVar}[1]{\ensuremath{\Fun{#1}}}

		\newcommand{\AlgoVarGets}[2]{\begingroup\ensuremath{\AlgoVar{#1}\gets} #2\endgroup}
		
		\let\AlgoVarIncr\AlgoIncrVar
		
		\let\AlgoVarDecr\AlgoDecrVar

		\newcommand{\Algo@EndStmtSemicolon}{\unskip;\par}
		\newcommand{\Algo@EndStmtNoSemicolon}{\unskip\par}

		\let\Algo@EndStmt\Algo@EndStmtSemicolon

		\newcommand{\AlgoHideSemicolon}{%
			\let\Algo@EndStmt\Algo@EndStmtNoSemicolon%
		}
		\newcommand{\AlgoShowSemicolon}{%
			\let\Algo@EndStmt\Algo@EndStmtSemicolon%
		}
		\AlgoHideSemicolon % Default

		\newcommand{\Algo@EndIndentShown}{%
			\par%\noindent
			\Algo@Kw{end}\par}
		\newcommand{\Algo@EndIndentHidden}{\par}
		\newcommand{\AlgoShowEnd}{%
			\let\Algo@EndIndent\Algo@EndIndentShown%
		}
		\newcommand{\AlgoHideEnd}{%
			\let\Algo@EndIndent\Algo@EndIndentHidden%
		}
		\AlgoHideEnd % Default

		\providecommand{\url}[1]{\texttt{#1}}
		\newcommand{\UrlEmail}[1]{\url{#1}}
		\newcommand{\UrlHttp}[1]{\url{#1}}

		\newcommand{\MCLabPaper@Minimal@IfEmpty@TestMacro}{}%
		\newcommand{\ifempty}[3]{%
			\ifx\MCLabPaper@Minimal@IfEmpty@TestMacro#1\MCLabPaper@Minimal@IfEmpty@TestMacro%
				#2%
			\else%
				#3%
			\fi%
		}

		\newcommand{\Util@For@Each@End}{\Util@For@Each@End}
		\newcommand{\Util@For@Each@Aux}[1]{%
		    \ifx\Util@For@Each@End #1% list terminated
		    \else
		      \Util@For@Each@do{#1}%
		      \expandafter\Util@For@Each@Aux
		    \fi
		  }
		\newcommand{\Util@For@Each}[2]{%
			\long\def\Util@For@Each@do##1{#2}%
			\Util@For@Each@Aux#1\Util@For@Each@End%
		}

		\@ifpackageloaded{xspace}{}{%
			\DeclareRobustCommand\xspace{\@xspace@firsttrue
			  \futurelet\@let@token\@xspace}
			\newif\if@xspace@first
			\def\@xspace@simple{\futurelet\@let@token\@xspace}
			\def\@xspace@exceptions@tlp{%
			  ,.'/?;:!~-)\ \/\bgroup\egroup\@sptoken\space\@xobeysp
			  \footnote\footnotemark
			  \xspace@check@icr
			}
			\begingroup
			  \text@command\relax
			  \global\let\xspace@check@icr\check@icr
			\endgroup
			\newcommand*\xspaceaddexceptions{%
			  \g@addto@macro\@xspace@exceptions@tlp
			}
			\newcommand*\xspaceremoveexception[1]{%
			  \def\reserved@a##1#1##2##3\@@{%
			    \@xspace@if@q@nil@NF##2{%
			      \def\reserved@a####1#1####2\@@{%
			        \gdef\@xspace@exceptions@tlp{####1####2}}%
			      \expandafter\reserved@a\@xspace@exceptions@tlp\@@
			    }%
			  }%
			  \expandafter\reserved@a\@xspace@exceptions@tlp#1\@xspace@q@nil\@@
			}
			\def\@xspace@break@loop#1\@nil{}
			\providecommand*\@xspace@hook{}
			\begingroup\expandafter\expandafter\expandafter\endgroup
			  \expandafter\ifx\csname eTeXversion\endcsname\relax
			  \begingroup
			    \catcode`\;=\active  \catcode`\:=\active
			    \catcode`\?=\active  \catcode`\!=\active
			    \catcode`\,=\active  \catcode`\'=\active  \catcode`\-=\active
			    \xspaceaddexceptions{;:?!,'-}
			  \endgroup
			  \let\@xspace@eTeX@setup\relax
			\else
			  \def\@xspace@eTeX@setup{%
			    \begingroup
			      \everyeof{}%
			      \endlinechar=-1\relax
			      \catcode`\ =10\relax
			      \makeatletter
			      \catcode`\\\z@
			      \catcode`\{\@ne
			      \catcode`\}\tw@
			      \scantokens\expandafter{\expandafter\gdef
			        \expandafter\@xspace@exceptions@tlp
			        \expandafter{\@xspace@exceptions@tlp}}%
			    \endgroup
			  }
			\fi
			\def\@xspace{%
			  \@xspace@lettoken@if@letter@TF \space{%
			    \if@xspace@first
			      \@xspace@firstfalse
			      \let\@xspace@maybespace\space
			      \@xspace@eTeX@setup
			    \fi
			    \expandafter\@xspace@check@token
			      \@xspace@exceptions@tlp\@xspace@q@nil\@nil
			    \@xspace@token@if@equal@NNT \space \@xspace@maybespace
			    {%
			      \@xspace@lettoken@if@expandable@TF
			      {\expandafter\@xspace@simple}%
			      {\@xspace@maybespace\@xspace@hook}%
			    }%
			  }%
			}
			\def\@xspace@check@token #1{%
			  \ifx\@xspace@q@nil#1%
			    \expandafter\@xspace@break@loop
			  \fi
			  \expandafter\ifx\csname @let@token\endcsname#1%
			    \let\@xspace@maybespace\relax
			    \expandafter\@xspace@break@loop
			  \fi
			  \@xspace@check@token
			}
			\def\@xspace@lettoken@if@letter@TF{%
			  \ifcat\noexpand\@let@token @% letter
			    \expandafter\@firstoftwo
			  \else
			    \expandafter\@secondoftwo
			  \fi}
			\def\@xspace@lettoken@if@expandable@TF{%
			  \expandafter\ifx\noexpand\@let@token\@let@token%
			    \expandafter\@secondoftwo
			  \else
			    \expandafter\@firstoftwo
			  \fi
			}
			\def\@xspace@token@if@equal@NNT#1#2{%
			  \ifx#1#2%
			    \expandafter\@firstofone
			  \else
			    \expandafter\@gobble
			  \fi}
			\def\@xspace@q@nil{\@xspace@q@nil}
			\def\@xspace@if@q@nil@NF#1{%
			  \ifx\@xspace@q@nil#1%
			    \expandafter\@gobble
			  \else
			    \expandafter\@firstofone
			  \fi}%
			\ProvidesFile{xspace.sty}%
		}

	\usepackage{xkeyval}
	\usepackage{etoolbox}
	\usepackage{booktabs}

	\newcommand{\MCLabPaper@Minimal@If@Pkg@Skipped}[3]{%
		\ifcsdef{MCLabPaper@Minimal@skip@#1}{#2}{#3}%
	}

	\newcommand*{\MCLabPaper@Minimal@RequirePackage}[2][]{%
		\Debug@Out{MCLabPaper_Minimal}{Require@Package}{[#1]{#2}}%
		\MCLabPaper@Minimal@If@Pkg@Skipped{#2}{%
			\Debug@Out{MCLabPaper_Minimal}{Require@Package}{ - skipping package}%
		}{%
			\Debug@Out{MCLabPaper_Minimal}{Require@Package}{ - loading package}%
			\usepackage[#1]{#2}%
		}%
	}

	\MCLabPaper@Minimal@RequirePackage{hyperref}
	\MCLabPaper@Minimal@RequirePackage{xparse}
	\MCLabPaper@Minimal@RequirePackage{subcaption}
	\MCLabPaper@Minimal@RequirePackage{tikz}

	\MCLabPaper@Minimal@If@Pkg@Skipped{cleveref}{}{%
		\MCLabPaper@Minimal@RequirePackage
			[italian, english, noabbrev, capitalize, nameinlink]
			{cleveref}
		\crefname{equation}{}{}
		\crefname{algocf}{\algorithmcfname}{\algorithmcfname s}	
		\Crefname{algocf}{\algorithmcfname}{\algorithmcfname s}	
		\crefname{paragraph}{\MCLabPaper@sectionname}{\MCLabPaper@sectionname@plural}
		\crefname{subparagraph}{\MCLabPaper@sectionname}{\MCLabPaper@sectionname@plural}
		\crefname{enumi}{item}{items}
		\Crefname{enumi}{Item}{Items}	
		\crefname{page}{page}{pages}% enforce lower case
		\Crefname{page}{Page}{Pages}%
		\crefname{lstlisting}{\lstlistingname}{\lstlistingname{s}}
		\Crefname{lstlisting}
			{\CapitaliseFirst{\lstlistingname}}
			{\CapitaliseFirst{\lstlistingname}{s}}
		\crefname{algocfline}{line}{lines}
		\Crefname{algocfline}{Line}{Lines}	
	}

	\MCLabPaper@Minimal@RequirePackage[inline, shortlabels]{enumitem}
	\def\List@Max@Depth{50}
	\setlistdepth{\List@Max@Depth}
	\SetEnumitemKey{nested}{label*=\arabic*.}

	\SetEnumitemKey{compact}{noitemsep}

	\SetEnumitemKey{supercompact}{nosep}

	\SetEnumitemKey{2columns}{
		before=\raggedcolumns\begin{multicols}{2},
		after=\end{multicols}
	}
	\SetEnumitemKey{3columns}{
		before=\raggedcolumns\begin{multicols}{3},
		after=\end{multicols}
	}
	\SetEnumitemKey{4columns}{
		before=\raggedcolumns\begin{multicols}{4},
		after=\end{multicols}
	}

	\MCLabPaper@Minimal@RequirePackage{array}

	\MCLabPaper@Minimal@If@Pkg@Skipped{tabularx}{}{%
		\MCLabPaper@Minimal@RequirePackage{tabularx}
		\newcolumntype{H}{>{\setbox0=\hbox\bgroup}c<{\egroup}@{}} % A tabular column-type to HIDE a column
		\newcolumntype{C}{>{\centering\arraybackslash}p} % A tabular column-type to CENTER a paragraph content
		\newcolumntype{L}{>{\raggedright\arraybackslash}p} % A tabular column-type to LEFT-ALIGN a paragraph content
		\newcolumntype{R}{>{\raggedleft\arraybackslash}p} % A tabular column-type to RIGHT-ALIGN a paragraph content
	}

	\MCLabPaper@Minimal@If@Pkg@Skipped{ifthen}{}{%
		\usepackage{ifthen}
		\DeclareDocumentCommand{\ifequals}{+m +m +m +m}{%
			\ifthenelse{\equal{#1}{#2}}{#3}{#4}%
		}
	}

	\NewDocumentCommand{\IncludeGraphics}{O{} m}{%
		\centering
		\includegraphics[#1]{\ThisFolder/#2}%
	}

	\csnumdef{Section@Curr@Depth}{0}%

	\newcommand{\SectionSetStart}[1]{%
		\ifequals{#1}{part}{%
			\csnumdef{Section@Curr@Depth}{0}%
		}{%
			\ifequals{#1}{chapter}{%
				\csnumdef{Section@Curr@Depth}{1}%
			}{%
				\ifequals{#1}{section}{%
					\csnumdef{Section@Curr@Depth}{2}%
				}{%
					\ifequals{#1}{subsection}{%
						\csnumdef{Section@Curr@Depth}{3}%
					}{%
						\ifequals{#1}{subsubsection}{%
							\csnumdef{Section@Curr@Depth}{4}%
						}{%
							\ifequals{#1}{paragraph}{%
								\csnumdef{Section@Curr@Depth}{5}%
							}{%
								\ifequals{#1}{subparagraph}{%
									\csnumdef{Section@Curr@Depth}{6}%
								}{%
									\PackageError{MCLabPaper_Minimal}%
		 								{SectionSetStart}{Starting from level '#1' not implemented.}%
		 						}%
		 					}%
		 				}%
		 			}%
				}%
			}%
		}%
	}
	\SectionSetStart{section} % Default

	\NewDocumentEnvironment{Section}{ s o m }{%
		\Debug@Out{MCLabPaper_Minimal}{Section}{curr depth: '\csuse{Section@Curr@Depth}'}%
		\ifcase \csuse{Section@Curr@Depth}%
			   \def\Section@cmd{part}%
			\or\def\Section@cmd{chapter}%
			\or\def\Section@cmd{section}%
			\or\def\Section@cmd{subsection}%
			\or\def\Section@cmd{subsubsection}%
			\or\def\Section@cmd{paragraph}%
			\or\def\Section@cmd{subparagraph}%
			\else\def\Section@cmd{subparagraph}% flattening from here
		\fi%
		\IfBooleanTF{#1}{%
			\csname\Section@cmd\endcsname*{#3}%		
		}{%
			\csname\Section@cmd\endcsname{#3}%
		}%
		\IfValueT{#2}{\label{#2}}%
		\csnumdef{Section@Curr@Depth}{\csuse{Section@Curr@Depth} + 1}%
	}{%
		\csnumdef{Section@Curr@Depth}{\csuse{Section@Curr@Depth} - 1}%
	}%

	\NewDocumentCommand{\SectionInput}{ s o m O{.} m}{%
		\IfBooleanTF{#1}%
			{ \begin{Section}*[#2]{#3} }%
			{ \begin{Section}[#2]{#3} }%
		\Input[#4]{#5}%
		\end{Section}%
	}

	\RenewDocumentCommand{\SetCardinality}{s m}{%
		\ensuremath{%
			\IfBooleanTF{#1}{
				\Fun{card}( #2 )%
			}{%
				\left| #2 \right|%
			}%
		}%
	}%
	\DeclarePairedDelimiter\MathParentheses{\lparen}{\rparen}

	\let\Core@DataGet\DataGet
	\RenewDocumentCommand{\DataGet}{o D(){} D<>{} m}{%
		\IfValueTF{#1}{%
			\ifempty{#3}%
				{\Core@DataGet[#1]{#4}}%
				{\Core@DataGet[#1]{#3/#4}}%
		}{%
			\ifempty{#3}%
				{\Core@DataGet{#4}}%
				{\Core@DataGet{#3/#4}}%
		}%
	}

	\MCLabPaper@Minimal@If@Pkg@Skipped{siunitx}{}{%
		\usepackage{siunitx}
		\RenewExpandableDocumentCommand{\Pretty}{O{} m}{\num{#2}\ifempty{#1}{}{ #1}}
	}

	\definecolor{MCLabPaper_CommentColor}{gray}{0.5}
	\MCLabPaper@Minimal@If@Pkg@Skipped{algorithm2e}{}{%
		\usepackage[noline,scleft,linesnumbered,noend]{algorithm2e}
		\SetAlgorithmName{Algorithm}{Algorithm}{List of Algorithms}
		\def\Algo@Comment@Sty#1{\textcolor{MCLabPaper_CommentColor}{\textit{#1}}}
		\SetCommentSty{Algo@Comment@Sty}
		\SetAlCapSkip{\abovecaptionskip} % same as figures (hence, they are tunable by using \SetCompact)
		\SetInd{0.3em}{0.3em}
		\SetKw{KwFunction}{function}
		\SetKw{KwProcedure}{procedure}
		\SetKw{KwMethod}{method}
		\SetKw{KwClass}{class}
		\SetKw{KwLet}{let}
		\SetKw{KwNew}{new}
		\SetKw{KwNil}{nil}
		\SetKw{KwBreak}{break}
		\SetKw{KwOutput}{output}
		\SetKw{KwField}{field}
		\SetKw{KwIn}{input}
		\SetKw{KwGlobal}{global}
		\SetKw{KwMethod}{method}
		\SetKw{KwOut}{output}
		\SetKw{KwParam}{param}
		\SetKw{KwAssert}{assert}
		\SetKw{KwReturns}{returns}
		\SetKw{KwEnd}{end}
		\SetKw{KwStatic}{static}
		\SetKwRepeat{DoWhile}{do}{while}
		\SetKwBlock{BeginNoEnd}{begin}{}
		\SetKwBlock{EndNoBegin}{}{end}
		\SetKwBlock{NoBeginEnd}{}{end}
		\SetKwBlock{NoBeginNoEnd}{}{}
		\SetKw{KwFrom}{from}
		\SetKw{KwTo}{to}
		\SetKw{KwDownto}{downto}
		\SetKw{KwSend}{send}
		\SetKw{KwRecv}{recv}
		\SetKw{KwWait}{wait}
		\RenewDocumentCommand{\Algo@Function}{m +m}{%
			\noindent%
			\KwFunction \AlgoFun{#1}%
			\EndNoBegin{#2}%
		}%
		\let\AlgoFunction\Algo@Function

	}

	\NewDocumentCommand{\Revision}{ o +d<> +m }{%
		#3%
	}

	\usepackage{bibunits}
	\newtoggle{Biblio@Is@Bib@Unit}
	\togglefalse{Biblio@Is@Bib@Unit}
	\newcounter{Biblio@BibUnit@Counter}
	\DeclareDocumentEnvironment{BibliographyUnit}{}{%
		\stepcounter{Biblio@BibUnit@Counter}%
		\ifcsundef{Biblio@BibUnit@style@\arabic{Biblio@BibUnit@Counter}}{%
			\cslet{Biblio@BibUnit@style@\arabic{Biblio@BibUnit@Counter}}%
				  {\Bibliography@style}%
		}{}%
		\bibunit[\csuse{Biblio@BibUnit@style@\arabic{Biblio@BibUnit@Counter}}]%	
		\global\toggletrue{Biblio@Is@Bib@Unit}%
	}{%
		\endbibunit%
		\togglefalse{Biblio@Is@Bib@Unit}%
	}

\usepackage{mdframed}
\newcommand{\ArticleDisclaimer}{%
	This article appears in IEEE Transactions on Software Engineering, 2023. 
	DOI: \href{http://doi.org/10.1109/TSE.2023.3298432}{10.1109/TSE.2023.3298432}
}

\newcommand{\BoxArticleDisclaimer}{%
	\begin{mdframed}
	\begin{center}
		\ArticleDisclaimer
	\end{center}
	\end{mdframed}
}
	\usepackage[nocompress]{cite}
	\usepackage{tabularx}

	\newcommand{\Title}{Optimising Highly-Parallel Simulation-Based Verification of \Aclp{CPS}}

		\NewAcronym{ALMA}
			{ALMA}{ALMAs}
			{Apollo Lunar Model Autopilot}
			{Apollo Lunar Model Autopilots}

		\NewAcronym{BDC}
			{BDC}{BDCs}
			{Buck DC-DC Converter}
			{Buck DC-DC Converters}

		\NewAcronym{BT}
			{LSPT}{LSPTs}
			{Longest Shared Prefix Tree}
			{Longest Shared Prefix Trees}

		\NewAcronym{CPS}
			{CPS}{CPSs}
			{Cyber-Physical System}
			{Cyber-Physical Systems}

		\NewAcronym{DFS}
			{DFS}{DFSs}
			{Depth First Search}
			{Depth First Searches}

		\NewAcronym{DS}
			{DS}{DSs}
			{Dynamical System}
			{Dynamical Systems}

		\NewAcronym{FCS}
			{FCS}{FCSs}
			{Fuel Control System}
			{Fuel Control Systems}

		\NewAcronym{FMU}
			{FMU}{FMUs}
			{Functional Mock-Up Unit}
			{Functional Mock-Up Units}

		\NewAcronym{FSA}
			{FSA}{FSA}
			{Finite State Automaton}
			{Finite State Automata}

		\NewAcronym{HPC}
			{HPC}{HPC}
			{High-Performance Computing}
			{High-Performance Computing}

		\NewAcronym{KPI}
			{KPI}{KPIs}
			{Key Performance Indicator}
			{Key Performance Indicators}

		\NewAcronym{LSP}
			{LSP}{LSPs}
			{Longest Shared Prefix}{Longest Shared Prefixes}

		\NewAcronym{SUV}
			{SUV}{SUVs}
			{System Under Verification}
			{Systems Under Verification}

		\NewAcronym{SLFV}
			{SLV}{SLV}
			{System-Level Verification}
			{System-Level Verification}

		\NewAcronym{tu}
			{t.u.}{t.u.}
			{time unit}
			{time units}		

		\SectionSetStart{section}
		
		\RenewDocumentCommand{\Fun}{m d()}{%
			\ensuremath{
				\textit{#1}
				\IfValueT{#2}{\MathParentheses{#2}}
			}\xspace%
		}

		\newtheorem*{theorem*}{Theorem}
		\newtheorem{lemma}{Lemma}
		
		\newtheorem*{notation*}{Notation}
		
		\newtheorem{proposition}{Proposition}
		
		\newtheorem*{observation*}{Observation}
		\crefname{observation}{Observation}{Observations}
		\newtheorem{definition}{Definition}

		\newlist{Requirements}{enumerate}{\List@Max@Depth}
		\setlist[Requirements]{label*=\arabic*.}

		\newlist{conditions}{enumerate}{10}
		\setlist[conditions]{nested}
		\crefname{conditionsi}{condition}{conditions}
		\Crefname{conditionsi}{Condition}{Conditions}

		\newlist{steps}{enumerate}{10}
		\setlist[steps]{nested,wide}
		\crefname{stepsi}{step}{steps}
		\Crefname{stepsi}{Step}{Steps}

		\newlist{points}{enumerate}{10}
		\setlist[points]{nested}
		\crefname{pointsi}{point}{points}
		\Crefname{pointsi}{Point}{Points}

		\SetInd{1.5ex}{1.5ex}
		\SetKw{KwIssue}{issue}

		\newcommand{\SmallHeader}[1]{\par\smallskip\noindent\textbf{#1.}\quad}

		\newcommand{\True}{\ensuremath{\Fun{true}}\xspace}%
		\newcommand{\False}{\ensuremath{\Fun{false}}\xspace}%

		\newcommand{\TimeStep}{\ensuremath{\tau}\xspace}

		\NewDocumentCommand{\GenericSet}{m o}{%
			\ensuremath{
				\mathbb{#1}
				\IfValueT{#2}{#2}
			}\xspace%
		}
		\NewDocumentCommand{\GenericValue}{m o}{%
			\ensuremath{
				#1\IfValueT{#2}{#2}%
			}\xspace%
		}

		\NewDocumentCommand{\TimeSet}{}{%
			\GenericSet{T}%
		}
		\NewDocumentCommand{\TimeSetEmpty}{}{%
			\emptyset%
		}
		\NewDocumentCommand{\TimePoint}{O{}}{%
			\ensuremath{\GenericValue{t}#1}\xspace%
		}

		\NewDocumentCommand{\Function}{m o d()}{%
			\ensuremath{%
				\mathbf{#1}%
				\IfValueT{#2}{#2}%
				\IfValueT{#3}{(#3)}%
			}\xspace%
		}
		\NewDocumentCommand{\InputFunction}{o d()}{%
			\Function{u}[#1](#2)%
		}
		\NewDocumentCommand{\InputFunctionEmpty}{}{%
			{\InputFunction[_{\emptyset}]}%
		}
		\NewDocumentCommand{\InputFunctionConstant}{O{\InputValue}}{%
			{\hat{\Function{#1}}}%
		}

		\define@cmdkey{InputFunctionSet}{inputset}[\InputSet]{}%
		\define@cmdkey{InputFunctionSet}{timeset}[\TimeSet]{}%
		\NewDocumentCommand{\InputFunctionSet}{ O{} }{%
			\setkeys{InputFunctionSet}{inputset, timeset, #1}%
			\ensuremath{%
				{\cmdKV@InputFunctionSet@inputset}^{\cmdKV@InputFunctionSet@timeset}%
			}\xspace%
		}

		\NewDocumentCommand{\RealRange}{s m m s}{%
			\ensuremath{
				\IfBooleanTF{#1}{\left[}{\left(}%
				#2 \MathComma #3
				\IfBooleanTF{#4}{\right]}{\right)}%
			}\xspace%
		}

		\NewDocumentCommand{\Restriction}{m m}{%
			\ensuremath{
				{#1}_{|{#2}}%
			}%
			\xspace%
		}

		\NewDocumentCommand{\TimeFunctionRestriction}{m m d()}{%
			\ensuremath{%
				\Restriction{#1}{#2}		
				\IfValueT{#3}{(#3)}%
			}\xspace%
		}

		\NewDocumentCommand{\TimeFunctionConcat}{m m d()}{%
			\ensuremath{%
				{#1} \cdot {#2}%
				\IfValueT{#3}{(#3)}%
			}\xspace
		}

		\NewDocumentCommand{\SUV}{o}{%
			\ensuremath{
				\mathcal{H}%
				\IfValueT{#1}{#1}%
			}\xspace%
		}

		\NewDocumentCommand{\InputSet}{o}{%
			\IfValueTF{#1}{%
				\GenericSet{U}[_{#1}]%
			}{%
				\GenericSet{U}%
			}%
		}

		\NewDocumentCommand{\InputValue}{o}{%
			\IfValueTF{#1}{%
				\GenericValue{u}[_{#1}]%
			}{
				\GenericValue{u}%	
			}%
		}

		\newcommand{\SUVstateSet}{\ensuremath{\GenericSet{X}}\xspace}
		\NewDocumentCommand{\SUVstate}{O{}}{%
			\ensuremath{%
				x
				\ifequals{#1}{}{}{#1}%
			}%
			\xspace%
		}
		\NewDocumentCommand{\SUVstateError}{}{%
			\ensuremath{\bot}\space%
		}

		\newcommand{\SUVinitState}{\ensuremath{\SUVstate_0}\xspace}
		\newcommand{\SUVinputSet}{\InputSet}
		\let\SUVinputValue\InputValue

		\let\SUVinputFunction\InputFunction
		\let\SUVinputFunctionSet\InputFunctionSet

		\newcommand{\SUVoutputSet}{\ensuremath{\GenericSet{Y}}\xspace}

		\NewDocumentCommand{\SUVtransition}{d()}{%
			\ensuremath{%
				\varphi\IfValueT{#1}{(#1)}%
			}\xspace%
		}
		\newcommand{\SUVobservation}{\ensuremath{\psi}\xspace}

		\newcommand{\SUVdef}{\ensuremath{%
		    (\TimeSet \MathComma \SUVstateSet \MathComma \SUVinitState \MathComma \SUVinputSet \MathComma \SUVoutputSet \MathComma \SUVtransition \MathComma \SUVobservation)%
		}\xspace}%

		\newcommand{\SUVobservationDef}{%
			\ensuremath{
				\SUVstateSet \to \SUVoutputSet
			}\xspace%
		}

		\DeclareDocumentCommand{\SLFV}{o}{%
			\ensuremath{
				\pi\IfValueT{#1}{#1}
			}\xspace%
		}

		\DeclareDocumentCommand{\SLFVdef}{o}{%
			\ensuremath{%
				\left(
					\SUV, 
					\TraceSet\IfValueT{#1}{#1}
				\right)
			}\xspace%
		}

		\DeclareDocumentCommand{\NbSlices}{}{\ensuremath{k}\xspace}

		\NewDocumentCommand{\TimeQuantum}{}{\ensuremath{\TimeStep}\xspace}
		\NewDocumentCommand{\Horizon}{}{\ensuremath{h}\xspace}
		
		\NewDocumentCommand{\TimeHorizonDef}{}{%
			\ensuremath{\TimeQuantum \Horizon}\xspace%
		}

		\NewDocumentCommand{\TraceProse}{t^ t+}{%
			\IfBooleanTF{#1}{I}{i}nput trace%
				\IfBooleanT{#2}{s}\xspace%
		}
		\NewDocumentCommand{\Trace}{o}{%
			\ensuremath{\InputFunction[#1]}\xspace%
		}
		\NewDocumentCommand{\TraceDef}{s o D<>{\Horizon}}{%
			\ensuremath{
				\IfBooleanF{#1}{(} 
					\InputValue[0]\IfValueT{#2}{#2}, 
					\ldots, 
					\InputValue[{#3-1}]\IfValueT{#2}{#2}
				\IfBooleanF{#1}{)} 
			}\xspace%
		}
		\NewDocumentCommand{\TraceSet}{o}{%
			\ensuremath{\mathcal{U}\IfValueT{#1}{#1}}\xspace%
		}
		\NewDocumentCommand{\TraceSetDef}{o D<>{\NbTraces}}{%
			\begingroup%
			\IfValueTF{#1}{%
				\def\Dist@Trace@Set@Def@Idx##1{#1_{##1}}%
			}{%
				\def\Dist@Trace@Set@Def@Idx##1{##1}%
			}%
			\ensuremath{
				\Trace[_{\Dist@Trace@Set@Def@Idx{0}}], 
				\ldots, 
				\Trace[_{\Dist@Trace@Set@Def@Idx{#2 - 1}}]%
			}%
			\endgroup%
			\xspace%
		}
		\NewDocumentCommand{\NbTraces}{o}{%
			\ensuremath{%
				n\IfValueT{#1}{#1}%
			}\xspace
		}

		\NewDocumentCommand{\TraceSeqInputFunctionSeq}{D(){\TraceSet}}{%
			\ensuremath{
				\Fun{U}({#1})
			}\xspace%
		}

		\let\TraceInputFunction\Trace

		\NewDocumentCommand{\TraceSeqInputFunctionSeqDef}
				{D<>{\NbTraces}}{%
			\ensuremath{
				\TraceInputFunction[_1]() \MathComma 
				\ldots \MathComma
				\TraceInputFunction[_{#1}]()
			}\xspace%
		}

		\newcommand{\IsLexLess}{\ensuremath{\prec}}%_{\SUVinputSet}}}
		\newcommand{\IsPrefixOf}{\ensuremath{\sqsubseteq}}
		\newcommand{\IsProperPrefixOf}{\ensuremath{\sqsubset}}

		\NewDocumentCommand{\SIM}{o}{\ensuremath{\mathcal{S}\IfValueT{#1}{#1}}\xspace}
		\NewDocumentCommand{\SIMstateSet}{o}{\ensuremath{\mathcal{W}\IfValueT{#1}{#1}}\xspace}
		\NewDocumentCommand{\SIMdef}{}{\ensuremath{(\SUV \MathComma \SIMstateSet)}\xspace}

		\NewDocumentCommand{\SIMstate}{o}{%
			\ensuremath{%
				w%
				\IfValueT{#1}{#1}%
			}%
			\xspace%
		}

		\NewDocumentCommand{\SIMinitState}{}{\SIMstate[_0]}

		\NewDocumentCommand{\SIMmemory}{o}{%
			\ensuremath{%
				\mathcal{M}%
				\IfValueT{#1}{#1}%
			}%
			\xspace%
		}
		\NewDocumentCommand{\SIMmemoryEmpty}{}{\ensuremath{\emptyset}\xspace}

		\NewDocumentCommand{\SIMmemorySize}{o}{%
			\ensuremath{m\IfValueT{#1}{#1}}\xspace%
		}
		\NewDocumentCommand{\SIMmemorySizeOpt}{}{%
			\ensuremath{\SIMmemorySize^*}\xspace%
		}

		\NewDocumentCommand{\SIMmemoryDom}{o}{%
			\SIMstateIDset%
		}
		\NewDocumentCommand{\SIMstateDef}{O{} o o}{%
			\ensuremath{%
				(\SUVstate[#1] \MathComma 
				 \InputFunction[\IfValueTF{#2}{#2}{#1}] \MathComma 
				 \SIMmemory[\IfValueTF{#3}{#3}{\IfValueTF{#2}{#2}{#1}}])%
			}%
			\xspace%
		}
		\NewDocumentCommand{\SIMinitStateDef}{}{%
			\ensuremath{
				(\SUVinitState \MathComma \InputFunctionEmpty \MathComma \SIMmemoryEmpty)}%
				\xspace%
		}

		\NewDocumentCommand{\SIMcmd}{m o d()}{%
			\ensuremath{%
				\textsc{#1\IfValueT{#2}{$_{#2}$}}%
				\IfValueT{#3}{(#3)}%
			}\xspace%
		}
		\NewDocumentCommand{\SIMload}{d()}{\SIMcmd{load}(#1)}
		\NewDocumentCommand{\SIMstore}{d()}{\SIMcmd{store}(#1)}
		\NewDocumentCommand{\SIMrun}{d()}{\SIMcmd{run}(#1)}
		\NewDocumentCommand{\SIMfree}{d()}{\SIMcmd{free}(#1)}
		\NewDocumentCommand{\SIMout}{d()}{\SIMcmd{output}(#1)}
		\NewDocumentCommand{\SIMgenericCmdArgs}{o}{%
			\Fun{args\IfValueT{#1}{$_{#1}$}}}
		\NewDocumentCommand{\SIMgenericCmd}{o}{%
			\SIMcmd{cmd}[#1]%
				(\SIMgenericCmdArgs[#1])%
		}
		\NewDocumentCommand{\SIMcmdLen}{d()}{%
			\ensuremath{
				\Fun{time\_adv}\IfValueT{#1}{(#1)}%
			}\xspace%
		}
		\NewDocumentCommand{\SIMtransition}{d()}{%
			\ensuremath{%
				\SUVtransition_{\SIM}%
				\IfValueT{#1}{(#1)}%
			}%
			\xspace%
		}

		\NewDocumentCommand{\SIMstateID}{o}{%
			\ensuremath{%
				\IfValueTF{#1}
					{\lambda(#1)}%
					{\lambda}%
			}\xspace%
		}
		\NewDocumentCommand{\SIMstateIDset}{}{%
			\ensuremath{\Lambda}\xspace%
		}

		\NewDocumentCommand{\SIMmemoryTuple}{O{} o o}{%
			\ensuremath{%
				[\SIMstateID#1 \mapsto %\MathComma 		 
				 (\SUVstate[\IfValueTF{#2}{#2}{#1}] \MathComma
				 \InputFunction[\IfValueTF{#3}{#3}{\IfValueTF{#2}{#2}{#1}}]
				)]
			}\xspace%
		}

		\NewDocumentCommand{\SIMcampaign}{o}{%
			\ensuremath{%
				\chi\IfValueT{#1}{#1}%
			}\xspace%
		}
		\newcommand{\SIMnbCampaignCmds}{\ensuremath{c}\xspace}

		\NewDocumentCommand{\SIMcampaignDef}{D<>{\SIMnbCampaignCmds}}{%
			\ensuremath{%
				\SIMgenericCmd[0] \ \ldots \ \allowbreak \SIMgenericCmd[#1 - 1]%		
			}\xspace%
		}
		\NewDocumentCommand{\SIMcampaignProse}{t^ t+}{%	
			\IfBooleanTF{#1}{S}{s}imulation campaign%
			\IfBooleanT{#2}{s}%
			\xspace%
		}

		\NewDocumentCommand{\SIMcampaignLen}{D(){\SIMcampaign}}{%
			\ensuremath{
				\Fun{len}(#1)%
			}\xspace%
		}
		\NewDocumentCommand{\SIMcampaignMaxMem}{D(){\SIMcampaign}}{%
			\ensuremath{
				\Fun{mem}(#1)%
			}\xspace%
		}
		\NewDocumentCommand{\SIMcampaignMaxMemProse}{s t+ O{\SIMmemorySize}}{%	
			\ensuremath{#3}-memory%
			\IfBooleanF{#1}{%
				\space% 
				\IfBooleanTF{#2}{\SIMcampaignProse+{}}{\SIMcampaignProse{}}%
			}%
			\xspace%
		}

		\NewDocumentCommand{\SIMcampaignStructureDef}{o D<>{\NbTraces}}{
			\ensuremath{
				\SIMcampaign[_{\IfValueT{#1}{#1, } 0}] ~
				\ldots ~
				\SIMcampaign[_{\IfValueT{#1}{#1, } #2 - 1}]
			}\xspace%
		}

		\NewDocumentCommand{\SIMseqStates}{D<>{\SIMnbCampaignCmds}}{%
			\ensuremath{%
				\SIMstate[_0] \MathComma \ldots \MathComma \SIMstate[_{#1}]%
			}\xspace%
		}

		\NewDocumentCommand{\SIMnbCampaignOutputCmds}{}{\NbTraces}

		\NewDocumentCommand{\SIMseqOutputs}{D<>{\SIMnbCampaignOutputCmds}}{%
			\ensuremath{%
				\SUVobservation(\SUVstate[_{j_0}]) \MathComma \ldots \MathComma \SUVobservation(\SUVstate[_{j_{#1 - 1}}])%
			}\xspace%
		}

		\NewDocumentCommand{\SIMinputFunctionSeq}{D(){\SIMcampaign}}{%
			\ensuremath{
				\TraceSet({#1})
			}\xspace%
		}

		\NewDocumentCommand{\SIMinputFunctionIndex}{}{%
			\ensuremath{
				j
			}\xspace%
		}

		\NewDocumentCommand{\SIMinputFunctionSeqDef}
				{D<>{\SIMinputFunctionIndex} O{\NbTraces}}{%
			\begingroup%
			\ifequals{#1}{}{%
				\def\SIM@inputFunction@index##1{##1}%
			}{%
				\def\SIM@inputFunction@index##1{{#1}_{##1}}%
			}%
			\ensuremath{
				\InputFunction[_{\SIM@inputFunction@index{0}}] \MathComma 
				\ldots \MathComma
				\InputFunction[_{\SIM@inputFunction@index{#2 - 1}}]
			}%
			\endgroup%
			\xspace%
		}

		\NewDocumentCommand{\SIMcampaignParallel}{o}{%
			\ensuremath{%
				\Xi\IfValueT{#1}{#1}%
			}\xspace%
		}
		\NewDocumentCommand{\SIMcampaignParallelDef}{s O{\NbSlices}}{%
			\ensuremath{%
				\IfBooleanF{#1}{\left(}
					\SIMcampaign[_0] \MathComma
					\ldots \MathComma 
					\SIMcampaign[_{#2 - 1}]%
				\IfBooleanF{#1}{\right)}%		
			}\xspace%
		}
		\NewDocumentCommand{\SIMcampaignParallelProse}{t^ D(){\NbSlices} t+}{%	
			\ifequals{#2}{}{}{\ensuremath{#2}-}%
			\IfBooleanTF{#1}{P}{p}arallel %
			\IfBooleanTF{#3}{\SIMcampaignProse+}{\SIMcampaignProse}%
		}
		\NewDocumentCommand{\SIMcampaignParallelMaxMemProse}{t^ D(){\NbSlices} O{\SIMmemorySize} t+}{%	
			\ifequals{#2}{}{}{\ensuremath{#2}-}%
			\IfBooleanTF{#1}{P}{p}arallel %
			\IfBooleanTF{#4}{\SIMcampaignMaxMemProse[#3]+}{\SIMcampaignMaxMemProse[#3]}%
		}

		\NewDocumentCommand{\SG}{d()}{% 
		    \ensuremath{%
		        \mathcal{G}\IfValueT{#1}{{#1}}
		    }%
		    \xspace%
		}% 

		\NewDocumentCommand{\SGNbTraces}{O{} d()}{%
		    \ensuremath{
		        \Fun{nb\_traces}
		        \ifequals{#1}{}{}{_{#1}}
		        \IfValueT{#2}{(#2)}
		    }\xspace%
		}

		\NewDocumentCommand{\SGTrace}{O{} d()}{%
		    \ensuremath{
		        \Fun{trace}
		        \ifequals{#1}{}{}{_{#1}}
		        \IfValueT{#2}{(#2)}
		    }\xspace%
		}

		\NewDocumentCommand{\TraceIndexSet}{o}{% 
		    \ensuremath{%
		        \mathcal{I}\IfValueT{#1}{{#1}}
		    }%
		    \xspace%
		}% 

		\NewDocumentCommand{\BT}{O{}}{%
			\ensuremath{%
				\mathcal{T}#1%
			}\xspace%
		}

		\NewDocumentCommand{\BTdef}{O{}}{%
			\ensuremath{%
				{(\BTvertexSet[#1], \BTparent[#1])}%
			}\xspace%
		}
		\NewDocumentCommand{\BTvertexSet}{O{}}{%
			\ensuremath{%
				{V#1}%
			}\xspace%
		}

		\NewDocumentCommand{\BTparent}{O{} d()}{%
			\ensuremath{%
				{\Fun{parent}#1\IfValueT{#2}{(#2)}}%
			}\xspace%
		}

		\NewDocumentCommand{\BTdepth}{d()}{%
			\Fun{depth}%
				\IfValueT{#1}{%
					\Fun{(}%
					{#1}%
					\Fun{)}%
				}%
				\xspace%
		}
		\NewDocumentCommand{\BTdepthProse}{t^ t+}{%	
			\IfBooleanTF{#1}{D}{d}epth%
			\IfBooleanT{#2}{s}%
			\xspace%
		}

		\NewDocumentCommand{\BTntraces}{d()}{%
			\Fun{ntraces}%
				\IfValueT{#1}{%
					\Fun{(}%
					{#1}%
					\Fun{)}%
				}%
				\xspace%
		}

		\NewDocumentCommand{\BTchildren}{d()}{%
			\Fun{children}%
				\IfValueT{#1}{%
					\Fun{(}%
					{#1}%
					\Fun{)}%
				}%
				\xspace%
		}
		\NewDocumentCommand{\BTsize}{d()}{%
			\Fun{size}%
				\IfValueT{#1}{%
					\Fun{(}%
					{#1}%
					\Fun{)}%
				}%
				\xspace%
		}
		\NewDocumentCommand{\BTsizeDef}{o}{%
			\ensuremath{
				\SetCardinality{
					\BTvertexSet%
						\IfValueT{#1}{#1}%
				}
			}%
			\xspace%
		}

		\NewDocumentCommand{\OPTbuildBT}{D(){}}{%
			\AlgoFun{{\Acs{BT}}(}{#1}\AlgoFun{)}\xspace%
		}
		\NewDocumentCommand{\OPTannotate}{D(){}}{%
			\AlgoFun{annotate(}{#1}\AlgoFun{)}\xspace%
		}

		\NewDocumentCommand{\OPTlastTraceInputValue}{o}{%
			\ensuremath{\InputValue[#1]^{\textup{prv}}}\xspace%
		}
		\NewDocumentCommand{\OPTlastTrace}{}{%
			\ensuremath{\Trace[_{\textup{prv}}]}\xspace%
		}

		\NewDocumentCommand{\OPTtracePortion}{m e{^_}}{%
			\def\OPTtracePortionBrace{}%
			\IfValueT{#2}{%
				\def\OPTtracePortionBrace{x}%
			}%
			\IfValueT{#3}{%
				\def\OPTtracePortionBrace{x}%
			}%
			\ensuremath{%
				\ifequals{\OPTtracePortionBrace}{}{%
					\framebox{#1}%
				}{%
					\IfValueTF{#2}{%
						{\overbrace{\framebox{#1}}^{#2}}%
					}{%
						{\underbrace{\framebox{#1}}_{#3}}%
					}%
				}%
			}\xspace%
		}

		\newcommand{\OPTlsp}{\AlgoVar{lsp}\xspace}
		\newcommand{\OPTpar}{\AlgoVar{par}\xspace}
		\newcommand{\OPTchild}{\AlgoVar{child}\xspace}

		\newcommand{\OPTnone}{\ensuremath{\textup{none}}\xspace}

		\newcommand{\OPTidxload}{\AlgoVar{load}\xspace}

		\NewDocumentCommand{\OPTstored}{D(){}}{%
			\Fun{stored(}{#1}\Fun{)}\xspace%
		}

		\NewDocumentCommand{\OPTrandomise}{D(){}}{%
			\AlgoFun{randomise(}{#1}\AlgoFun{)}\xspace%
		}

		\NewDocumentCommand{\OPTstateID}{o d()}{%
			\ensuremath{%
				\AlgoFun{ann}%
					\IfValueT{#1}{#1}%
					\IfValueT{#2}{%
						\AlgoFun{(}%
						{#2}%
						\AlgoFun{)}%
					}%
			}\xspace%
		}

		\NewDocumentCommand{\OPTsimCmds}{D(){}}{%
			\AlgoFun{sim\_cmds(}{#1}\AlgoFun{)}\xspace%
		}

		\NewDocumentCommand{\OPTstoreState}{D(){}}{%
			\AlgoFun{do\_store(}{#1}\AlgoFun{)}\xspace%
		}
		\NewDocumentCommand{\OPTisWorthStoringState}{D(){}}{%
			\AlgoFun{worth\_storing(}{#1}\AlgoFun{)}\xspace%
		}

		\NewDocumentCommand{\TraceLabelled}{O{}}{%
			\ensuremath{%
				\Trace[#1^{\Fun{id}}]%
			}\xspace%
		}
		\NewDocumentCommand{\TraceLabelledDef}{O{} D(){0} D<>{\Horizon}}{%
			\ensuremath{
				( \SIMstateID[#1_{#2}], \InputValue[#1_{#2}], \ldots, \SIMstateID[#1_{#3-1}], \InputValue[#1_{#3-1}] )%
			}\xspace%
		}

		\NewDocumentCommand{\TraceSetLabelled}{O{}}{%
			\ensuremath{\TraceSet[^{\Fun{id}}#1]}\xspace%
		}

		\NewDocumentCommand{\TraceSetLabelledDef}{D<>{\NbTraces}}{%
			\ensuremath{
				\TraceLabelled[_1], \ldots, \TraceLabelled[_{#1}]%
			}\xspace%
		}

		\NewDocumentCommand{\TraceSetLabelledRnd}{o}{%
			\ensuremath{%
				\TraceSet[^{\Fun{id}}_{\IfValueT{#1}{{#1},}\textup{rnd}}]%
			}\xspace%
		}

		\DataPut
			{experiments/results/stats/operational scenarios/max}
			{\Pretty{195869671}}
		\DataPut
			{experiments/results/stats/operational scenarios/max/prose}
			{almost 200 million}
		\DataPut
			{experiments/results/stats/computational cores/max}
			{\Pretty{65536}}
		\DataPut
			{experiments/results/stats/machines/max}
			{\Pretty{1024}}
		\DataPut
			{experiments/results/stats/machines/cores}
			{\Pretty{64}}

		\DataPut
			{experiments/scenarios/buck/horizon}
			{\Pretty{60}}
		\DataPut
			{experiments/scenarios/buck/number}
			{\Pretty{49971109}}
		\DataPut
			{experiments/scenarios/buck/input space/size}
			{\Pretty{25}}
		\DataPut
			{experiments/scenarios/buck/simulation time/avg/seconds}
			{\Pretty{80}}
		\DataPut
			{experiments/scenarios/buck/simulation time/overall/prose}
			{around 126 years}
		
		\DataPut
			{experiments/scenarios/apollo/horizon}
			{\Pretty{100}}
		\DataPut
			{experiments/scenarios/apollo/number}
			{\Pretty{107535209}}
		\DataPut
			{experiments/scenarios/apollo/input space/size}
			{\Pretty{432}}
		\DataPut
			{experiments/scenarios/apollo/simulation time/avg/seconds}
			{\Pretty{60}}
		\DataPut
			{experiments/scenarios/apollo/simulation time/overall/prose}
			{around 205 years} 
		
		\DataPut
			{experiments/scenarios/fcs/horizon}
			{\Pretty{100}}
		\DataPut
			{experiments/scenarios/fcs/number}
			{\Pretty{195869671}}
		\DataPut
			{experiments/scenarios/fcs/input space/size}
			{\Pretty{6}}
		\DataPut
			{experiments/scenarios/fcs/simulation time/avg/seconds}
			{\Pretty{40}}
		\DataPut
			{experiments/scenarios/fcs/simulation time/overall/prose}
			{around 248 years}

	\NewDocumentEnvironment{SupplMatStatement}{m o}{%
		\par\medskip\noindent
		\textbf{\Cref{#1}}\IfValueT{#2}{\space(#2)}.
	}{\par\medskip}

\makeatother

\begin{document}
	% Main article
	\title{\Title}

	\author{Toni~Mancini,
	        Igor~Melatti,
	        and~Enrico~Tronci% <-this '%' stops a space
		\IEEEcompsocitemizethanks{%
			\IEEEcompsocthanksitem
			Authors are with the Computer Science Department, Sapienza University of Rome, Italy.
			\protect\\
			% note need leading \protect in front of \\ to get a newline within \thanks as \\ is fragile and will error, could use \hfil\break instead.
			E-mail:
				\UrlEmail{tmancini@di.uniroma1.it},
				\UrlEmail{melatti@di.uniroma1.it},
				\UrlEmail{tronci@di.uniroma1.it}%
		}%
		\thanks{%
			Manuscript received MMMMM DD, 2022%; revised August 26, 2015.
		}%
	}

	\IEEEtitleabstractindextext{%
		\AcronymResetAll%
		\AcronymUsed{HPC}%
		\begin{abstract}
			\Acp{CPS}, \ie systems comprising both software and physical components, arise in many industry-relevant application domains and often mission- or safety-critical.

			\Ac{SLFV} of \Acp{CPS} aims at certifying that given (\eg safety \Revision{or liveness}) specifications are met, or at estimating the value of some \Aclp{KPI}, when the system runs in its operational environment, that is in presence of inputs (from the user or other systems) and/or of additional, uncontrolled disturbances. 

			In order to enable\Revision<(both exhaustive and statistical)>{} \Ac{SLFV} of complex systems from the early design phases, the currently most adopted approach envisions the \emph{simulation} of a \emph{system model} under the \Revision{(time bounded)} \emph{operational scenarios} deemed of interest. 

			Unfortunately, simulation-based \Ac{SLFV} can be computationally prohibitive (years of sequential simulation), since system model simulation is computationally intensive and the set of scenarios of interest can be extremely large.

			In this article, we present a technique that, given a collection of scenarios of interest (extracted from mass-storage databases or from symbolic structures like constraint-based scenario generators), computes \emph{parallel shortest simulation campaigns}, which drive a possibly large number of system model simulators running in parallel in a \Ac{HPC} infrastructure through all (and only) those scenarios in the user-defined (possibly random) order, by wisely avoiding multiple simulations of repeated trajectories, and thus minimising the overall completion time, compatibly with the available simulator memory capacity.

			Our experiments on \Ac{SLFV} of Modelica/FMU and Simulink case study models with up to almost \emph{200 million scenarios} show that our optimisation yields \emph{speedups as high as $8\times$}. 
			This, together with the enabled \emph{massive parallelisation}, makes practically viable (a few weeks in a \Ac{HPC} infrastructure) verification tasks (both statistical and exhaustive\Revision{, with respect to the given set of scenarios}) which would otherwise take \emph{inconceivably} long time.
		\end{abstract}

		% Note that keywords are not normally used for peerreview papers.
		%\begin{IEEEkeywords}
		% Model Checking, Hybrid Systems, Formal Verification
		%\end{IEEEkeywords}
	}

	\maketitle

	\AcronymResetAll%

	\BoxArticleDisclaimer

\begin{BibliographyUnit}
	\begin{Section}
	 	[sec:introduction]
	  	{Introduction}

	 	\Acp{CPS} consist of interconnected hardware (the physical part) and software (the cyber part).
		\Acp{CPS} are ubiquitous in many industry-relevant application domains, \eg aerospace, automotive, energy, biology, healthcare, among many others.
		In many \Acp{CPS} (\eg in embedded systems), the software part consists of a (typically microprogrammed) controller which continuously senses the state of the system and sends commands to the hardware actuators in order to achieve an envisioned goal condition while satisfying some requirements. 

		\Ac{SLFV} of a \Ac{CPS} aims at verifying that the whole system (\ie the software and the hardware working together) meets the given specifications when running in its operational environment, \ie in presence of inputs and/or additional limited uncontrolled (but possible) events (such as faults, noise signals, or changes in system parameters, collectively referred to as \emph{disturbances}).

		Since industry-relevant \Acp{CPS} are often mission- or safety-critical, their \Ac{SLFV} is of paramount importance to build confidence on their robustness and, ultimately, to perform their qualification.
		To this end, \Ac{SLFV} of \Acp{CPS} is supported from the early design stages by well-known model-based design software tools, \eg among the others, Simulink, VisSim, Dymola, ESA Satellite Simulation Infrastructure SIMULUS.
		Such tools allow the user to mathematically model the physical parts of a \Ac{CPS} (the hardware model), by means of, \eg differential equations and/or algorithmic snippets to manage, \eg the occurrence of events, and enable their numerical simulation, both open-loop and closed-loop.
		In particular, during closed-loop simulation, the actual software for the controller continuously reads values from the connected hardware model and decides control actions.
		During simulation of \Ac{CPS} models, the above model-based design tools also allow the user to inject a time series of inputs and other disturbances stemming from the environment, representing an actual \emph{operational scenario}.

		By designing a proper set of scenarios deemed plausible (given the operational environment), \Ac{SLFV} of the system is performed either by verifying that the \Ac{CPS} model satisfies the given specifications under \emph{all} of them (\Revision{aka }\emph{exhaustive model checking}%
			\Revision{, where exhaustiveness is intended with respect to such set of scenarios}), or, when they are too many to be simulated exhaustively, the residual probability of errors or expected values of suitable \Acp{KPI} are estimated by simulating the system on a randomly chosen subset of scenarios (\emph{statistical model checking} \cite{agha-etal:2018:surveysmc}).

		\begin{Section}[sec:intro:motivations]{Background and Motivations}
			Unfortunately, models of industry-relevant \Acp{CPS} are often defined as systems of highly non-linear and possibly stiff differential equations, and their complexity hinders the possibility of any symbolic reasoning (via, \eg model checkers for hybrid systems). 
			As a result, the main workhorse for \Ac{SLFV} of such system models is their black-box simulation on each single scenario, in order to check whether the required system-level specifications are satisfied on all of them or to estimate the values for the \Acp{KPI} of interest. 

			Simulating a \Ac{CPS} model on a single scenario can take from seconds to minutes, depending on the requested simulation time horizon and on the complexity of the system model. For example, simulating our case study models on a single scenario takes around \DataGet<experiments/scenarios/apollo>{simulation time/avg/seconds} (\Ac[, ]{ALMA}[,] by Simulink), \DataGet<experiments/scenarios/buck>{simulation time/avg/seconds} (\Ac[, ]{BDC}[,] by JModelica/\Acs{FMU}) and \DataGet<experiments/scenarios/fcs>{simulation time/avg/seconds} seconds (\Ac[, ]{FCS}[,] by Simulink) on average. This is due to the frequent injections of disturbances and/or changes of parameters, as prescribed by the scenario being simulated.
			All this makes simulation-based \Ac{SLFV} campaigns of such \Acp{CPS} a \emph{complex and extremely time-consuming activity}.

			There are two major sources of complexity to deal with when carrying out simulation-based \Ac{SLFV} of \Acp{CPS}.

			\emph{The first source of complexity} stems right from the definition of the set of scenarios deemed plausible or worth of interest, against which the system model must be verified.
			Traditionally, such scenarios (which collectively define the \Ac{CPS} \emph{operational environment}) are manually defined by system verification teams together with domain experts, and stored in large databases.
			When a new version of the \Ac{CPS} model has to be verified, such scenarios are injected during simulation and the resulting model trajectories are evaluated.
			Beyond being extremely time-consuming (possibly requiring months of work from expert designers), this na{\"i}ve operational environment definition activity is extremely fragile, as it is hard to assure that the successful verification of the \Ac{CPS} model against such scenarios is sufficient to certify absence of errors. 
			This is because it would be impossible to state whether the defined set of scenarios is representative of \emph{all} the possible situations of interest.

			To overcome this obstruction, previous work~\cite{mancini-etal:2013:cav,%
			%mancini-etal:2016:fundam,
			mancini-etal:2021:tse-supervisory} proposed to lift the hand-crafted definition of operational scenarios into the definition of a \emph{declarative constraint-based specification} of the system operational environment via an automaton encoded in a high-level language.
			The set of possible scenarios against which to verify the \Ac{CPS} model is then defined as the set of time series of inputs and other uncontrollable events encoded by accepting computation paths on such an automaton. Also, one such form of automaton, named \emph{scenario generator} in \cite{mancini-etal:2021:tse-supervisory}, allows the efficient extraction of any of its entailed scenarios from their unique indices.
			The definition of the \Ac{CPS} operational environment by such high-level models greatly eases the task of the verification engineers to capture all scenarios deemed plausible, also allowing them to dynamically focus on those scenarios satisfying additional constraints (see \cite{mancini-etal:2021:tse-supervisory} for examples), thus enabling prioritisation of the (typically very long) verification activity.

			Also, the availability of an environment model entailing a possibly large, yet finite number of scenarios enables the \emph{exhaustive} (\wrt such an environment model) \emph{verification} of the \Ac{CPS} at hand.
			Indeed, when the \Ac{CPS} model is exercised on \emph{all} the (finite number of) scenarios entailed by the environment model, a clear \emph{degree of assurance} is attained at the end of the verification process.
			Furthermore, by properly \emph{randomising} the scenario verification order, suitable information on the probability that a yet-to-be-simulated scenario exists for which the \Ac{CPS} shows an error (\emph{omission probability}) can be returned \emph{any-time} during verification~\cite{mancini-etal:2016:micpro}.
			This allows the user to halt the verification process when the residual probability of an error goes below a given threshold (graceful degradation).
			Similar advantages can be achieved when the environment model yields a too large or an infinite number of scenarios, which would hinder \Revision<exhaustive verification>{the possibility to verify the system on all of them}. 
			In such cases, statistical model checking can be exploited by randomly sampling a finite number of scenarios from the environment model, and a statistically-sound degree of assurance that the property under verification holds, or a statistically sound estimation of the value of some \Acp{KPI} can be generated at the end of the (finite) verification process.

			\emph{The second major source of complexity} to deal with when performing simulation-based \Ac{SLFV} of \Acp{CPS} is carrying out the \emph{actual simulation} of the system model on all the selected scenarios (regardless on how they are selected).
			This is because, to achieve a high-enough level of assurance on its correctness, the system must be typically simulated on a very large number of scenarios (\eg in our case studies we tackle verification processes on up to \DataGet<experiments/results/stats>{operational scenarios/max/prose} scenarios), yielding \emph{prohibitive} simulation times.
			\Revision[focus]{%
				Tackling this last issue is the main focus of this article.
			}
		\end{Section}

		\begin{Section}[sec:intro:contributions]{Contributions}
			\Revision[focus:2]{
				We present an approach to compute \emph{optimised} simulation campaigns to perform \Ac{SLFV} of \Acp{CPS} in a \emph{highly parallel} environment (\eg a large \Ac[, ]{HPC}[,] infrastructure), given identical simulators of the \Ac{CPS} model and a (possibly large yet finite) collection of operational scenarios.
			}
			Our contributions are as follows.

			\SmallHeader
				{Shortest simulation campaigns for 
					highly-parallel \texorpdfstring{\Ac{CPS}}{CPS} verification}
				We present an algorithm that, given as input a (typically very large) set of operational scenarios (either generated from a high-level environment model or extracted from a database), computes a set of \emph{optimised simulation campaigns} out of them, which drive multiple simulators of the \Ac{CPS} (running in parallel) through all (and only) such scenarios in the (possibly random) order chosen by the user, while aiming at \emph{minimising} the verification \emph{completion time}.

				Since the parallel execution of the computed simulation campaigns requires no inter-process communication, very large \Ac{HPC} infrastructures can be seamlessly exploited to greatly shorten the overall verification activity.

			\SmallHeader
				{Case studies}
				We show the applicability of our algorithm on three case studies of industry-scale \Acp{CPS}, by performing their verification against \emph{very large} sets of scenarios (up to \DataGet{experiments/results/stats/operational scenarios/max/prose} scenarios), using up to (virtually) \DataGet{experiments/results/stats/computational cores/max} cores of a \Ac{HPC} infrastructure (that is \DataGet{experiments/results/stats/machines/max} \DataGet{experiments/results/stats/machines/cores}-core machines), and evaluate the benefits of our optimised simulation campaign computation algorithm and the scalability of our overall approach.

				Thanks to our overall architecture which envisions a \emph{simulator-independent} campaign computation algorithm and \emph{simulator-specific drivers}, we can virtually control any available simulation engine. 
				We have currently developed (and successfully used in our experiments) drivers for the widely popular simulation platforms Simulink and JModelica/\Ac{FMU}.
		\end{Section}
	\end{Section} 	

	\begin{Section}
		[sec:framework]
		{Formal Framework}
		%{20-framework/content.tex}

		We denote with \Reals, \RealsGEZ and \RealsGZ the sets of, \respectively, all, non-negative, and strictly positive real numbers, and with $\IntegersGZ$ and $\IntegersGEZ$ the sets of, \respectively, strictly positive and non-negative integer numbers. 
		Also, given two sets $A$ and $B$, we denote by $A^B$ the set of functions from $B$ to $A$.

		We now briefly describe how we model our \Ac{SUV}, its operational environment, and the property to be verified. 
		For brevity, formal definitions \Revision{(including notation recap)} and statements as well as their proofs are delayed to 
			%\cref{app:sec:framework}.
			Appendix~A.

		\SmallHeader{\Acf{SUV}} 
			%\Input{suv.tex}
			We assume our (black-box) \Ac{SUV} $\SUV$ to be a deterministic, time-invariant, causal, state-input-output dynamical system (\eg \cite{sontag:1998:book,mancini-etal:2015:gandalf,mancini-etal:2017:ipl}) over a continuous or discrete time set $\TimeSet$ (hence $\TimeSet$ is either $\IntegersGEZ$ or $\RealsGEZ$, or an interval thereof), and whose input space $\InputSet$ defines the set of possible values for the user inputs and the other uncontrollable events $\SUV$ is subject to, \eg faults in sensors and actuators or changes in system parameters.
			Thus, $\SUV$ takes as input a time function $\InputFunction \in \InputFunctionSet$, defining the \Ac{SUV} input values at all time points ($\InputFunction$ is called an \emph{operational scenario}, or just \emph{scenario}).

		\SmallHeader{Property to be verified} 
			%\Input{property.tex}
			\Revision[monitor]{%
				For maximum generality, we assume that the property to be verified and/or the \Acp{KPI} to be computed under each scenario are encoded as a monitor within $\SUV$.
				The monitor observes the state of the system and checks whether the property under verification is satisfied and/or computes the values of the \Acp{KPI} of interest.
			}
			\Revision{%
				The use of monitors as black boxes gives us maximum flexibility and it allows us to abstract away the actual formalism used to define the property (\eg it is immediate to 
				define as monitors bounded safety and bounded liveness properties as well as checkers for formulas in any temporal logic; see, \eg \cite{maler-etal:2004:monitoring,rajhans-2021:specification} and citations thereof).
				Since the monitor output is all we need to carry out our verification task, in the sequel we assume that the only outputs of the \Ac{SUV} are those from the monitor.
			}

		\SmallHeader{\Acf{SLFV}} 
			%\Input{slfv.tex}
			An \emph{\Ac{SLFV} problem} is a pair $\SLFV = \SLFVdef$, where $\SUV$ is a \Ac{SUV} (with an embedded monitor) and $\TraceSet$ is a set of scenarios for it.
			The \emph{answer to \Ac{SLFV} problem} $\SLFV$ is the collection of the outputs of the \Ac{SUV} (monitor) produced at the end of each \Revision{scenario} $\SUVinputFunction \in \TraceSet$, where $\SUVinputFunction$ is injected in $\SUV$ starting from its initial state.

			Our definition of (answer to an) \Ac{SLFV} problem is very general and, depending on how the \Ac{SUV} monitor is defined, seamlessly accounts for verification activities aimed at either checking whether a scenario in $\TraceSet$ exists which raises an error in the \Ac{SUV}\Revision{ (\emph{error scenario})}, or at computing \emph{statistics} on (\eg expected values of) some \Acp{KPI}.
			Namely, to find an error scenario, it is enough to define the monitor to return \Pass or \Fail at the end of each of them, depending on whether the scenario satisfies or violates the property under verification.
			Conversely, to compute any sort of statistics on any \Acp{KPI} of interest, it is enough to define the monitor to compute and output such \Ac{KPI} values at the end of each scenario.

		\SmallHeader{%
				\texorpdfstring
					{\Acs{SUV}}
					{SUV}
				operational environment%
			}
			%\Input{environment/content.tex}
			According to our focus on verification tasks where numerical simulation is the only means to get the trajectory of the \Ac{SUV} when fed with an input scenario, we will assume that the set $\TraceSet$ is finite and finitely representable, and that each scenario is time bounded.
			Hence, we assume that the set of values taken by input scenarios in $\TraceSet$ (actually, for simplicity, the set $\InputSet$ itself) is finite (and, without loss of generality, ordered) and scenarios in $\TraceSet$ are defined via \emph{piecewise constant} input time functions having discontinuities at time points multiple of a given (arbitrarily small) \emph{time quantum} $\TimeQuantum \in \TimeSet \setminus \Set{0}$. Such scenarios can be conveniently represented as \emph{\TraceProse+} (\cref{def:trace}).

			\begin{definition}[\TraceProse^]
			\label{def:trace}
				%\Input{def_trace_informal.tex}
				An \TraceProse $\Trace$ with values in $\InputSet$ is a finite sequence $\TraceDef$ where all %, for each $i \in [0, \Horizon-1]$, 
				$\InputValue[_i]$ belong to $\InputSet$. Value $\Horizon$ %\in \IntegersGZ$ 
				is the trace \emph{horizon}.
			\end{definition}

			Given time quantum $\TimeQuantum \in \TimeSet \setminus \Set{0}$,
			an \TraceProse $\Trace = \TraceDef$ is interpreted as the bounded-horizon piecewise constant time function $\InputFunction \in \InputFunctionSet[timeset={\RealRange*{0}{\TimeQuantum \Horizon}}]$ defined as
				$\InputFunction(\TimePoint) =
					\InputValue[_{\Floor{\frac{\TimePoint}{\TimeQuantum}}}]$
					for $\TimePoint \in \RealRange*{0}{\TimeHorizonDef}$.
			From now on we assume that a time quantum $\TimeQuantum$ is given, and thus interchangeably refer to input traces and to their uniquely associated time-bounded piecewise constant time functions.

			Our assumptions above naturally apply to scenarios whose values denote \emph{events} such as user requests or faults.
			However, scenarios encoding input time functions assuming continuous values can be tackled by means of a suitable discretisation of their domains, whilst smooth continuous-time input functions (\eg additive noise signals) can be managed as long as they can be cast into (or suitably approximated by) finitely parametrisable functions, in which case the input space actually defines such a (discrete or discretised) parameter space. 
			Examples of finite parameterisations of the \Ac{SUV} input space are those defining limited, quantised Taylor expansions of continuous-time inputs, or those defining quantised values for the first coefficients (those carrying out the most information) of the Fourier series of a finite-bandwidth noise, see, \eg \cite{abbas-etal:2013:tecs,mancini-etal:2021:jlamp}. 

			As argued in \cite{mancini-etal:2021:tse-supervisory}, our assumptions are in line with an engineering (rather than purely mathematical) point of view, where man-made \Acp{CPS} need to satisfy the properties under verification with some degree of \emph{robustness} with respect to the actual input time functions (see, \eg 
			\cite{fainekos-etal:2009:robustness,abbas-etal:2013:tecs} 
			and references thereof).
			Our case studies in \cref{sec:expres} contain uses of several of such features, and show that our setting can be easily met in practice.
	\end{Section}	

	\begin{Section}
		[sec:simulator]
		{\Acs{SUV} simulators}
		%{30-simulator/content.tex}
		In our simulation-based setting, we aim at performing \Ac{SLFV} of our \Ac{SUV} by driving the execution of a \emph{simulator} of the \Ac{SUV} model (in \eg Simulink, Modelica) via the simulation engine scripting language, which also takes care of injecting piecewise constant input time functions representing scenarios.
		By extending the formal notion of \emph{\Ac{SUV} simulator} in~\cite{mancini-etal:2015:gandalf,mancini-etal:2021:jlamp}, we provide a general mathematical framework that allows us to link scenarios given as input to a \Ac{SUV} $\SUV$ (as input traces encoding piecewise-constant input time functions) to inputs for a simulator of $\SUV$ (\emph{simulation campaigns}). 

		Formal definitions as well as statements and their proofs in this section are delayed to 
			%\cref{app:sec:simulator}.
			Appendix~B.

		A simulator for \Ac{SUV} \SUV is a tuple $\SIMdef$, where $\SIMstateSet$ denotes the set of simulator \emph{states}. Each $\SIMstate \in \SIMstateSet$ has the form $\SIMstate = \SIMstateDef$, where: $\SUVstate$ is a state of \SUV; 
		$\InputFunction$ is an input time function (an input trace in our setting) for \SUV; 
		$\SIMmemory$ (simulator \emph{memory}) is a finite map whose elements are of the form: $\SIMmemoryTuple[][']$, with $\SIMstateID \in \SIMstateIDset$ being an identifier from a countable set (\emph{unique} in $\SIMmemory$), $\SUVstate[']$ a state of \SUV, and $\InputFunction[']$ is an input trace. 

		A simulator \SIM can take the following \emph{commands}:
			\SIMout, which reads the output of \SIM in the current state;
			\SIMload(\SIMstateID), which loads from memory the state associated to identifier $\SIMstateID$ and makes it the current simulator state (and raises an error if such a state is not in memory);
			\SIMstore(\SIMstateID), which stores into memory the current simulator state under identifier $\SIMstateID$ (and raises an error if $\SIMstateID$ already occurs in memory);
			\SIMfree(\SIMstateID), which frees simulator memory entry $\SIMstateID$ (and raises an error if no such entry exists);
			\SIMrun(\SUVinputValue, \TimePoint), which injects input $\SUVinputValue$ and advances simulation by time $\TimePoint \in \TimeSet$.
		The \emph{time advancement} due to a command is the time simulated by $\SIM$ when executing it, and is $\TimePoint$ for $\SIMrun(\SUVinputValue, \TimePoint)$ and $0$ for all the other commands.

		A \emph{simulation campaign} $\SIMcampaign$ for $\SIM$ is a sequence $\SIMcampaignDef$ of simulator commands (with their arguments).
		To $\SIMcampaign$ we can univocally associate: 
		(i)~the \emph{sequence of states} traversed by the simulator while executing it;
		(ii)~the \emph{length} $\SIMcampaignLen(\SIMcampaign)$, which is the sum of the time advancements of its commands;
		(iii)~the \emph{required simulator memory} $\SIMcampaignMaxMem(\SIMcampaign)$, which is the maximum number of entries in the simulator memory among the traversed states;
		(iv)~the \emph{output sequence}, which is the sequence of the results of its \SIMout commands (\ie the outputs of \SIM in the states where an \SIMout command is issued).
		A simulation campaign is \emph{executable} if it does not raise errors.

		\Cref{thm:simulation-campaign:input-functions} links inputs to a simulator $\SIM$ for $\SUV$ (\ie simulation campaigns) to inputs for $\SUV$ (input time functions), and lays the foundations to our \Ac{SLFV} approach, ensuring that we can carry out a verification activity on $\SUV$ by properly driving its simulator $\SIM$. This will be the focus of \cref{sec:simcampaigns}.

		\begin{proposition}
			\label{thm:simulation-campaign:input-functions}
			%\Input{thm-input-functions_stmt_informal.tex}
			Let
				$\SIM$ be a simulator for $\SUV$,
				$\SIMcampaign$ an executable simulation campaign for $\SIM$,
				and $\SIMseqStates$ the associated sequence of simulator states.
			For each $i \in [0, \SIMnbCampaignCmds]$, the input time function $\InputFunction[_i]$ in $\SIMstate[_i] = \SIMstateDef[_i]$ 
			is such to drive $\SUV$ from its initial state to $\SUVstate[_i]$.
		\end{proposition}
	\end{Section}

	\begin{Section}
		[sec:simcampaigns]
		{Simulation-based \Acs{SLFV}}
		%{40-simulation-based-slfv/content.tex}
		To perform simulation-based \Ac{SLFV} of $\SUV$ over $\NbTraces$ input traces $\TraceSet$ we need a simulator $\SIM = \SIMdef$ for $\SUV$ and an executable \SIMcampaignProse $\SIMcampaign$ for $\SIM$ that somewhat drives $\SIM$ along the scenarios for $\SUV$ encoded by traces of $\TraceSet$ and collects the simulator outputs at the end of each scenario.

		To this end, \cref{def:simulation-campaign:input-functions} allows us to associate to any executable \SIMcampaignProse $\SIMcampaign$ for $\SIM$ the sequence $\SIMinputFunctionSeq$ of \Ac{SUV} scenarios (as piecewise constant input time functions) for $\SUV$ actually explored by $\SIMcampaign$.
		Full definitions and statements in this section as well as their proofs are delayed to 
			%\cref{app:sec:simcampaigns}.
			Appendix~C.

		\begin{definition}[Sequence of input time functions associated to a \SIMcampaignProse]
			\label{def:simulation-campaign:input-functions}
			%\Input{def_sim_campaign_input_function_informal.tex}
			The \emph{sequence of input time functions} associated to \SIMcampaignProse $\SIMcampaign$ containing $\NbTraces$ \SIMout commands is $\SIMinputFunctionSeq = \SIMinputFunctionSeqDef$, where $\InputFunction[_{\SIMinputFunctionIndex_{i}}]$ is the input time function associated to the state where the simulator executes the $i$-th \SIMout command of $\SIMcampaign$.
		\end{definition}

		\Cref{def:simulation-campaign:slfv} formalises the notion of a \SIMcampaignProse aimed at computing the answer to an \Ac{SLFV} problem.

		\begin{definition}[\SIMcampaignProse^ for an \Ac{SLFV} problem]
		\label{def:simulation-campaign:slfv}
			%\Input{def_sim_campaign_for_slfv_problem.tex}
			A \SIMcampaignProse $\SIMcampaign$ \emph{for} \Ac{SLFV} problem $\SLFV = \SLFVdef$ is an executable campaign for a simulator \SIM of \SUV, such that the sequence $\SIMinputFunctionSeq(\SIMcampaign) = \SIMinputFunctionSeqDef$ of its associated input time functions is a \emph{permutation} of $\TraceSet$.
		\end{definition}

		A \SIMcampaignProse $\SIMcampaign$ for 
		$\SLFV = \SLFVdef$ 
		can be used to compute the answer to $\SLFV$ by executing $\SIMcampaign$ on a simulator \SIM for \SUV and by collecting the simulator outputs during $\SIMcampaign$.
		If $\SLFV$ aims at finding scenarios witnessing a property violation, the input function associated to any simulator state whose output is $\Fail$ constitutes such an error scenario. 
		Conversely, \Revision<is>{if} $\SLFV$ amounts to compute statistics on some \Acp{KPI} of interest, the \Ac{KPI} values returned as the outputs of $\SIMcampaign$ (at the end of each simulated scenario) can be used to build such statistics.

		\begin{Section}
			[sec:simcampaigns:shortest]
			{Shortest \SIMcampaignProse+}
			%{shortest/content.tex}
				Among all the simulation campaigns for a given \Ac{SLFV} problem, the shortest campaigns whose required simulator memory is bounded by a given constant $\SIMmemorySize \in \IntegersGZ$ (the simulator memory capacity) have a special interest (\cref{def:simulation-campaign:shortest}).

			\begin{definition}[Shortest (\SIMcampaignMaxMemProse*) \SIMcampaignProse for a \Ac{SLFV} problem]
			\label{def:simulation-campaign:shortest}
				%\Input{def_sim_campaign_shortest.tex}
				Let $\SIMmemorySize \in \IntegersGZ \cup \Set{\infty}$.
				A \emph{shortest \SIMcampaignMaxMemProse} \SIMcampaign for \SLFV is a \SIMcampaignProse for \SLFV such that $\SIMcampaignMaxMem(\SIMcampaign) \leq \SIMmemorySize$ and for which no other \SIMcampaignProse \SIMcampaign['] for \SLFV exists %and \SIM 
				such that
				$\SIMcampaignLen(\SIMcampaign[']) < \SIMcampaignLen(\SIMcampaign)$ and $\SIMcampaignMaxMem(\SIMcampaign[']) \leq \SIMmemorySize$.
				When $\SIMmemorySize = \infty$ (\ie we do not put any limitation on the required simulator memory capacity to execute \SIMcampaign), we call \SIMcampaign simply a \emph{shortest \SIMcampaignProse} for \SLFV.
			\end{definition}

			By definition, any shortest \SIMcampaignMaxMemProse for \SLFV is not shorter than any shortest \SIMcampaignMaxMemProse[(\SIMmemorySize+1)] for \SLFV. 
			Also, any shortest \SIMcampaignMaxMemProse[\infty] for \SLFV would actually require only a finite simulator memory capacity, which is upper bounded by the number \SIMmemorySizeOpt of the distinct longest sequences of disturbances occurring as prefixes of multiple traces of \TraceSet (\Acp[, ]{LSP}[], see 
				%forthcoming \cref{def:bt}%
				Definition~13 in Appendix~D.1%
			). 
			Hence, any shortest \SIMcampaignMaxMemProse[\infty] for \SLFV would actually be a \SIMcampaignMaxMemProse[\SIMmemorySizeOpt].

			Computation of shortest \SIMcampaignProse+ can be pursued by recalling that our \Ac{SUV} \SUV is deterministic and needs to be simulated, for each scenario (\TraceProse), starting from its initial state \SUVinitState. 
			Hence, if two \TraceProse+ $\Trace[_a], \Trace[_b] \in \TraceSet$ have a common prefix, the \Ac{SUV} state at the end of such a prefix may be stored during simulation of the first simulated trace (\eg $\Trace[_b]$) and loaded back before simulating the other (\eg $\Trace[_a]$), whose inputs could then be injected from that point on only. 
			This avoids repeated simulation of the common prefix.

			This form of compression is particularly effective in practice, as the occurrence of multiple scenarios sharing a common prefix is very frequent when defining \Ac{SUV} operational environments.
			For example, in our case studies, the shortest \SIMcampaignProse+, as computed by our algorithm, are shorter than na{\"i}ve ones by a factor of 5 to 8. This translates in similar speed-ups of the required overall simulation time (see \cref{sec:expres}).
		\end{Section}

		\begin{Section}
			[sec:simcampaigns:randomised]
			{Randomised \SIMcampaignProse+}
			%{randomised/content.tex}
			\Cref{thm:simulation-campaigns:shortest-any-order} states that, if we put no limitation on the required memory capacity, a shortest \SIMcampaignProse exists for \emph{any} ordering of the scenarios of the \Ac{SLFV} problem at hand.

			\begin{proposition}
				\label{thm:simulation-campaigns:shortest-any-order}
					%\Input{thm-simulation-campaigns-shortest-any-order/content_stmt.tex}
					Let $\SLFV = \SLFVdef$ be an \Ac{SLFV} problem ($\SetCardinality{\TraceSet} = \NbTraces$) and $\SIM$ be a simulator for $\SUV$.
				For any permutation $\TraceSetDef[\SIMinputFunctionIndex]$ of \TraceProse+ of $\TraceSet$, 
				there exists an executable shortest \SIMcampaignProse $\SIMcampaign$ for $\SLFV$ on $\SIM$, such that
				$\SIMinputFunctionSeq(\SIMcampaign) = 
					\TraceInputFunction[_{\SIMinputFunctionIndex_0}], 
					%\TraceInputFunction[_{\SIMinputFunctionIndex_1}], 
					\ldots, 
					\TraceInputFunction[_{\SIMinputFunctionIndex_{\NbTraces - 1}}]$. 
			\end{proposition}

			However, the choice of the scenario verification order is an important issue.
			For example, as long as this order is deterministic, no partial conclusion can be drawn, during simulation, about the absence of error scenarios.
			This is because in a verification setting we need to adopt an adversarial model in which the adversary will place the single error scenario of $\TraceSet$ as the last scenario simulated by $\SIMcampaign$.
			Previous work~\cite{mancini-etal:2016:micpro,mancini-etal:2017:ipl} shows that the upfront availability of all scenarios to be verified (set \TraceSet) allows us to adopt a simple yet very effective approach to draw, at \emph{any time} during simulation, mathematically-sound partial conclusions on the probability that a property violation will be witnessed by a yet-to-be-simulated scenario. 
			The idea is to choose our scenario verification order \emph{uniformly at random} among all possible orders. With such a \emph{randomised \SIMcampaignProse}, after having verified the absence of errors on the first $j \in [0, \NbTraces]$ scenarios of \TraceSet in the generated random order (where $\NbTraces = \SetCardinality{\TraceSet}$), the probability that an error will be found in a yet-to-be-simulated scenario (\emph{omission probability}) is upper-bounded by $1 - \frac{j}{\NbTraces}$.
			With this approach, we effectively conjugate \emph{exhaustiveness} with \emph{randomness}.

			Randomising the scenario verification order is also required when approximations of statistics (\eg expected values) of \Acp{KPI} for each scenario are to be computed with guaranteed accuracy via statistical model checking.

			Efficiently computing a shortest, possibly randomised \SIMcampaignProse for our \Ac{SLFV} problem is the purpose of \cref{sec:algo}.
		\end{Section}		

		\begin{Section}
			[sec:simcampaigns:parallel]
			{Parallel \SIMcampaignProse+}
			%{parallel/content.tex}
			As anticipated in \cref{sec:intro:motivations}, a major efficiency bottleneck for simulation-based \Ac{SLFV} of industry-relevant \Acp{CPS} is simulation time.
			This is due both to the typically very large number of scenarios to simulate (\eg up to \DataGet<experiments/results/stats>{operational scenarios/max/prose} in our case studies) and to the time needed to numerically simulate the \Ac{CPS} model (our \Ac{SUV}) on each such scenario 
			(up to \DataGet<experiments/scenarios/buck>{simulation time/avg/seconds} seconds in our case studies).

			The answer to an \Ac{SLFV} problem $\SLFV = \SLFVdef$ (\ie the collection of the simulator outputs at the end of each scenario) can be computed by arbitrarily partitioning $\TraceSet$ into $\NbSlices \in \IntegersGZ$ subsets (\emph{slices}) $\TraceSet[_0], \ldots, \TraceSet[_{\NbSlices - 1}]$ (where $\NbSlices$ is the number of available computational nodes), and by computing and taking the union of the answers to the $\NbSlices$ smaller \Ac{SLFV} problems $\SLFV[_i] = \SLFVdef[_i]$, $i \in [0, \NbSlices - 1]$.
			In our simulation-based setting, this can be achieved using \NbSlices simulators 
			for \SUV running as \NbSlices independent processes (\eg in parallel in a \Ac{HPC} infrastructure) and independently driven by \NbSlices \SIMcampaignProse+ \SIMcampaign[_1], \ldots, \SIMcampaign[_{\NbSlices}], 
			where, for all $i$, \SIMcampaign[_i] is a \SIMcampaignProse for \SLFV[_i]. 
			\Cref{def:simulation-campaign:slfv:parallel} formalises this concept.

			\begin{definition}[\protect{\SIMcampaignParallelProse^()} for an \Ac{SLFV} problem]
			\label{def:simulation-campaign:slfv:parallel}
				%\Input{def_sim_campaign_parallel.tex}
				A \emph{\SIMcampaignParallelProse} for \Ac{SLFV} problem $\SLFV = \SLFVdef$ is a tuple 
					$\SIMcampaignParallel = \SIMcampaignParallelDef$
				%$\NbSlices \in \IntegersGZ$,	
				%
				such that there exists a partition of \TraceSet into sets
					$\TraceSet[_0], \ldots, \TraceSet[_{\NbSlices - 1}]$ 
				such that, 
				for all $i$, %\in [0, \NbSlices - 1]$, 
				\SIMcampaign[_i] is a \SIMcampaignProse for $\SLFV[_i] = \SLFVdef[_i]$.

				The \emph{length} of \SIMcampaign is $\SIMcampaignLen(\SIMcampaign) = \max_{i=0}^{\NbSlices - 1} \SIMcampaignLen(\SIMcampaign[_i])$.
				Given $\SIMmemorySize \in \IntegersGZ \cup \Set{\infty}$, \SIMcampaignParallel is a \emph{\SIMcampaignParallelMaxMemProse} if all \SIMcampaign[_i]{}s are \SIMcampaignMaxMemProse+.
			\end{definition}

			The concepts of \emph{shortest} and shortest \SIMcampaignMaxMemProse are straightforwardly extended to \SIMcampaignParallelProse()+.

			As shown in \cite{mancini-etal:2016:micpro}, when the \Ac{SLFV} activity seeks to certify absence of error scenarios, if all \SIMcampaign[_i]{}s of a \SIMcampaignParallelProse() $\SIMcampaignParallel = \SIMcampaignParallelDef$ are \emph{randomised} (\ie each \SIMcampaign[_i]{} implements a verification order of the scenarios in \TraceSet[_i] chosen independently and uniformly at random among all possible orders), then, at any time during the parallel simulation-based \Ac{SLFV} activity, where $\SIMcampaign[_i]$ has verified the absence of errors on the first $j_i \in [0, \NbTraces[_i]]$ scenarios of \TraceSet[_i] in the generated random order (where $\NbTraces[_i] = \SetCardinality{\TraceSet[_i]}$), the omission probability (\ie the probability that an error will be found in a yet-to-be-simulated scenario) is upper-bounded by $1 - \min_{i=0}^{\NbSlices - 1}\left(\frac{j_i}{\NbTraces[_i]}\right)$.
		\end{Section}
	\end{Section}

	\begin{Section}
		[sec:algo]
		{Parallel computation of \SIMcampaignParallelProse()+}
		%{50-algorithm/content.tex}
		\AcronymReset{LSP}%
		We are now ready to present our algorithm to compute a 
		\SIMcampaignParallelProse() $\SIMcampaignParallel = \SIMcampaignParallelDef$ 
		for the \Ac{SLFV} problem $\SLFV = \SLFVdef$ at hand.
		The computed $\SIMcampaignParallel$ can be executed on \NbSlices simulators for \SUV running independently on \NbSlices nodes of a \Ac{HPC} infrastructure.

		Full definitions, additional pseudocode and its description, as well as proofs of statements in this section are delayed to 
			%\cref{app:sec:algo}.
			Appendix~D.

		\begin{Section}[sec:algo:input]{Input}
			%{10-input/content.tex}
			Our algorithm takes as input a collection \TraceSet of $\NbTraces \in \IntegersGZ$ \TraceProse+ (encoding the scenarios on which the \Ac{SUV} must be verified) and 
			the memory capacity $\SIMmemorySize \in \IntegersGZ$ of each of the \NbSlices simulators 
			in terms of the maximum number of states that each simulator can keep simultaneously stored.

			\TraceProse^+ are given either explicitly in the form of a database in mass memory, or symbolically, by means of a \emph{scenario generator}, as designed in \cite{mancini-etal:2021:tse-supervisory}. 
			In particular, a scenario generator $\SG$ is a symbolic data structure built from a set of requirements (or \emph{constraints}), in turn defined by means of multiple automata (\emph{monitors}). From $\SG$, \TraceProse+ of any horizon satisfying those requirements can be efficiently extracted from their unique indices. 
			Namely, a scenario generator $\SG$ offers two main functions: $\SGNbTraces()$ and $\SGTrace()$.
			Function $\SGNbTraces(\SG)$ returns the number $\NbTraces$ of \TraceProse+ entailed by $\SG$, while, for $j \in [0, \NbTraces - 1]$, $\SGTrace(\SG, j)$ extracts the $j$-th trace (in \emph{lexicographic order}) from $\SG$.
			When the set of scenarios is given via a scenario generator, the \TraceProse+ 
			are given as a \emph{set of integers} $\TraceIndexSet$ 
			representing unique indices of traces to extracted from $\SG$. In other words, when a scenario generator are involved, our set of \TraceProse+ is defined as $\TraceSet = \Set{ \SGTrace(\SG, j) \mid j \in \TraceIndexSet }$.
		\end{Section}

		\begin{Section}[sec:algo:slicing]{Enabling parallelism}
			%{20-slicing/content.tex}
			The typically very large number of \TraceProse+ implies that $\TraceSet$ cannot be represented explicitly in central memory, and any form of global optimisation to find a shortest \SIMcampaignParallelMaxMemProse() would be unviable.
			Hence, our algorithm makes wise use of the available RAM and parallel computational nodes, and exploits suitable heuristics in order to compute an as-short-as-possible randomised \SIMcampaignMaxMemProse.
			However, when the available capacity \SIMmemorySize of each simulator memory is above a certain threshold which depends on \TraceSet, the algorithm will indeed compute \NbSlices \emph{provably shortest \SIMcampaignMaxMemProse+} \SIMcampaignParallelDef* (\cref{thm:correctness}).

			Computing an as-short-as-possible \SIMcampaignParallelMaxMemProse() needs to heavily exploit the presence of multiple traces sharing a prefix. 
			Hence, when splitting $\TraceSet$ into slices, 
			it is important to 
			keep as much as possible in the same slice traces sharing long common prefixes.

			To this end, our algorithm works best when \TraceProse+ can be accessed in lexicographic order (according to the total order defined over the \Ac{SUV} input space $\InputSet$), since in this case it can easily keep in the same slice traces that are close together according to the lexicographic order.

			Accessing traces in lexicographic order is immediate when they are extracted from a scenario generator, since it would be enough to access them in ascending order of their indices.
			Hence, in this case slicing is performed by simply partitioning of the set of indices $\TraceIndexSet$ of the traces selected for \Ac{SLFV} into $\NbSlices$ evenly-long sequences $\TraceIndexSet[_0], \ldots, \TraceIndexSet[_{\NbSlices - 1}]$, where each such sequence defines trace indices in ascending order and, for each $i > 0$ the trace indices in the $i$-th slice are all larger than those in the $(i-1)$-th slice.
			The $i$-th slice of traces would then simply be: $\TraceSet[_i] = \Set{ \SGTrace(\SG, j) \mid j \in \TraceIndexSet[_i] }$, $i \in [0, \NbSlices - 1]$.

			Conversely, when \TraceProse+ are extracted from a database, standard mass-memory sorting algorithms are exploited to reorder them lexicographically.
			Even when the number of traces is very large, 
			such mass-memory sorting algorithms offer good scalability and can be effectively used for this purpose.
			In particular, as shown in \cref{sec:expres:results}, the advantages (in terms of savings in the simulation time) achieved by performing \Ac{SLFV} using optimised simulation campaigns heavily outperform the additional cost of ordering them if needed, and this justifies investing computation time in such a preprocessing.

			For each slice, a desired, possibly randomised, verification order can be easily defined by the user. 
			For example, a uniformly random verification order can be computed by computing a random permutation of trace indices (when traces are extracted from a scenario generator) or of their keys (when pre-sorted in mass-memory databases, see, \eg \cite{mancini-etal:2016:micpro}).
		\end{Section}

		\begin{Section}[sec:algo:optimiser]
			{Computing a \SIMcampaignProse from each slice}
			%{30-optimiser/content.tex}
			From this point on, computation of the \SIMcampaignParallelProse() $\SIMcampaignParallel = \SIMcampaignParallelDef$ proceeds \emph{embarrassingly in parallel}, using up to \NbSlices \emph{independent} computational nodes, one for each slice.
			\let\tmpSIM\SIM%
			\def\SIM{\tmpSIM[_i]}%
			\let\tmpOPTstateID\OPTstateID%
			\def\OPTstateID{\tmpOPTstateID[_{i}]}%
			\let\tmpSIMcampaign\SIMcampaign%
			\def\SIMcampaign{\tmpSIMcampaign[_i]}%
			\let\tmpWholeTraceSet\TraceSet%
			\def\TraceSet{\tmpWholeTraceSet[_i]}%
			\let\tmpWholeTraceSetLabelled\TraceSetLabelled%
			\def\TraceSetLabelled{\tmpWholeTraceSetLabelled[_i]}%
			\let\tmpWholeTraceSetLabelledRnd\TraceSetLabelledRnd%
			\def\TraceSetLabelledRnd{\tmpWholeTraceSetLabelledRnd[i]}%
			Our algorithm to compute a 
			\SIMcampaignProse for a single slice \TraceSet 
			is sketched as \cref{algo:optimiser}.

			\begin{algorithm}
				%\Input{pseudocode/main.tex}
				\KwIn{\mbox{$\TraceSet$, slice of traces in desired (\eg random) order}}
				
				\KwIn{$\TimeQuantum \in \TimeSet$, time quantum}
				
				\KwIn{$\SIMmemorySize \in \IntegersGZ$, the simulator memory capacity}

				\KwOut{\SIMcampaign, the output simulation campaign}
				
				\BlankLine
				\AlgoVarGets
					{$\SIMcampaign$}
					{an empty sequence of commands}%
					\;

				\AlgoVarGets
					{\BT} 
					{\OPTbuildBT(\TraceSet)}%
					\tcc*{build \Acl{BT}}%

				\AlgoVarGets{$j$}{0}\tcc*{trace counter} 

				\ForEach{$\Trace \in \TraceSet$ (in the given order)}{%
					\nllabel{algo:optimiser:line:foreachTrace}	
					append \OPTsimCmds(%
						\Trace, 
						$j$,
						\BT%
						)%
					to \SIMcampaign
					;
					\AlgoVarIncr{$j$}\;
				}
				\Return \SIMcampaign\;
				
				\caption{\SIMcampaignProse^ computation for a slice.}
				\label{algo:optimiser}	
			\end{algorithm}

			\begin{Section}[sec:algo:bt]{\Acl{BT}}
				%{bt/content.tex} 
				\AcronymReset{LSP}%
				The first step of \cref{algo:optimiser} (function \OPTbuildBT()) is to build a data structure called \emph{\Acf{BT}}, representing the longest prefixes shared by multiple traces.

				In the following, given two (possibly empty) sequences of inputs $\Trace[_a]$ and $\Trace[_b]$ (\ie sequences of values of $\InputSet$), we denote by $\Trace[_a] \IsPrefixOf \Trace[_b]$ (\respectively, $\Trace[_a] \IsProperPrefixOf \Trace[_b]$) the fact that $\Trace[_a]$ is a prefix (\respectively, proper prefix) of $\Trace[_b]$.

				A \emph{\Ac{LSP}} for \TraceSet is a (possibly empty) sequence \Trace of inputs such that there exist two traces $\Trace[_a]$ and $\Trace[_b]$ in \TraceSet such that: 
					$\Trace \IsPrefixOf \Trace[_a]$,
					$\Trace \IsPrefixOf \Trace[_b]$,
					and there exists no $\Trace[']$ in \TraceSet such that
						$\Trace \IsProperPrefixOf \Trace[']$,
						$\Trace['] \IsPrefixOf \Trace[_a]$, and
						$\Trace['] \IsPrefixOf \Trace[_b]$.
				The intelligent storing of the states reached by the simulator after having executed such \Acp{LSP} (under the available simulator memory capacity constraints) would avoid their recomputation, thus producing shorter \SIMcampaignProse+. 

				A \emph{\Ac{BT}} for $\TraceSet$ (see 
					%\cref{app:sec:algo:optimiser:bt} 
					Appendix~D.1.1
				for formal statements, details, and pseudocode) is a tree $\BT = \BTdef$. 
				Nodes (set \BTvertexSet) denote distinct \Acp{LSP} of \TraceSet and the parent node $\BTparent(\Trace)$ of node $\Trace$ (if one exists) is such that
					$\BTparent(\Trace) \IsProperPrefixOf \Trace$, 
					and no sequence $\Trace[']$ exists as a node of $\BT$ such that $\BTparent(\Trace) \IsProperPrefixOf \Trace['] \IsProperPrefixOf \Trace$.
				The latter condition implies that a \Ac{BT} is a rooted tree.

				The \emph{\BTdepthProse} of \Ac{BT} node $\Trace = \TraceDef<d>$ is $\BTdepth(\Trace) = d$, which represents the time point $d \TimeQuantum$ reached by the simulator (starting from its initial state) after having injected input sequence $\Trace$. 
				The \BTdepthProse of the node associated to the empty sequence is zero. 
				To each node $\TraceDef<d> \in \BTvertexSet$, the number of traces in $\TraceSet$ having $\TraceDef<d>$ as a (proper or non-proper) prefix is stored as 
					$\BTntraces(\TraceDef*<d>)$.

				A \Ac{BT} $\BT$ for \TraceSet is \emph{complete} if no \Ac{BT} for \TraceSet exists whose nodes are a proper subset of those of $\BT$.
				The \emph{size} of a \Ac{BT} is the number of its nodes.

				To compute a complete \Ac{BT} for $\TraceSet$ in central memory, function \OPTbuildBT() scans $\TraceSet$ in lexicographic order, since, under this ordering, deciding which trace prefixes are nodes of the tree is straightforward and memory-efficient. 

				To keep an as small as possible RAM footprint of the \Ac{BT}, the algorithm represents in central memory each of its nodes $\TraceDef<d>$ by a unique identifier $\SIMstateID(\TraceDef*<d>)$.
				Unique identifiers for each trace prefix are available for free when traces are extracted from a scenario generator. 
				If traces are taken from a database, any efficiently computable injective function of finite sequences of input values (or even a cryptographic hash function, when the probability of conflicts is small enough) can be used.
			\end{Section}

			\begin{Section}[sec:algo:simcampaign]{Generation of \SIMcampaignProse commands}
				%{simcampaign/content.tex}
				\Cref{algo:optimiser} proceeds at generating an optimised \SIMcampaignProse \SIMcampaign which would drive simulator \SIM along all the \TraceProse+ according to the chosen (possibly random) order, still trying to save as many simulation steps as possible, compatibly with simulator memory capacity constraints.

				To this end, the \TraceProse+ \TraceSet are considered sequentially in the given order. 
				For each trace $\Trace$, function \OPTsimCmds() is invoked to append to \SIMcampaign a sequence of commands to simulate it from the \emph{best} intermediate state available in the simulator memory (see below). 
				During generation of simulator commands, for each \Ac{BT} node \SIMstateID, the algorithm keeps a boolean flag \OPTstored(\SIMstateID) (initialised to \False) whose value reflects, at any point during the computation of \SIMcampaign, the fact that state \SIMstateID would be available or not in the memory of \SIM at that point during the execution of \SIMcampaign. 
				Namely, \OPTstored(\SIMstateID) is set to \True (\respectively, \False) when issuing a \SIMstore(\SIMstateID) (\respectively, \SIMfree(\SIMstateID)) command.

				\SmallHeader{Generating trace simulation commands}
					%\Input{sim_cmds.tex}
					\cref{algo:optimiser:sim_cmds} shows the pseudocode of function \OPTsimCmds which issues the actual commands aimed at simulating trace $\Trace$, which are appended to $\SIMcampaign$.
					The function proceeds as follows:
					\begin{steps}
						\item
						Selects $\SIMstateID_{\OPTidxload}$, the state corresponding to the longest prefix of \Trace that, at the current point of the prospective simulation, would be available in the simulator memory and appends command $\SIMload(\SIMstateID_{\OPTidxload})$ to $\SIMcampaign$, to load it back.

						\item Revises the nodes of the \Ac{BT} associated to prefixes of \Trace (proceeding backwards from the full $\Trace$). 
						For each such \Ac{BT} node $\SIMstateID_{q}$, value $\BTntraces(\SIMstateID_q)$ is decremented (thus memorising the fact that such prefix will occur in one less future trace). 
						If $\BTntraces(\SIMstateID_q)$ becomes zero, the algorithm knows that the input sequence associated to $\SIMstateID_q$ will not occur as a prefix in any future trace, and removes $\SIMstateID_q$ from the \Ac{BT} (which, since prefixes of $\Trace$ are processed backwards from the entire $\Trace$, is a leaf of the \Ac{BT}). 
						Also, if $\SIMstateID_q$ is known to be stored in the simulator memory at this point of the execution of \SIMcampaign (\ie $\OPTstored(\SIMstateID_q) = \True$), it appends to \SIMcampaign command $\SIMfree(\SIMstateID_q)$ to free-up the simulator memory.

						\item 
						Appends to \SIMcampaign a \SIMrun command for each maximally long constant portion of \Trace such that no intermediate state traversed by the simulator needs to be stored to shorten simulation of future traces (\ie function \OPTisWorthStoringState() returns \False for it). 

						\item
						If the state reached by the simulator after each \SIMrun command is worth to be stored as it can shorten simulation of a later trace (this implies it is a node of the \Ac{BT}), the function proceeds at storing it (see below).
					\end{steps}

					\begin{algorithm}
						%\Input{pseudocode/sim_cmds.tex}
						\AlgoFunction{\OPTsimCmds(%
										\Trace,
										$j$,
										\BT%, 
									)%
						}{
							\KwIn{%
								$\Trace = \TraceDef$, 
									current ($j$-th) trace%
							}

							\KwIn{$\BT = \BTdef$, \Acl{BT}}

							\KwOut{sequence of sim.\ commands for $\Trace$}

							\lIf(\tcc*[f]{first trace}){$j = 0$}{%		
								\AlgoVarGets{\OPTidxload}{0}%\;
							}
							\Else(\tcc*[f]{not first trace}){%		
								\AlgoVarGets
									{\OPTidxload}
									{max $q \in [0, \Horizon]$ s.t.\\ 
										$\qquad\SIMstateID_{\OPTidxload} = \SIMstateID(\TraceDef*<q>) \in \BTvertexSet \land \OPTstored(\SIMstateID_{\OPTidxload})$}%
									\nllabel{algo:optimiser:sim_cmds:line:load:idx}%
								\;
								\KwIssue $\SIMload(\SIMstateID_{\OPTidxload})$% 
								\nllabel{algo:optimiser:sim_cmds:line:load}%
								\;%
							}
							
							\For(\tcc*[f]{revise \BT}){%
								$q$ \KwFrom $\Horizon-1$ \KwDownto $0$ s.t.\
								$\SIMstateID_{q} = \SIMstateID(\TraceDef*<q>) \in \BTvertexSet$
							}{%
								\AlgoVarDecr{$\BTntraces(\SIMstateID_{q})$}\;
								\If{%
									$\BTntraces(\SIMstateID_{q}) = 0$
								}{%
									\tcc{$\SIMstateID_{q}$ won't occur in future traces}
									\If{$\OPTstored(\SIMstateID_{q})$}{%
										\KwIssue $\SIMfree(\SIMstateID_{q})$; % to $\SIMcampaign$\;%
										\nllabel{algo:optimiser:sim_cmds:line:free}%
										\AlgoVarGets{\OPTstored($\SIMstateID_{q}$)}{\False}\;
									}			
									remove $\SIMstateID_{q}$ from \BTvertexSet{}%
									%\tcc*{$\BTchildren(\SIMstateID_{q}) = \emptyset$}
									\tcc*{$\SIMstateID_{q}$ is leaf in \BT}
								}
							}%

							\tcc{All nodes still in \BT will occur in future traces}

							\AlgoVarGets
								{start}
								{\OPTidxload}%
							\;

							\While{$\AlgoVar{start} < \Horizon$}{
								\AlgoVarGets
									{end}
									{max $e \in [\AlgoVar{start}, \Horizon-1]$ s.t.\ 
										\CompactMath{%
											$\forall q \in [\AlgoVar{start}+1, e]$
											\\ 
											\mbox{$\qquad \InputValue[_q] = \InputValue[_{q-1}] 
												\land
												\lnot \OPTisWorthStoringState(
													\SIMstateID(\InputValue[_0], \ldots, \InputValue[_{q-1}]),
													\BT%, 
												)
											$}%
										}%
									}%
								\;		
								\KwIssue \SIMrun(\InputValue[_{\AlgoVar{start}}],~(\AlgoVar{end} - \AlgoVar{start} + 1)\TimeQuantum)% 
								\nllabel{algo:optimiser:sim_cmds:line:run}%
								\;

								\AlgoVarGets
									{start}
									{$\AlgoVar{end}+1$}%
								\;
								
								\If{$\AlgoVar{start} \leq \Horizon \land 
										\SIMstateID_{\AlgoVar{start}} = \SIMstateID(\TraceDef*<\AlgoVar{start}>) \in \BTvertexSet \land$
									$\OPTisWorthStoringState(
										\SIMstateID_{\AlgoVar{start}},
										\BT 
									  )
									$%
								}{
									\OPTstoreState(%
										$\SIMstateID_{\AlgoVar{start}}$,
										\BT, 
										\SIMmemorySize,
										\SIMcampaign
									)
										\tcc*{possibly issues \SIMfree(\SIMstateID') for some $\SIMstateID'$ s.t.\ $\OPTstored(\SIMstateID')$ and sets $\OPTstored(\SIMstateID')$ to \False, before issuing $\SIMstore(\SIMstateID_{\AlgoVar{start}})$}
										\nllabel{algo:optimiser:sim_cmds:line:possibly_store}%

									\AlgoVarGets{$\OPTstored(\SIMstateID_{\AlgoVar{start}})$}{\True}\;
								}		
											
							} % end while

							\KwIssue \SIMout\; 
						} % end function

						\caption{Function \OPTsimCmds.}
						\label{algo:optimiser:sim_cmds}	
					\end{algorithm}

				\SmallHeader{Storing intermediate simulation states}
					%\Input{store_state.tex}
					Given the limited capacity \SIMmemorySize of the simulator memory, the decision of which \Ac{BT} simulator states will be actually stored must be taken wisely.
					This is charge of function \OPTisWorthStoringState(). 

					Since the \Ac{BT} has no information on the \emph{order} with which simulator states represented by \Ac{BT} nodes will occur in \TraceSet (such data would be too large to be kept in RAM), any approach to compute an optimal plan to decide which intermediate state to store and free (and when to do that during the execution of the simulation campaign) is clearly not viable. 
					Hence, the function proceeds \emph{heuristically}.

					In particular, 
						\OPTisWorthStoringState(\SIMstateID, \BT)
					works as follows.
					If $\SIMstateID$ is not a \Ac{BT} node or is expected to be already stored in memory at that point of the execution of the \SIMcampaignProse (\ie $\OPTstored(\SIMstateID) = \True$), then \OPTisWorthStoringState() returns \False;
					otherwise, if the simulator memory is expected to have room to accommodate an additional state (\ie the number of \Ac{BT} nodes $\SIMstateID'$ such that $\OPTstored(\SIMstateID') = \True$ is $< \SIMmemorySize$), then \OPTisWorthStoringState() returns \True.

					In case the simulator memory is expected to be full at that point of the execution of the simulation campaign, then the function decides whether it is best to make space for $\SIMstateID$ by freeing up another simulator state $\SIMstateID'$ already in memory, or to rather ignore the request of storing $\SIMstateID$ in the first place. 

					To this end, the function searches for a currently stored state $\SIMstateID'$ whose associated node in the \Ac{BT} is not the root node and has the smallest depth-difference \wrt its parent node, where the depth-difference of $\SIMstateID'$ is $\BTdepth(\SIMstateID') - \BTdepth(\BTparent(\SIMstateID')) > 0$.
					Since the depth-difference of a simulator state defines the additional number of \TimeQuantum-simulation steps needed by the simulator to reach that state when starting from the state represented by its parent node in the \Ac{BT}, $\SIMstateID'$ is a currently stored state which could be used to shorten simulation of a future trace, but whose removal from simulator memory minimises the number of additional $\TimeQuantum$-simulation steps needed to recompute it (from the state associated to its parent node in the \Ac{BT}).

					In case the depth-difference of $\SIMstateID'$ is less than that of $\SIMstateID$, then the function decides that it is worth removing $\SIMstateID'$ from the simulator memory to make room for $\SIMstateID$, and returns \True.
					Otherwise, the function knows that freeing-up $\SIMstateID'$ to make room for $\SIMstateID$ would cost more (in terms of additional \TimeQuantum-long simulation steps to recompute $\SIMstateID'$ from the state represented by its parent) than simply ignoring the request to store $\SIMstateID$, and returns \False.

					When \OPTisWorthStoringState() returns \True, function \OPTstoreState() appends $\SIMstore(\SIMstateID)$ to $\SIMcampaign$, preceded by $\SIMfree(\SIMstateID')$ in case \OPTisWorthStoringState() has selected $\SIMstateID'$ as the state to be freed-up (in which case $\OPTstored(\SIMstateID')$ is set to \False as well).

					In order to efficiently 
					find $\SIMstateID'$, the currently stored \Ac{BT} nodes are indexed so as to retrieve efficiently those having minimal depth-difference \wrt their parents.
			\end{Section}
		\end{Section}

		\medskip
		%\Input{40-correctness/content.tex}
		The following result holds (see 
			%\cref{app:sec:algo:correctness} 
			Appendix~D.2
		for the full statement and proof).

		\begin{proposition}[Correctness of \cref{algo:optimiser}]
			\label{thm:correctness}
			%\Input{thm-correctness_stmt_informal.tex}
			Let $\SLFV = \SLFVdef$ be an \Ac{SLFV} problem for \Ac{SUV} \SUV,
			with \TraceProse+ $\TraceSet$ being associated to time quantum $\TimeQuantum$, and let $\SIMmemorySize$ be a positive integer.
			Given any partition $\Set{\TraceSet[_0], \ldots, \TraceSet[_{\NbSlices-1}]}$ of $\TraceSet$, let $\SIMcampaignParallel = \SIMcampaignParallelDef$ be the \SIMcampaignParallelProse such that $\SIMcampaign[_i]$ is computed by \cref{algo:optimiser} on inputs $\TraceSet[_i]$ (under any user-defined order), $\TimeQuantum$, and $\SIMmemorySize$.
			We have that:
			\begin{points}
				\item \label{item:thm:correctness:sequence}
				For all $i \in [0, \NbSlices-1]$, the sequence $\SIMinputFunctionSeq(\SIMcampaign[_i])$ is $\TraceSet[_i]$;
				\item There exists $\SIMmemorySizeOpt \in \IntegersGZ$ such that, if $\SIMmemorySize \geq \SIMmemorySizeOpt$, all \SIMcampaign[_i]{}s are \emph{shortest} \SIMcampaignMaxMemProse+.
			\end{points}
		\end{proposition}

		\Cref{item:thm:correctness:sequence}
		implies that \SIMcampaignParallel is a \SIMcampaignParallelMaxMemProse for \SLFV.
		Each $\SIMcampaign[_i]$ drives an independent copy 
		of a simulator of \Ac{SUV} \SUV along the scenarios in $\TraceSet[_i]$ in the chosen, possibly random, order. 
		In the latter case, an upper bound to the omission probability can be computed at any time during parallel simulation (\cref{sec:simcampaigns:parallel}).
	\end{Section}

	\begin{Section}
		[sec:expres]
		{Implementation and experimental results}
		%{60-expres/content.tex}
		In this section we outline our implementation of the parallel 
		algorithm of \cref{sec:algo} and analyse its performance and scalability on three real case studies. 

		\begin{Section}
			[sec:expres:implementation]
			{Implementation}
			%{10-implementation/content.tex}
			We implemented our algorithm as a C-language tool which takes as input positive integers \NbSlices (number of slices) and \SIMmemorySize (memory capacity of each simulator), and the set of \TraceProse+ \TraceSet for which a parallel campaign is sought. In our experiments we extracted the set of \TraceProse+ from scenario generators, defined as discussed in \cite{mancini-etal:2021:tse-supervisory}.
			The computed campaign can be executed on $\NbSlices$ simulators for the \Ac{SUV} \SUV, running independently on $\NbSlices$ computational nodes.
			Each simulator is steered by a \emph{driver} which receives the simulation campaign as input. 
			This driver is the only simulator-dependent component of our tool pipeline.
			We implemented drivers for two popular simulators, namely: 
			Simulink and JModelica/\Ac{FMU}.
			Additional drivers can be easily written along the same lines.
		\end{Section}

		\begin{Section}
			[sec:expres:case-studies]
			{Case studies}
			%{20-case-studies/content.tex}
			We selected three industry-relevant \Ac{SUV} models defined in the language of two popular simulators, namely Simulink and Modelica.

			\begin{Section}
				[sec:expres:case-studies:buck]
				{\Acf{BDC}}
				%{buck/content.tex}
				It is a mixed-mode analog circuit converting the DC input voltage (denoted as $V_i$) to a desired DC output voltage ($V_o$), often used off-chip to scale down the typical laptop battery voltage (12--24 V) to the few volts needed by, \eg a laptop processor (the \emph{load}) as well as on-chip to support dynamic voltage and frequency scaling in multicore processors (see, \eg \cite{mari-etal:2014:tosem}).
				A \Ac{BDC} converter is self-regulating, \ie it is able to maintain the desired output voltage $V_o$ notwithstanding variations in the input voltage $V_i$ or in the load $R$.
				We used a Modelica model of the fuzzy logic--based \Ac{BDC} controller 
				of \cite{so-etal:1996:buck}, converted into an FMU 2.0 object via the JModelica extension in \cite{sinisi-etal:2021:fmu}.
			\end{Section}

			\begin{Section}
				[sec:expres:case-studies:apollo]
				{\Acf{ALMA}}
				%{apollo/content.tex}
				It is a Simulink/Stateflow model
				defining the logic that implements the phase-plane control algorithm of the autopilot of the lunar module used in the Apollo 11 mission.
				The Module is equipped with actuators (16 reaction jets to rotate the Module along the three axes) subject to temporary unavailabilities. 
				The controller takes as input requests to change the Module attitude (\ie to perform a rotation along the three axes) 
				and computes which reaction jets to fire to obey each request.
			\end{Section}

			\begin{Section}
				[sec:expres:case-studies:fcs]
				{Fault Tolerant \Acf{FCS}}
				%{fcs/content.tex}
				It is a Simulink/Stateflow model of a controller for a fault tolerant gasoline engine, which has also been used as a case study in~\cite{clarke-etal:2010:probabilistic,kim-kim:2012:hybrid,zuliani-etal:2013:fmsd,kim-etal:2013:validating,mancini-etal:2013:cav,mancini-etal:2016:micpro,mancini-etal:2016:fundam}.
				The \Ac{FCS} has four sensors 
				subject to temporary faults, and the whole control system is expected to tolerate single sensor faults.
			\end{Section}
		\end{Section}

		\begin{Section}
			[sec:expres:setting]
			{Experimental setting}
			%{30-setting/content.tex}
			We defined a scenario generator for each \Ac{SUV}, entailing \TraceProse+ (time quantum $\TimeQuantum = 1$ \Ac[, ]{tu}[]) with the properties listed in \cref{tab:exres:scenario-generators}.
			Several constraints have been enforced on the \TraceProse+. This allows us to focus the \Ac{SLFV} activities on clearly selected portions of the space of inputs and to keep the overall number of traces under control. 
			The enforced constraints are detailed in 
				%\cref{app:sec:expres:case-studies}.
				Appendix~E.
			Here, we just point out that we experimented with the optimisation of \SIMcampaignParallelProse()+ for up to 
			around 50 (\Ac{BDC}), 100 (\Ac{ALMA}) and 200
			(\Ac{FCS}) million traces.

			\begin{table}
			\begin{tabular}{ccccl}
			\toprule
				  \Acs{SUV}
				& $\SetCardinality{\InputSet}$
				& horizon
				& n.\ traces
				& constraints on traces
			\\
			\midrule
				    \Acs{BDC} 
				  & \DataGet<experiments/scenarios/buck>{input space/size}
				  & \DataGet<experiments/scenarios/buck>{horizon} \Acsp{tu}
				  & \DataGet<experiments/scenarios/buck>{number}
				  & %\cref{app:sec:expres:case-studies:buck}
				  	Appendix~E.1
			\\
				    \Acs{ALMA} 
				  & \DataGet<experiments/scenarios/apollo>{input space/size}
				  & \DataGet<experiments/scenarios/apollo>{horizon} \Acsp{tu}
				  & \DataGet<experiments/scenarios/apollo>{number}
				  & %\cref{app:sec:expres:case-studies:apollo}
				  	Appendix~E.2
			\\
				    \Acs{FCS} 
				  & \DataGet<experiments/scenarios/fcs>{input space/size}
				  & \DataGet<experiments/scenarios/fcs>{horizon} \Acsp{tu}
				  & \DataGet<experiments/scenarios/fcs>{number}
				  & %\cref{app:sec:expres:case-studies:fcs}
				  	Appendix~E.3
			\\
			\bottomrule
			\end{tabular}

			\caption{Scenario generators for our case studies.}
			\label{tab:exres:scenario-generators}
			\end{table}

			To show scalability of our algorithm when computing optimised \SIMcampaignParallelProse()+ as well as the overall savings in simulation time provided by our approach to \Ac{SLFV}, we exploited (virtually) up to \DataGet<experiments/results/stats>{machines/max} identical \DataGet<experiments/results/stats>{machines/cores}-core machines (CPU: AMD EPYC 7301, RAM: 256GB) of our \Ac{HPC} infrastructure, thus our maximum number of slices $\NbSlices$ has been set to \DataGet<experiments/results/stats>{computational cores/max}.

			Since actual simulation of the generated campaigns in all the considered settings would be prohibitively long, simulation time of each campaign has been estimated as follows. 
			For each \Ac{SUV}, we generated and actually simulated a random campaign of 100k commands, where each command (\SIMload, \SIMstore, \SIMfree, \SIMout, $\SIMrun$ for all needed durations) was evenly represented. We then computed the average time needed by the simulator to execute each single command, and used such expected values (standard deviation showed to be negligible) to estimate the completion time of each campaign.
		\end{Section}

		\begin{Section}
			[sec:expres:results]
			{Experimental results}
			%{40-results/content.tex}
			Our experimental results are summarised in \cref{fig:expres:results:buck} (\Ac{BDC}), \cref{fig:expres:results:apollo} (\Ac{ALMA}) and \cref{fig:expres:results:fcs} (\Ac{FCS}).
			We computed several \emph{randomised} \SIMcampaignParallelProse()+ for each case study, one of each of several random subsets of all the traces entailed by our scenario generator. 

			In order to show performance and effectiveness of simulation campaign optimisation in contexts ranging from statistical model checking to random exhaustive verification, we sampled trace subsets by fixing their size from 25\% to 100\% of the overall number of traces. 

			Each experiment has been repeated for various amounts of simulator memory available (1 state, meaning no optimisation at all, since only one simulator state --typically the initial state-- can be be stored and loaded back, up to $\SIMmemorySizeOpt$, the maximum number of states required in each experiment for maximum optimisation).
			Given the presence of randomisation, all experiments have been repeated with 5 different random seeds, and all results have been averaged.

			\begin{Section}
				{Scalability of the campaign computation algorithm}
				%{10-optimisation/content.tex}
				The first (left-most) column of \cref{fig:expres:results:buck,fig:expres:results:apollo,fig:expres:results:fcs} shows the time (in seconds) needed by our algorithm to compute a \SIMcampaignParallelProse() for each \Ac{SUV} and each combination of values for the number of traces (row), the number of parallel processes (slices), and the amount of simulator memory (different line shapes).

				The plots show that the computation time ranges from a few seconds to a few hours, and this time is \emph{always negligible} when compared to the time savings that such optimisation yields in terms of simulation time (see the corresponding plots on right-most column, where time is expressed in days of parallel computation).
			\end{Section}

			\begin{Section}{Campaigns efficiency \wrt parallelisation}
				%{20-efficiency-wrt-k/content.tex}
				The second column of \cref{fig:expres:results:buck,fig:expres:results:apollo,fig:expres:results:fcs} shows how efficiency of the computed campaigns is preserved when a higher number of parallel processes are expected to be used in the verification process (hence, the \TraceProse+ are split in a higher number of slices).

				Namely, for each \Ac{SUV} and each combination of values for the number of traces $\NbTraces$ (row), the number of parallel processes (slices) $\NbSlices$, and the amount of simulator memory $\SIMmemorySize$ (different line shapes), the charts plot the average value (among our randomised experiments) of the following quantity: 
				$
					\frac
						{\Fun{sim\_time}(\SIMcampaign_{\NbTraces, 1024, \SIMmemorySize}) \times 1024}
						{\Fun{sim\_time}(\SIMcampaign_{\NbTraces, \NbSlices, \SIMmemorySize}) \times \NbSlices}
				$,
				which measures, in terms of (estimated) simulation time (\Fun{sim\_time}), the efficiency of the \SIMcampaignParallelProse() $\SIMcampaign_{\NbTraces, \NbSlices, \SIMmemorySize}$ (which verifies \NbTraces random traces in parallel on $\NbSlices$ processes assuming that each simulator can keep $\SIMmemorySize$ states simultaneously stored) with respect to the corresponding \SIMcampaignParallelProse() $\SIMcampaign_{\NbTraces, 1024, \SIMmemorySize}$ (which verifies the same traces under the same assumptions regarding the simulator memory, but running on just $1024$ parallel processes, our minimum value).

				The plots show how efficiency is \emph{always very high}, and, even when it degrades to a bit less than 90\%, the induced overhead in simulation time is \emph{always negligible} when compared to the \emph{very large time savings} yielded by exploiting a higher number of parallel simulators.
			\end{Section}

			\begin{Section}{Campaigns efficiency \wrt available simulator memory}
				%{30-efficiency-wrt-m/content.tex}
				The third column of \cref{fig:expres:results:buck,fig:expres:results:apollo,fig:expres:results:fcs} shows how efficiency of the computed campaigns is preserved when reducing the memory available on each simulator.

				Namely, for each \Ac{SUV} and each combination of values for the number of traces $\NbTraces$ (row), the number of parallel processes (slices) $\NbSlices$, and for each value for the amount of simulator memory $\SIMmemorySize$ (different line shapes), the charts plot the average value of the following quantity: 
				$
					\frac
						{\Fun{sim\_time}(\SIMcampaign_{\NbTraces, \NbSlices, \SIMmemorySizeOpt})}
						{\Fun{sim\_time}(\SIMcampaign_{\NbTraces, \NbSlices, \SIMmemorySize})}
				$,
				which measures, in terms of simulation time (\Fun{sim\_time}), the efficiency of the \SIMcampaignParallelProse() $\SIMcampaign_{\NbTraces, \NbSlices, \SIMmemorySize}$ (which verifies \NbTraces random traces in parallel on $\NbSlices$ processes assuming that each simulator can keep only $\SIMmemorySize$ states simultaneously stored) with respect to the corresponding \SIMcampaignParallelProse() $\SIMcampaign_{\NbTraces, \NbSlices, \SIMmemorySizeOpt}$ (which verifies the same traces with the same number of parallel processes, but assuming maximum simulator memory, \ie $\SIMmemorySize = \SIMmemorySizeOpt$).

				The plots show how efficiency is \emph{very well preserved} when reducing the value for $\SIMmemorySize$ to up to $\SIMmemorySizeOpt \times 50\%$, unsurprisingly degrading for lower values of $\SIMmemorySize$.
				We also point out that the maximum memory required to each simulator (\ie when $\SIMmemorySize = \SIMmemorySizeOpt$) is always very limited, and easily met in practice. 
				Namely, since simulator states occupy at most a few dozens of Kilobytes, the memory requirements are always less than (upper limits reached for 1024 parallel processes/slices): 2GB for \Ac{BDC} ($\SIMmemorySizeOpt \leq \Pretty{15681}$); 4GB for \Ac{ALMA} ($\SIMmemorySizeOpt \leq \Pretty{62050}$); 8GB for \Ac{FCS} ($\SIMmemorySizeOpt \leq \Pretty{156115}$).
			\end{Section}

			\begin{Section}{Simulation speedups and time savings}
				%{40-speedup/content.tex}
				The fourth column of \cref{fig:expres:results:buck,fig:expres:results:apollo,fig:expres:results:fcs} shows the speedups in simulation time achieved by our computed optimised campaigns under different settings regarding the memory available on each simulator.

				Namely, for each \Ac{SUV} and each combination of values for the number of traces $\NbTraces$ (row), the number of parallel processes (slices) $\NbSlices$, and for each value for the amount of simulator memory $\SIMmemorySize$ (different line shapes), the charts plot the average value of the following quantity: 
				$
					\frac		
						{\Fun{sim\_time}(\SIMcampaign_{\NbTraces, \NbSlices, 1})}
						{\Fun{sim\_time}(\SIMcampaign_{\NbTraces, \NbSlices, \SIMmemorySize})}
				$,
				which measures, in terms of simulation time (\Fun{sim\_time}), the speedup of each \SIMcampaignParallelProse() $\SIMcampaign_{\NbTraces, \NbSlices, \SIMmemorySize}$ (which verifies \NbTraces random traces in parallel on $\NbSlices$ processes assuming that each simulator can keep only $\SIMmemorySize$ states simultaneously stored) with respect to the corresponding campaign $\SIMcampaign_{\NbTraces, \NbSlices, 1}$ (which verifies the same traces with the same number of parallel processes, but assuming that each simulator can keep simultaneously stored only one state, that is no optimisation at all).
				The plots show how our simulation campaign optimiser always achieves \emph{very significant speedups}, up to more than $8\times$.

				The fifth column of \cref{fig:expres:results:buck,fig:expres:results:apollo,fig:expres:results:fcs} shows how these speedups translate in \emph{huge reductions in simulation time} (in days).
				Namely, for each \Ac{SUV} and each combination of values for $\NbTraces$, $\NbSlices$, and $\SIMmemorySize$, the charts plot the average value of the overall simulation time of the \SIMcampaignParallelProse()+ $\SIMcampaign_{\NbTraces, \NbSlices, \SIMmemorySize}$, which verify the given \Ac{SUV} on \NbTraces random traces under simulator memory setting $\SIMmemorySize$.
				The plots clearly show that \emph{our simulation campaign optimiser makes practically viable} (in some days or at most weeks of parallel simulation) \emph{verification tasks that would take an inconceivable long time without optimisation} (\ie when $\SIMmemorySize = 1$).
			\end{Section}

			\begin{figure*}
				\includegraphics[width=\linewidth]{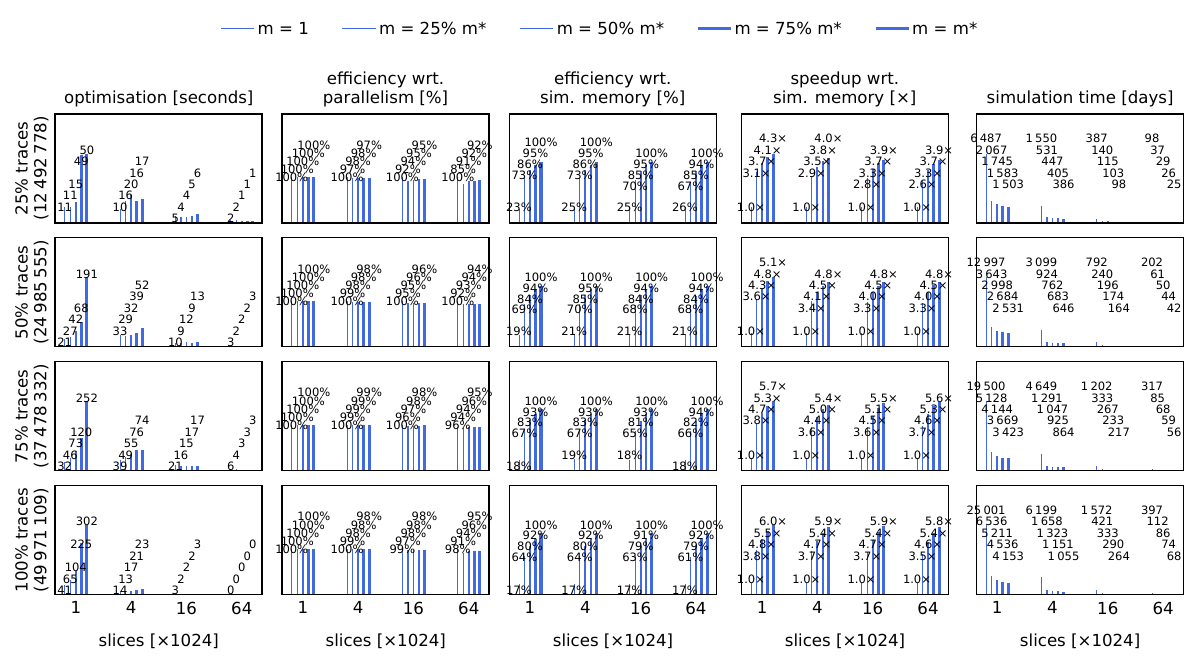}
				\caption{Experimental results: \Acf{BDC}.}
				\label{fig:expres:results:buck}
			\end{figure*}

			\begin{figure*}
				\includegraphics[width=\linewidth]{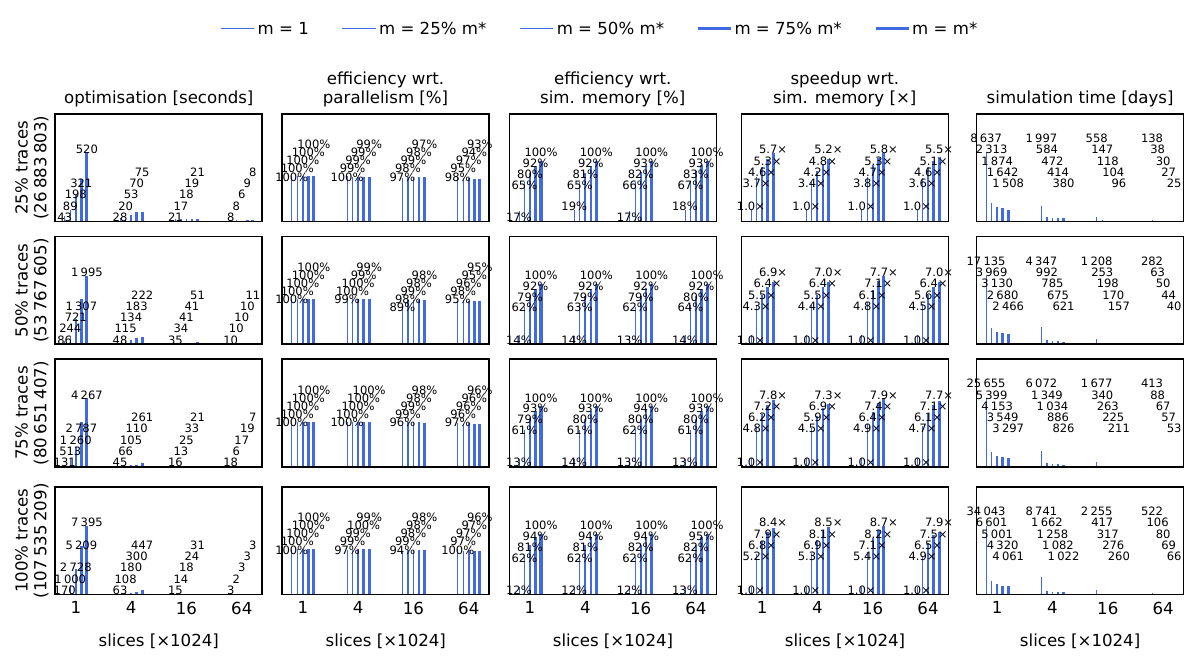}
				\caption{Experimental results: \Acf{ALMA}.}
				\label{fig:expres:results:apollo}
			\end{figure*}

			\begin{figure*}
				\includegraphics[width=\linewidth]{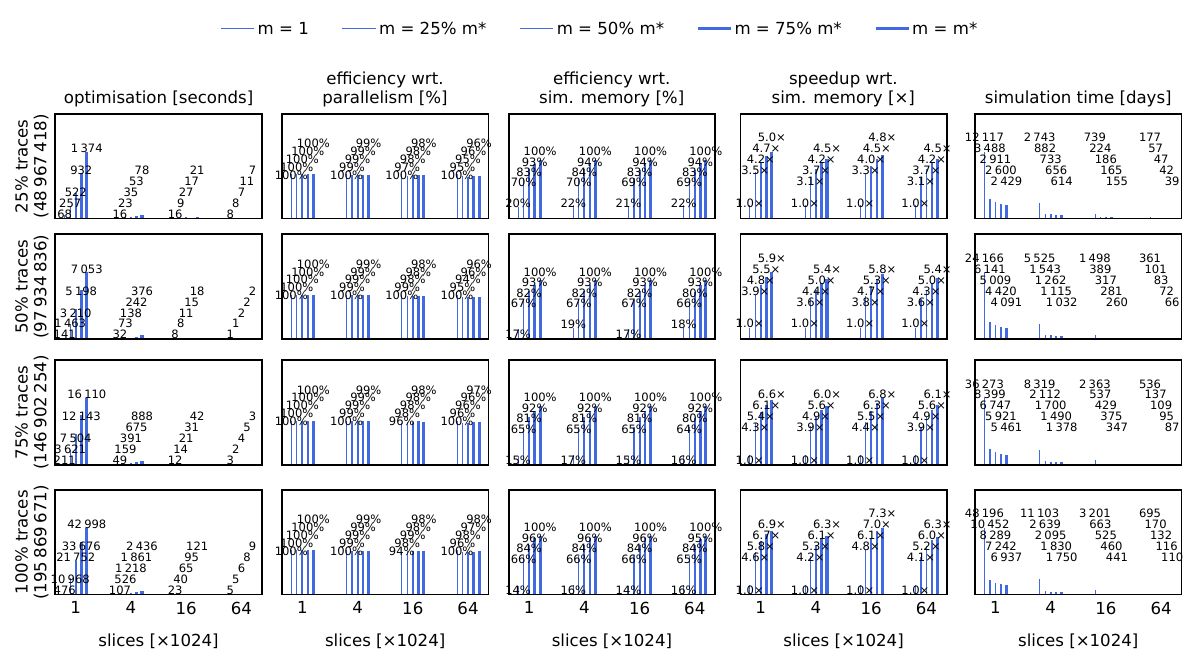}
				\caption{Experimental results: \Acf{FCS}.}
				\label{fig:expres:results:fcs}
			\end{figure*}
		\end{Section}

		\begin{Section}
			[sec:expres:limitations]
			{Limitations}
			%{50-limitations/content.tex}
			\Revision[ls-inflation]{%
				Our optimised campaigns heavily rely on storing and loading back intermediate simulator states to avoid simulating common prefixes of different traces multiple times.
				Hence, for \Ac{SUV} models exhibiting very large states (\eg those defined via partial differential equations, transport delays, or variable delay blocks), the time to execute \SIMstore and \SIMload commands may become substantial, and this raises a question on whether it would be faster to skip optimisation altogether and just run the non-optimised campaigns. 
				Here we briefly discuss this issue.

				In the case of \Ac{SUV} models showing larger states than ours, but which are also proportionally slower to advance, the speedups enabled by the campaign optimisation would be somewhat preserved.
				Thus, the problematic situations for our optimiser occur when dealing with \Ac{SUV} models whose states are larger, but whose simulation is only sub-proportionally slower to advance. 
				
				To assess to what extent our optimised campaigns still grant time savings \wrt the non-optimised campaigns, we reconsidered our experiments by \emph{artificially inflating} the duration of $\SIMstore$ and $\SIMload$ commands by a factor $f$ ranging from 1 to 100, keeping unchanged the duration of \SIMrun commands. 
				Thus, we placed ourselves in the most hostile setting, \ie the verification of variations of our \Ac{SUV} models that, although requiring the \emph{same} time to be advanced, have larger states which need \emph{$f$ times} the time need by our original \Ac{SUV} models to be stored and loaded back.

				Unsurprisingly, the speedups achieved by optimised campaigns gradually decrease when $f$ increases, but \emph{still typically grant substantial savings in simulation time}.
				For example, the speedups achieved for our case studies (100\% traces) fall to: 
					$2.2$--$2.4\times$ (\Ac{BDC}),
					$4.3$--$6.0\times$ (\Ac{ALMA}), 
					$3.7$--$5.5\times$ (\Ac{FCS})
					for $f=10$;
					$1.0$--$1.3\times$ (\Ac{BDC}), 
					$2.5$--$3.2\times$ (\Ac{ALMA}), 
					$2.5$--$3.1\times$ (\Ac{FCS}) 	
					for $f=50$;
					$0.9$--$1.0\times$ (\Ac{BDC}), 
					$1.6$--$2.4\times$ (\Ac{ALMA}), 
					$1.7$--$2.3\times$ (\Ac{FCS}) 	
					for $f=100$. 
			}
		\end{Section}
	\end{Section}	

	\begin{Section}
		[sec:related-work]
		{Related work}
		%{70-related-work/content.tex}
		Black-box simulation-based \Ac{SLFV} of cyber-physical systems has been widely addressed in the literature. 
		\Revision[breach]{%
			For example, simulation-based reachability analysis for large linear continuous-time dynamical systems has been investigated in \cite{donze:2010:breach,bak-etal:2017:cav}.
		}
		A simulation-based data-driven approach to verification of hybrid control systems described by a combination of a black-box simulator for trajectories and a white-box transition graph specifying mode switches has been investigated in \cite{fan-etal:2017:cav}.
		Formal verification of discrete time Simulink models (\eg Stateflow or models restricted to discrete time operators) with small domain variables has been investigated in, \eg \cite{tripakis-etal:2005:translating,meenakshi-etal:2006:simulink,whalen-etal:2007:integration,bostrom:2016:contract}.
		However, none of the approaches above supports simulation-based bounded model checking of arbitrary simulation models on a (typically extremely large) set of operational scenarios given as input, and none of them addresses the issue of simulation campaign optimisation.

		To the best of our knowledge, the only available literature which deals with simulation campaign optimisation is our previous works \cite{mancini-etal:2013:cav,mancini-etal:2014:dsd-anytime}, where preliminary versions of our algorithm have been presented.
		With respect to those conference papers, the current article presents a new, more scalable algorithm which guarantees to compute a shortest simulation campaign when enough simulator memory is allowed, and exploits various heuristics to compute an as short campaign as possible even when such memory requirements are not met. 
		Our algorithm computes simulation campaigns that obey to the verification order decided by the user, possibly randomised so as to compute, at any time during simulation, an upper bound to the omission probability, using the results of \cite{mancini-etal:2016:micpro}. 

		Our algorithm takes as input a set of operational scenarios that can be provided in several ways, \eg from a high-level constraint-based model as discussed in \cite{mancini-etal:2021:tse-supervisory}, or as a mass-memory database of scenarios. This allows us to seamlessly support both (random) exhaustive verification (when the given scenarios completely define the set of operational scenarios of interest for the verification task) and statistical model checking (when the given scenarios are a random sample of such scenarios).

		When exhaustive verification is not a viable option, given the huge number of scenarios of interest, simulation-based statistical model checking is often preferred, in order to compute statistically-sound information about the \Ac{SUV} properties of interest from a random sample of the possible scenarios, see, \eg \cite{younes-etal:2002:probabilistic,younes-etal:2006:sttt,basu-etal:2010:bookChapter,bogdoll-etal:2011:bookChapter,boyer-etal:2013:qest,jegourel-etal:2012:tacas,jegourel-etal:2013:cav,gonschorek-etal:2017:memocode,grosu-etal:2004:isola,grosu-etal:2005:tacas,jegourel-etal:2012:tacas}. 
		Simulation-based statistical model checking has been successfully applied in several domains, \eg Simulink \Ac{CPS} models \cite{clarke-etal:2011:atva,zuliani-etal:2013:fmsd},
		mixed-analog circuits \cite{clarke-etal:2010:probabilistic};
		smart grid control policies 
		\cite{mancini-etal:2014:smartgridcomm,hayes-etal:2016:tsg,mancini-etal:2015:dsd,mancini-etal:2018:smartgridcomm}; biological models 
		\cite{miskov-etal:2013:bcb,mancini-etal:2014:fmcad,mancini-etal:2015:iwbbio,sinisi-etal:2020:bioinf}. 
		\Revision[falsification]{%
			Finally, simulation-based \emph{falsification} of \Ac{CPS} properties (\eg for Simulink models) has been extensively investigated. Examples are in \cite{annpureddy-etal:2011:tacas,deshmukh-etal:2015:stochastic,abbas-etal:2013:tecs,dreossi-etal:2015:efficient,hoxha-etal:2017:sttt,sankaranarayanan-etal:2017:SIGBED,adimoolam-etal:2017:cav,tuncali-etal:2019:rapidly} and citations thereof. Some of such works also propose suitable data-structures (\eg tree-like) to represent the set of possible traces, as we do.
		}

		Our simulation campaign optimisation algorithm is \emph{independent} of the chosen verification technique, and the computed campaigns would bring \emph{significant speedups} in terms of simulation time to all of them. For example, the first row of \cref{fig:expres:results:buck,fig:expres:results:apollo,fig:expres:results:fcs} shows that speedups up to around $6\times$ in simulation time can be achieved even when a small random sample (only 25\%) of the entire sets of scenarios is chosen to perform statistical model checking.

		The ability to perform \emph{parallel} verification of the \Ac{SUV} is also a key enabler to make simulation-based \Ac{SLFV} of industry-scale \Acp{CPS} practically viable.
		Parallel approaches have been investigated, see \eg \cite{alturki-etal:2011:calco} in the context of probabilistic properties.
		Our approach seamlessly allows \emph{massive embarrassingly parallel} verification\Revision< (both statistical and exhaustive)>{}. 
		This is because, once the input set of scenarios has been split into slices, a parallel simulation campaign is computed, which is used to feed \emph{independent} verification processes to be run in parallel. 
	\end{Section}	

	\begin{Section}
		[sec:conclusions]
		{Conclusions}
		%{80-conclusions/content.tex}
		In this article we focused on the generation of \emph{optimised \SIMcampaignProse+} to carry out\Revision< (both exhaustive and statistical)>{} \Ac{SLFV} of \Acp{CPS} using arbitrarily many simulators of the system model running in parallel in a large \Ac{HPC} infrastructure, with the goal of \emph{minimising the overall completion time}.

		By taking as input a user-defined collection of (a random sample of) operational scenarios of interest from either a mass-storage database or a symbolic structure such as a constraint-based scenario generator in a (possibly random) user-defined order, our optimiser computes shortest parallel campaigns which exercise the system model on all (and only) the given scenarios. Our campaigns greatly speed-up verification by wisely avoiding the repeated computation of recurrent system trajectories as much as possible, compatibly with simulator memory constraints.

		Our experiments on\Revision< exhaustive as well as statistical>{} \Ac{SLFV} of Modelica/\Ac{FMU} and Simulink case study models with up to 
		\emph{\DataGet<experiments/results/stats>{operational scenarios/max/prose} scenarios} show that our optimisation yields \emph{speedups as high as $8\times$} and scales very well to large \Ac{HPC} infrastructures (efficiency almost always $\geq 90\%$ even when using \DataGet<experiments/results/stats>{computational cores/max} computational nodes, \ie \DataGet<experiments/results/stats>{machines/max} \DataGet<experiments/results/stats>{machines/cores}-core parallel machines). 

		The conjoint exploitation of \SIMcampaignProse optimisation and massive parallelism makes practically viable (a few weeks in a \Ac{HPC} infrastructure) verification tasks (both exhaustive and statistical) which would otherwise take \emph{inconceivably} long time.
	\end{Section}	

	\begin{Section}*
		{Acknowledgements}
		%{acknowledgements.tex}
		The authors are grateful to the anonymous reviewers for their comments and suggestions.

		This work was partially supported by: 
		Italian Ministry of University and Research under grant ``Dipartimenti di eccellenza 2018--2022'' of the Department of Computer Science of Sapienza University of Rome; 
		EC FP7 project PAEON (Model Driven Computation of Treatments for Infertility Related Endocrinological Diseases, 
		600773);
		EC FP7 project SmartHG (Energy Demand Aware Open Services for Smart Grid Intelligent Automation, 
		317761);
		INdAM ``GNCS Project 2022''; % da aggiornare all'ultima edizione.
		Sapienza University projects 
		  RG12117A8B393BDC% 2021-2024 - VPH grande (T. Mancini)
		%, RG11816436BD4F21% 2018-2021 - VPH grande (T. Mancini) 
		, RG11916B892E54DB% 2019-2022 - grande AI+psichiatria (G. Angeletti PI, E. Tronci)
		%, RP11916B8665242F% 2019-2022 - piccolo vulnerability
		, RG120172B9329D33% 2020-2023 - grande AI+psichiatria (PI: Ferracuti)
		;
		Lazio POR FESR projects 
		  E84G20000150006% SDFS
		, F83G17000830007% SCAPR
		;
		NRRP Mission 4, Comp.\ 2, Inv.\ 1.5, NextGenEU, MUR CUP B83C22002820006, Rome Technopole.
	\end{Section}	
 
	\BoxArticleDisclaimer

	\providecommand{\MCLabProceedingsOf}[1]{#1}
	
\end{BibliographyUnit}

\BoxArticleDisclaimer

\clearpage
\appendices 
\begin{BibliographyUnit}
	%\Input{90-appendix/content.tex}
		\begin{Section}{Formal framework}
			\label[appendix]{app:sec:framework}

			\begin{Section}{Modelling the 
					\texorpdfstring
						{\Acl{SUV}}
						{System Under Verification}
				}
				\label[appendix]{app:framework:suv}

				Our \Ac{SUV} is a continuous- or discrete-time \Acl{DS} whose 
				inputs are operational scenarios defined in terms of time functions (\cref{def:time-functions}) of values in the \Ac{SUV} input space, defining possible values for the user inputs and other \emph{uncontrollable events}, \eg faults in sensors and actuators or changes in system parameters. 
				%\Ac{SUV} inputs are collectively referred to as \emph{disturbances}.

				\begin{definition}[Time set, time function]
				\label{def:time-set}
				\label{def:time-functions}
				A \emph{time set} $\TimeSet$ is \RealsGEZ (for continuous-time systems) or \IntegersGEZ (for discrete-time systems) or an interval thereof. 
				Given a time set $\TimeSet$ and a set of values $\InputSet$, a \emph{time function} on \TimeSet with values in $\InputSet$ is a function $\InputFunction : \TimeSet \to \InputSet$ which associates to each time point $\TimePoint \in \TimeSet$ a value $\InputFunction(\TimePoint) \in \InputSet$.

				Given a time set $\TimeSet$ and a set of values $\InputSet$, 
				we denote by $\InputFunctionSet$ the set of time functions on $\TimeSet$ with values in $\InputSet$.
				When $\TimeSet$ is the empty interval $\TimeSetEmpty$, we conventionally assume that $\InputFunctionSet[timeset=\TimeSetEmpty]$ consists of a single time function $\InputFunctionEmpty$ (the \emph{empty} time function, having duration zero). 
				\end{definition}

				In the following, we sometimes find convenient to denote the empty time interval $\TimeSetEmpty$ by $\RealRange*{\TimePoint}{\TimePoint}$, where $\TimePoint$ is any value in $\TimeSet$.
				\Cref{def:restriction-concatenation} defines two operations on time functions we use in the following.

				\begin{definition}[Restriction and concatenation of time functions]
				\label{def:restriction-concatenation}
				\def\tmpTimeRangeA{\ensuremath{\TimeSet_1}\xspace}
				\def\tmpTimeRangeAdef{\RealRange*{\TimePoint[_1]}{\TimePoint[_2]}}%
				\def\tmpTimeRangeB{\ensuremath{\TimeSet_2}\xspace}
				\def\tmpTimeRangeAB{\ensuremath{\tmpTimeRangeA \cup \tmpTimeRangeB}\xspace}
				Given a time function $\InputFunction \in \InputFunctionSet$ and a time set $\tmpTimeRangeA \subseteq \TimeSet$, 
				the \emph{restriction} of $\InputFunction$ to $\tmpTimeRangeA$ is function
					$\TimeFunctionRestriction
								{\InputFunction}
								{\tmpTimeRangeA}  
						\in \InputFunctionSet[timeset=\tmpTimeRangeA]$ defined in \tmpTimeRangeA and such that
							$\TimeFunctionRestriction
							  	{\InputFunction}
							  	{\tmpTimeRangeA}(\TimePoint) 
							  	= 
							  	\InputFunction(\TimePoint)$ 
							for all $\TimePoint \in \tmpTimeRangeA$.

				Given two time sets $\tmpTimeRangeA$ and $\tmpTimeRangeB$ such $\tmpTimeRangeA \cap \tmpTimeRangeB = \emptyset$ and $\tmpTimeRangeA \cup \tmpTimeRangeB$ is a time set (\ie \RealsGEZ, \IntegersGEZ, or an interval thereof), and  
				time functions $\InputFunction[_1] \in \InputFunctionSet[timeset=\tmpTimeRangeA]$ and $\InputFunction[_2] \in \InputFunctionSet[timeset=\tmpTimeRangeB]$,
				the \emph{concatenation} of 
					$\InputFunction[_1]$ 
				and $\InputFunction[_2]$ 
				is function	
					$\TimeFunctionConcat
						{\InputFunction[_1]}
						{\InputFunction[_2]}$ 
							in $\InputFunctionSet[timeset=\tmpTimeRangeAB]$ such that,
							for all $\TimePoint \in \tmpTimeRangeAB$:
							$\TimeFunctionConcat
								{\InputFunction[_1]}
								{\InputFunction[_2]}(\TimePoint) 
							= 
							\InputFunction[_1](\TimePoint)$ 
							if $\TimePoint \in \tmpTimeRangeA$ and 
							$\InputFunction[_2](\TimePoint)$ otherwise.
				\end{definition}

				\Cref{def:ds} recalls the notion of deterministic, causal \Acl{DS} from~\cite{sontag:1998:book} (see also~\cite{mancini-etal:2015:gandalf,mancini-etal:2017:ipl}), which
				take as input time functions with values in its \emph{input space} and output time functions with values in its \emph{output space}.

				\begin{definition}[\Acl{DS}]
					\label{def:suv}
					\label{def:ds}	
					\AcronymReset{DS}%
					A \emph{deterministic, causal} \Ac{DS} $\SUV$ is a tuple $\SUVdef$, where:
					\begin{itemize}
						\item $\TimeSet$ is the time set;
						\item $\SUVstateSet$, the \emph{state space} of $\SUV$, is a non-empty set whose elements denote \emph{states}; 
						\item $\SUVinitState \in \SUVstateSet$ is the \emph{initial state} of $\SUV$;
						\item $\SUVinputSet$, the \emph{input space} of $\SUV$, is the set of its possible input values;
						\item $\SUVoutputSet$, the \emph{output space} of $\SUV$, is the non-empty set of its possible output values;
						\item $\SUVtransition$ is the \emph{transition map} of $\SUV$. 
							  Given $\TimePoint[_1], \TimePoint[_2] \in \TimeSet$ such that $\TimePoint[_1] \leq \TimePoint[_2]$, $\SUVstate \in \SUVstateSet$, $\SUVinputFunction \in \SUVinputFunctionSet$,  
							  $\SUVtransition(\TimePoint[_2], \TimePoint[_1], \SUVstate, \SUVinputFunction)$ denotes the state reached by \SUV at time $\TimePoint[_2]$ when starting from state $\SUVstate$ at time $\TimePoint[_1]$ and given input time function $\SUVinputFunction$.

						Function $\SUVtransition$ must satisfy the following properties:

						\begin{itemize}
							\item \emph{Causality}: 
								for all 		
								$\TimePoint[_1], \TimePoint[_2], \TimePoint[_3] \in \TimeSet$ such that $\TimePoint[_1] \leq \TimePoint[_2] \leq \TimePoint[_3]$, 
								$\SUVinputFunction \in \SUVinputFunctionSet$, 
								and 
								$\SUVstate \in \SUVstateSet$,
								$\SUVtransition(\TimePoint[_2], \TimePoint[_1], \SUVstate, \SUVinputFunction)
								= \SUVtransition(\TimePoint[_2], \TimePoint[_1], \SUVstate, 
										\TimeFunctionRestriction{\SUVinputFunction}{
											\RealRange*{\TimePoint[_1]}{\TimePoint[_2]}
										}
								  )$.

							\item \emph{Consistency}: 
							for all 
								$\TimePoint \in \TimeSet$, 
								$\SUVinputFunction \in \SUVinputFunctionSet$, 
								and
								$\SUVstate \in \SUVstateSet$,
								$\SUVtransition(\TimePoint, \TimePoint, \SUVstate, \SUVinputFunction)
									= \SUVstate
								$.

							\item \emph{Semigroup}: 
								for all 		
									$\TimePoint[_1], \TimePoint[_2], \TimePoint[_3] \in \TimeSet$ such that $\TimePoint[_1] \leq \TimePoint[_2] \leq \TimePoint[_3]$, 
									$\SUVinputFunction[_{1,2}] \in 
										\SUVinputFunctionSet[timeset={\RealRange*{\TimePoint[_1]}{\TimePoint[_2]}}]$,
									$\SUVinputFunction[_{2,3}] \in 
										\SUVinputFunctionSet[timeset={\RealRange*{\TimePoint[_2]}{\TimePoint[_3]}}]$,
									and 
									$\SUVstate \in \SUVstateSet$,
									$ 
									\SUVtransition(
										\TimePoint[_3], \TimePoint[_2],
										\SUVtransition(
											\TimePoint[_2], \TimePoint[_1],
											\SUVstate,
											\InputFunction[_{1,2}]
										),
										\InputFunction[_{2,3}]
									)
									=
									\SUVtransition(
										\TimePoint[_3], 
										\TimePoint[_1], \SUVstate, 
										\TimeFunctionConcat
											{\SUVinputFunction[_{1,2}]}
											{\SUVinputFunction[_{2,3}]}
									).
								$		
						\end{itemize}  
						\item $\SUVobservation : \SUVobservationDef$ is the \emph{observation function} of $\SUV$. 
					\end{itemize}

					A \Ac{DS} is \emph{time invariant} if its time set $\TimeSet$ is right-unbounded and, for any 
						$\TimePoint[_1], \TimePoint[_2], \TimeStep \in \TimeSet$ such that $\TimePoint[_1] \leq \TimePoint[_2]$, 
						$\SUVstate \in \SUVstateSet$,
						and for any input time function
						$\SUVinputFunction \in 
							\SUVinputFunctionSet[timeset={\RealRange*{\TimePoint[_1]}{\TimePoint[_2]}}]$, 
						we have:
						$\SUVtransition(
							\TimePoint[_1], 
							\TimePoint[_2],
							\SUVstate,
							\SUVinputFunction
						 ) =
						 \SUVtransition(
							\TimePoint[_1] + \TimeStep, 
							\TimePoint[_2] + \TimeStep,
							\SUVstate,
							\SUVinputFunction[']
						 )
						$, where $\SUVinputFunction['] \in 
							\SUVinputFunctionSet[timeset={\RealRange*{\TimePoint[_1] + \TimeStep}{\TimePoint[_2] + \TimeStep}}]$ is such that $\SUVinputFunction['](\TimePoint) = \SUVinputFunction(\TimePoint - \TimeStep)$ for all $\TimePoint \in \RealRange*{\TimePoint[_1] + \TimeStep}{\TimePoint[_2] + \TimeStep}$.
				\end{definition}

				In the following, we assume that our \Ac{SUV} is a discrete- or continuous-time input-state-output deterministic, causal, time-invariant \Ac{DS}, whose state can undertake continuous as well as discrete changes, and whose output ranges on any combination of discrete and continuous values.
			\end{Section}

			\begin{Section}
				{\Acl{SLFV}}
				\label[appendix]{sec:framework:slfv}

				Given a \Ac{SUV} $\SUV$ and a set of time functions $\TraceSet$ on its input space (defining operational scenarios), verifying $\SUV$ on $\TraceSet$ means to collect the outputs of the \Ac{SUV} monitor at the end of each scenario in $\TraceSet$ (\cref{def:slfv}).

				\begin{definition}[{\Ac[, ]{SLFV}[]}]
				\label{def:slfv}
				A \Ac{SLFV} problem is a pair $\SLFV = \SLFVdef$, where $\SUV$ is a \Ac{SUV} (with an embedded monitor) having input space $\InputSet$, and $\TraceSet$ is a set of input time functions in $\SUVinputFunctionSet$.

				The answer to $\SLFV$ is the collection of the outputs of the \Ac{SUV} monitor produced at the end of each $\SUVinputFunction \in \TraceSet$, where $\SUVinputFunction$ is injected in $\SUV$ starting from its initial state.
				That is:
				$\Set{ \SUVobservation(\SUVtransition(\TimePoint, 0, \SUVinitState, \SUVinputFunction)) ~\mid~ \SUVinputFunction \in \TraceSet, \SUVinputFunction \in \SUVinputFunctionSet[timeset={\RealRange*{0}{\TimePoint}}] }$.
				\end{definition}  
			\end{Section}

			\begin{Section}
				{Modelling the 
					\texorpdfstring
						{\Acs{SUV}}
						{SUV}
					operational environment
				}
				\label[appendix]{app:framework:environment}
				Given our focus on verification tasks where numerical simulation is the only means to get the trajectory of the \Ac{SUV} when fed with an input scenario, we will assume that the set $\TraceSet$ is \emph{finite} and \emph{finitely representable}, and that each scenario is \emph{time-bounded}.
				Hence, in the following, we assume that the set of values taken by input scenarios in $\TraceSet$ (actually, for simplicity, the set $\SUVinputSet$ itself) is \emph{finite} (and, without loss of generality, \emph{ordered}) and scenarios in $\TraceSet$ are defined via \emph{piecewise constant} input time functions having discontinuities at time points multiple of a given (arbitrarily small) \emph{time quantum} $\TimeQuantum \in \TimeSet \setminus \Set{0}$. Such scenarios can be conveniently represented as \emph{\TraceProse+} (\cref{def:trace}).

				\begin{SupplMatStatement}{def:trace}[\TraceProse^]
					%\Input(absolute){20-framework/environment/def_trace.tex}
					Let $\InputSet$ be a finite set of values (the \Ac{SUV} input space) and $\TimeQuantum \in \TimeSet \setminus \Set{0}$ be a \emph{time quantum}.

					An \TraceProse $\Trace$ with values in $\InputSet$ is a finite sequence $\TraceDef$ where all, for each $i \in [0, \Horizon-1]$, 
					$\InputValue[_i]$ belongs to $\InputSet$. Value $\Horizon \in \IntegersGZ$ 
					is the trace \emph{horizon}.

				\end{SupplMatStatement}

				An \TraceProse $\Trace = \TraceDef$ is interpreted as the bounded-horizon piecewise constant time function $\InputFunction \in \InputFunctionSet[timeset={\RealRange*{0}{\TimeQuantum \Horizon}}]$ defined as
					$\InputFunction(\TimePoint) =
						\InputValue[_{\Floor{\frac{\TimePoint}{\TimeQuantum}}}]$
						for $\TimePoint \in \RealRange*{0}{\TimeHorizonDef}$.
				Thus, in the following, we will assume that a time quantum $\TimeQuantum \in \TimeSet \setminus \Set{0}$ is given, and interchangeably refer to input traces and to their uniquely associated piecewise constant time functions.
				We recall (see the main article) that smooth continuous-time functions can be managed as long as they can be cast into (or suitably approximated by) \emph{finitely parametrisable} functions (\eg via quantised values of a bounded number of coefficients of their Fourier series), in which case the input space actually defines such a (discrete or discretised) parameter space. 
			\end{Section}
		\end{Section}

		\begin{Section}
			{\Acs{SUV} simulators}
			\label[appendix]{app:sec:simulator}

			In this \namecref{app:sec:simulator}, we formalise the notion of \Ac{SUV} simulator (\cref{def:simulator}, which extends the definition of~\cite{mancini-etal:2015:gandalf,mancini-etal:2021:jlamp}).
			This notion offers a general mathematical framework that will allow us to link inputs to a \Ac{SUV} $\SUV$ (piecewise-constant input time functions representing \Ac{SUV} scenarios and encoded as input traces) to inputs for a simulator of $\SUV$ (\emph{simulation campaigns}, \cref{def:simulation-campaign}).

			\begin{definition}[\Acs{SUV} simulator]
			\label{def:simulator}
			Let $\SIMstateIDset$ be a countable set of identifiers (\eg $\IntegersGEZ$).
			A \Ac{SUV} simulator $\SIM$ is a tuple $\SIMdef$ where $\SUV = \SUVdef$ is a time-invariant \Ac{DS} (our \Ac[, ]{SUV}[]), and $\SIMstateSet$ is a set whose elements denote simulator \emph{states}. 
			Each simulator state $\SIMstate \in \SIMstateSet$ has the form $\SIMstate = \SIMstateDef$, where:
			\begin{itemize}
				\item $\SUVstate \in \SUVstateSet$ defines a state of $\SUV$ or a distinguished state $\SUVstateError$;
				\item $\InputFunction \in \InputFunctionSet[timeset={\RealRange*{0}{\TimePoint}}]$ denotes an input time function over $\SUVinputSet$ defined over a (possibly empty) bounded time set interval \RealRange*{0}{\TimePoint} (for some $\TimePoint \in \TimeSet$);
				\item $\SIMmemory$ is a finite \emph{map} defining the content of the simulator \emph{memory}. Each element of $\SIMmemory$ is of the form: $\SIMmemoryTuple[][']$, where $\SIMstateID \in \SIMstateIDset$ is an identifier (\emph{unique} in $\SIMmemory$), $\SUVstate[']$ a \Ac{SUV} state, and $\InputFunction['] \in \InputFunctionSet[timeset={\RealRange*{0}{\TimePoint[']}}]$ ($\TimePoint['] \in \TimeSet$) is an input time function over a bounded (possibly empty) interval of the time set $\TimeSet$.
			\end{itemize}

			The simulator \emph{initial state} is $\SIMinitState = \SIMinitStateDef$, where $\InputFunctionEmpty \in \InputFunctionSet[timeset={\RealRange*{0}{0}}]$ is the empty input time function, having zero duration.
			\end{definition}

			As an extension to \cite{mancini-etal:2015:gandalf,mancini-etal:2021:jlamp}, each \Ac{SUV} state $\SUVstate \in \SUVstateSet$ occurring in a simulator state $\SIMstate = \SIMstateDef \in \SIMstateSet$ or in the simulator memory $\SIMmemory$ is always accompanied by the input time function (piecewise constant in our setting) $\InputFunction$ which would drive the \Ac{SUV} from its initial state $\SUVinitState$ to $\SUVstate$.
			This is enforced by \cref{def:simulator:commands}, which gives the semantics of simulator commands and of the simulator transition function, and formally stated in \cref{thm:simulation-campaign:input-functions}.

			Our extension eases the presentation of the forthcoming results. The definitions is \cite{mancini-etal:2015:gandalf,mancini-etal:2021:jlamp} can be obtained back by ignoring input time functions in simulator states and in the simulator memory.

			\begin{definition}[Simulator commands and transition function]
			\label{def:simulator:commands}
			Let $\SIM = \SIMdef$ be a simulator for \Ac{SUV} $\SUV = \SUVdef$.
			The \emph{commands} of $\SIM$ are:
				\SIMout,
				\SIMload(\SIMstateID),
				\SIMstore(\SIMstateID),
				\SIMfree(\SIMstateID),
				\SIMrun(\SUVinputValue, \TimeQuantum),
			where $\SIMstateID$ is an identifier, $\SUVinputValue \in \SUVinputSet$ is an input value for $\SUV$, and $\TimeQuantum \in \TimeSet \setminus \Set{0}$ is a non-zero time duration ($\SIMstateID$, $\SUVinputValue$, and $\TimeQuantum$ are \emph{command arguments}).

			\label{def:simulator:transition}
			The transition function of $\SIM$, $\SIMtransition$, defines how the internal simulator state changes upon execution of each command.
			Namely, $\SIMtransition(\SIMstate, \SIMgenericCmd) = \SIMstate[']$ where $\SIM$ moves from state $\SIMstate$ to state $\SIMstate[']$ upon processing command \SIMgenericCmd.

			For each $\SIMstate = \SIMstateDef$, let $\InputFunction \in \InputFunctionSet[timeset={\RealRange*{0}{\TimePoint}}]$ for some $\TimePoint \in \TimeSet$.

			Function $\SIMtransition$ is defined as follows:
			\begin{itemize}
				\item $\SIMtransition(\SIMstate, \SIMout) = \SIMstate$, as the $\SIMout$ command only reads the output of $\SIM$ in the current state, which is the output of \Ac{SUV} $\SUV$ (\cref{def:suv}) when in state $\SUVstate$ associated to $\SIMstate$.

				\item $\SIMtransition(\SIMstate, \SIMload(\SIMstateID)) =
						(\SUVstate['], \InputFunction['], \SIMmemory)$
					if $\SUVstate \neq \SUVstateError$ and $\SIMmemoryTuple[]['] \in \SIMmemory$. 

				\item $\SIMtransition(\SIMstate, \SIMstore(\SIMstateID)) =
						(\SUVstate, \InputFunction, \SIMmemory \cup \Set{\SIMmemoryTuple})$
					if $\SUVstate \neq \SUVstateError$ and $\not\exists \InputFunction['], \SUVstate['] ~ \SIMmemoryTuple[]['] \in \SIMmemory$.

				\item $\SIMtransition(\SIMstate, \SIMfree(\SIMstateID)) =
						(\SUVstate, \InputFunction, \SIMmemory \setminus \Set{\SIMmemoryTuple})$
					if $\SUVstate \neq \SUVstateError$ and $\SIMmemoryTuple \in \SIMmemory$.

				\item $\SIMtransition(\SIMstate, \SIMrun(\hat{\SUVinputValue}, \TimeStep)) =
						(\SUVtransition(\TimeQuantum, \SUVstate, \hat{\InputFunction}), 
							\TimeFunctionConcat
								{\InputFunction}
								{\hat{\InputFunction}},
						 \SIMmemory
						)$
					if $\SUVstate \neq \SUVstateError$; in the formula, 
						${\hat{\InputFunction}}$					
					is the input time function in $\InputFunctionSet[timeset={\RealRange*{\TimePoint}{\TimePoint+\TimeStep}}]$ having constant value $\hat{\SUVinputValue}$
					and $\TimeFunctionConcat
								{\InputFunction}
								{\InputFunctionConstant}
								 \in \InputFunctionSet[timeset={\RealRange*{0}{\TimePoint+\TimeStep}}]$ is the concatenation of $\InputFunction$ and 
								 	$\hat{\InputFunction}$.

				\item 
					$\SIMtransition(\SIMstate, \SIMgenericCmd) =
					(\SUVstateError, \InputFunction, \SIMmemory)$
					in all the other cases.
			\end{itemize}

			The \emph{time advancement} of command $\SIMgenericCmd$ is the time simulated by $\SIM$ when executing $\SIMgenericCmd$. Namely:
			$\SIMcmdLen(\SIMgenericCmd) = \TimeStep$ if $\SIMgenericCmd = \SIMrun(\hat{\SUVinputValue}, \TimeStep)$ and is $0$ for all the other commands.
			\end{definition}

			Given a sequence of scenarios (formally represented as piecewise constant input time functions encoded as \TraceProse+), we can build a sequence of commands (\emph{simulation campaign}, \cref{def:simulation-campaign}) driving the simulator through those scenarios.
			Conversely, given a simulation campaign, we can compute the sequence of scenarios (piecewise constant input time functions) simulated by it (\cref{def:simulation-campaign:input-functions}).

			\begin{definition}[Simulation campaign, state and output sequences]
			\label{def:simulation-campaign}
			Let $\SIM = \SIMdef$ be a simulator for \Ac{SUV} $\SUV$ and let $\SIMtransition$ be the transition function for $\SIM$.

			\begin{itemize}
			\item A \emph{simulation campaign} $\SIMcampaign$ for $\SIM$ is a sequence of simulator commands $\SIMcampaign = \SIMcampaignDef$ along with their arguments, where $\SIMnbCampaignCmds \in \IntegersGEZ$.

			\item The \emph{length} $\SIMcampaignLen(\SIMcampaign)$ of simulation campaign $\SIMcampaign$ is the sum of the time advancements of commands in $\SIMcampaign$. Namely:
			$
				\SIMcampaignLen(\SIMcampaign) =
					\sum_{i=0}^{\SIMnbCampaignCmds-1} \SIMcmdLen(\SIMgenericCmd[_i])
			$.

			\item To simulation campaign $\SIMcampaign$ we can univocally associate the \emph{sequence of simulator states} $\SIMseqStates$ traversed by the simulator while executing it, namely: $\SIMinitState = \SIMinitStateDef$, \ie the initial simulator state (\cref{def:simulator}), and, for each $i \in [1, \SIMnbCampaignCmds]$, $\SIMstate[_i] = \SIMtransition(\SIMstate[_{i-1}], \SIMgenericCmd[_{i-1}])$.

			\item Simulation campaign $\SIMcampaign$ is \emph{executable} if and only if $\SIMstate[_{\SIMnbCampaignCmds}] = \SIMstateDef[_{\SIMnbCampaignCmds}]$ is such that $\SUVstate[_{\SIMnbCampaignCmds}] \neq \SUVstateError$.

			\item The \emph{required simulator memory} $\SIMcampaignMaxMem(\SIMcampaign)$ of simulation campaign $\SIMcampaign$ is the maximum number of entries in the simulator memory among $\SIMseqStates$ (\ie the states traversed by $\SIMcampaign$, where $\SIMstate[_i] = \SIMstateDef[_i]$ for $i \in [0, \SIMnbCampaignCmds]$). Namely:
			$
				\SIMcampaignMaxMem(\SIMcampaign) =
					\max_{i=0}^{\SIMnbCampaignCmds} \SetCardinality{\SIMmemory[_i]}
			$.

			\item The \emph{output sequence} associated to executable simulation campaign $\SIMcampaign$ containing $\SIMnbCampaignOutputCmds \in \IntegersGEZ$ \SIMout commands is $\SIMseqOutputs$, where, for each $i \in [0, \SIMnbCampaignOutputCmds-1]$, $\SUVobservation(\SUVstate[_{j_i}])$ is the output of \Ac{SUV} $\SUV$ (\cref{def:suv}) when in the state $\SUVstate[_{j_i}]$ associated to the simulator state $\SIMstateDef[_{j_i}]$ corresponding to the $i$-th \SIMout command.
			\end{itemize}
			\end{definition}

			\Cref{thm:simulation-campaign:input-functions} links inputs to a simulator $\SIM$ for $\SUV$ (\ie simulation campaigns) to inputs for $\SUV$ (input time functions): for each simulation campaign $\SIMcampaign$, the (piecewise constant) input time function $\InputFunction$ of any simulator state $\SIMstateDef$ traversed by $\SIM$ while executing $\SIMcampaign$ drives $\SUV$ from its initial state $\SUVinitState$ to $\SUVstate$.

			\begin{SupplMatStatement}{thm:simulation-campaign:input-functions}
				Let
					$\SIM = \SIMdef$ be a simulator for $\SUV$,
					$\SIMcampaign = \SIMcampaignDef$ an executable simulation campaign for $\SIM$,
					and $\SIMseqStates$ the sequence of simulator states associated to $\SIMcampaign$.

				For each $i \in [0, \SIMnbCampaignCmds]$, the input time function $\InputFunction[_i]$ in $\SIMstate[_i] = \SIMstateDef[_i]$ belongs to
					$\InputFunctionSet[timeset={\RealRange*{0}{\TimeStep_i}}]$ (for some $\TimeStep_i \in \TimeSet$)
				and is such that
						$\SUVtransition(\TimeStep_i, 0, \SUVinitState, \InputFunction[_i]) = \SUVstate[_i]$.
					
				\begin{proof}
					Let $\SUV = \SUVdef$.
					For each $i \in [0, \SIMnbCampaignCmds]$, let the $i$-th state traversed by \SIM during execution of \SIMcampaign be $\SIMstate[_i] = \SIMstateDef[_i]$.

					We prove that, for each $i$:
						\begin{enumerate}
						\item $\InputFunction[_i] \in \InputFunctionSet[timeset={\RealRange*{0}{\TimeStep_i}}]$ for some $\TimeStep_i \in \RealsGEZ$; and
						\item $\SUVtransition(\TimeStep_i, 0, \SUVinitState, \InputFunction[_i]) = \SUVstate[_i]$.
						\end{enumerate}

					The proof is by induction on $i$.

					\begin{description}
					\item[Base case] ($i=0$)
					By \cref{def:simulator}, $\SIMinitState = \SIMinitStateDef$, where $\InputFunctionEmpty \in \InputFunctionSet[timeset={\RealRange*{0}{\TimeStep_0}}]$ with $\TimeStep_0 = 0$, \ie $\InputFunctionEmpty$ is the input time function having zero duration. By \cref{def:suv} (consistency), $\SIMtransition(\TimeStep_0, \TimeStep_0, \SUVinitState, \InputFunctionEmpty) = \SUVinitState$. The thesis follows.

					\item[Inductive case] ($i \in [1, \SIMnbCampaignCmds]$)
					Assume, by inductive hypothesis, that the $(i-1)$-th state traversed by the simulator when executing \SIMcampaign, $\SIMstate[_{i-1}] = \SIMstateDef[_{i-1}]$ is such that $\InputFunction[_{i-1}] \in \InputFunctionSet[timeset={\RealRange*{0}{\TimeStep_{i-1}}}]$ for some $\TimeStep_{i-1} \in \RealsGEZ$ and
					$\SIMtransition(\TimeStep_{i-1}, 0, \SUVinitState, \InputFunction[_{i-1}]) = \SUVstate[_{i-1}]$.
					We now prove that
					$\InputFunction[_{i}] \in \InputFunctionSet[timeset={\RealRange*{0}{\TimeStep_{i}}}]$ for some $\TimeStep_{i} \in \RealsGEZ$ and that
					$\SIMtransition(\TimeStep_{i}, 0, \SUVinitState, \InputFunction[_{i}]) = \SUVstate[_{i}]$.

					The proof is by cases, depending on the type of command \SIMgenericCmd[_{i}] of \SIMcampaign, which moves \SIM from state $\SIMstate[_{i-1}]$ to state $\SIMstate[_{i}] = \SIMstateDef[_{i}]$ (see the definition of \SIMtransition in \cref{def:simulator:transition}).

						\begin{itemize}
							\item If $\SIMgenericCmd[_{i}] = \SIMstore(\SIMstateID)$, $\SIMfree(\SIMstateID)$ for some $\SIMstateID \in \SIMstateIDset$ or $\SIMout$, then the thesis trivially follows, as, from executability of \SIMcampaign and the definition of \SIMtransition, $\SUVstate[_i] = \SUVstate[_{i-1}]$ and $\InputFunction[_{i}] = \InputFunction[_{i-1}]$.

							\item If $\SIMgenericCmd[_{i}] = \SIMload(\SIMstateID)$ for some $\SIMstateID \in \SIMstateIDset$, then executability of \SIMcampaign and the definition of \SIMtransition imply that tuple $\SIMmemoryTuple[][_b]$ exists in $\SIMmemory[_{i-1}]$, with $\SUVstate[_b] = \SUVstate[_i]$ and $\InputFunction[_b] = \InputFunction[_i]$.
							The thesis follows from the inductive hyphotesis.

							\item If $\SIMgenericCmd[_{i}] = \SIMrun(\SUVinputValue, \TimeQuantum)$ for some $\SUVinputValue \in \SUVinputSet$ and $\TimeQuantum \in \RealsGZ$, the definition of \SIMtransition implies that
								$\InputFunction[_i] =
									\TimeFunctionConcat
										{\InputFunction[_{i-1}]}
										{\InputFunctionConstant}					
								$, \ie the concatenation of the input time function associated to simulator state \SIMstate[_{i-1}], \ie $\InputFunction[_{i-1}] \in \InputFunctionSet[timeset={\RealRange*{0}{\TimeStep_{i-1}}}]$, and the input time function $\InputFunctionConstant \in \InputFunctionSet[timeset={\RealRange*{\TimeStep_{i-1}}{\TimeStep_{i-1}+\TimeQuantum}}]$ having constant value $\hat{\InputValue}$ and defined in time set $\RealRange*{\TimeStep_{i-1}}{\TimeStep_{i-1}+\TimeQuantum}$.
								Thus, $\InputFunction[_i] \in \InputFunctionSet[timeset={\RealRange*{0}{\TimeStep_{i-1}+\TimeQuantum}}]$ and
								from \cref{def:suv} (semigroup), the second point of the thesis follows from the inductive hypothesis.
						\end{itemize}
					\end{description}
				\end{proof}
			\end{SupplMatStatement}
		\end{Section}

		\begin{Section}
			%[app:sec:simcampaigns]
			{Simulation-based \Acs{SLFV}}
			%{40-simulation-based-slfv/content.tex}

			\label[appendix]{app:sec:simcampaigns}
			To perform \emph{simulation-based} \Ac{SLFV} of $\SUV$ over $\NbTraces$ input traces $\TraceSet$ we need a simulator $\SIM = \SIMdef$ for $\SUV$ and an executable \SIMcampaignProse $\SIMcampaign$ for $\SIM$ that somewhat drives $\SIM$ along the \NbTraces scenarios for $\SUV$ encoded by traces of $\TraceSet$ and collects the simulator outputs at the end of each scenario.

			To this end, \cref{def:simulation-campaign:input-functions} allows us to associate to any executable \SIMcampaignProse $\SIMcampaign$ for $\SIM$ the sequence $\SIMinputFunctionSeq$ of \Ac{SUV} scenarios (as piecewise constant input time functions) for $\SUV$ actually explored by $\SIMcampaign$.

			\begin{SupplMatStatement}{def:simulation-campaign:input-functions}[Sequence of input time functions associated to a \SIMcampaignProse]	
				%\Input(absolute){40-simulation-based-slfv/def_sim_campaign_input_function.tex}
				Let 
					$\SIM = \SIMdef$ be a simulator for \Ac{SUV} $\SUV$, 
					$\SIMtransition$ the transition function of $\SIM$, 
					$\SIMcampaign = \SIMcampaignDef$ 
					an executable \SIMcampaignProse for $\SIM$ containing $\NbTraces \in \IntegersGEZ$ \SIMout commands, 
					and \SIMseqStates the sequence of simulator states associated to $\SIMcampaign.$

				The \emph{sequence of input time functions} associated to \SIMcampaignProse $\SIMcampaign$ containing $\NbTraces$ \SIMout commands is $\SIMinputFunctionSeq = \SIMinputFunctionSeqDef$, where, for all $i \in [0, \NbTraces-1]$, $\InputFunction[_{\SIMinputFunctionIndex_{i}}]$ is the input time function associated to the state where the simulator executes the $i$-th \SIMout command of $\SIMcampaign$.
			\end{SupplMatStatement}

			\Cref{def:simulation-campaign:slfv} formalises the notion of a \emph{\SIMcampaignProse aimed at computing the answer to a \Ac{SLFV} problem}.

			\begin{SupplMatStatement}{def:simulation-campaign:slfv}[\SIMcampaignProse for an \Ac{SLFV} problem]
				%\Input(absolute){40-simulation-based-slfv/def_sim_campaign_for_slfv_problem.tex}
				A \SIMcampaignProse $\SIMcampaign$ \emph{for} \Ac{SLFV} problem $\SLFV = \SLFVdef$ is an executable campaign for a simulator \SIM of \SUV, such that the sequence $\SIMinputFunctionSeq(\SIMcampaign) = \SIMinputFunctionSeqDef$ of its associated input time functions is a \emph{permutation} of $\TraceSet$.
			\end{SupplMatStatement}

			\begin{Section}
				%[app:sec:simcampaigns:randomised]
				{Randomised \SIMcampaignProse+}
				%{randomised/content.tex}

				\label[appendix]{app:sec:simcampaigns:randomised}
				\Cref{thm:simulation-campaigns:shortest-any-order} states that, if we put no limitation on the required memory capacity, a shortest \SIMcampaignProse exists for \emph{any} ordering of the scenarios of the \Ac{SLFV} problem at hand.

				\begin{SupplMatStatement}{thm:simulation-campaigns:shortest-any-order}
					%\Input(absolute){40-simulation-based-slfv/randomised/thm-simulation-campaigns-shortest-any-order/content_stmt.tex}
					Let $\SLFV = \SLFVdef$ be a \Ac{SLFV} problem ($\SetCardinality{\TraceSet} = \NbTraces$) and $\SIM$ be a simulator for $\SUV$.
					For any permutation $\TraceSetDef[\SIMinputFunctionIndex]$ of \TraceProse+ of $\TraceSet$, 
					there exists an executable shortest \SIMcampaignProse $\SIMcampaign$ for $\SLFV$ on $\SIM$, such that
					$\SIMinputFunctionSeq(\SIMcampaign) = 
						\TraceInputFunction[_{\SIMinputFunctionIndex_0}], 
						%\TraceInputFunction[_{\SIMinputFunctionIndex_1}], 
						\ldots, 
						\TraceInputFunction[_{\SIMinputFunctionIndex_{\NbTraces - 1}}]$. 

					\begin{proof}
						%\Input(absolute){40-simulation-based-slfv/randomised/thm-simulation-campaigns-shortest-any-order/content_proof.tex}
						Proof sketch.
						Arrange \TraceProse+ of $\TraceSet$ (with time quantum $\TimeQuantum \in \TimeSet \setminus \Set{0}$) as a rooted tree whose nodes are trace \emph{prefixes} (including the empty prefix, which is the root of the tree) and whose edges connect nodes $(\InputValue[_0], \ldots, \InputValue[_{d}])$ ($d \geq 0$) with their parents $(\InputValue[_0], \ldots, \InputValue[_{d-1}])$ and are labelled with $\InputValue[_{d}]$ (when $d = 0$, $(\InputValue[_0], \ldots, \InputValue[_{d-1}])$ is conventionally assumed to be the empty prefix).
						Every leaf node (or, equivalently, the sequence of edge labels along the unique path from root to it) is uniquely associated to a complete trace of $\TraceSet$. 

						Clearly, a shortest \SIMcampaignProse for $\TraceSet$ must be long at least $\TimeQuantum l$, where $\TimeQuantum \in \TimeSet$ is the time quantum associated to traces in $\TraceSet$ and $l$ is the number of edges of the tree.

						A simulation campaign long exactly $\TimeQuantum l$ can be easily generated for any ordering $\TraceSetDef[j]$ of $\TraceSet$, by considering paths of the tree connecting the root to the leaves (corresponding to the traces) in the required order. 
						The campaign is very simple, given that for this proof we can rely on unlimited simulator memory.

						Namely, we want to traverse each edge of the tree from its parent to its child node exactly once, with the aim to reach all the leaves (representing all the traces in $\TraceSet$) in the required order.
						When traversing edge from $(\InputValue[_0], \ldots, \InputValue[_{d-1}])$ (parent, $d \geq 0$) to $(\InputValue[_0], \ldots, \InputValue[_{d}])$ (child node), we issue commands:
						$\SIMstore(\SIMstateID)$ (for a fresh identifier $\SIMstateID$), $\SIMrun(\InputValue[_{d}], \TimeQuantum)$. 
						When reaching a leaf node, we issue command $\SIMout$. 
						If the trace just considered is not the last one in the given order, we start the new trace by issuing command $\SIMload(\SIMstateID)$, where $\SIMstateID$ is the \emph{deepest} state of the tree already saved by a previous \SIMstore command along the root-to-leaf path identifying the new trace, and continue from there.
					\end{proof}
				\end{SupplMatStatement}
			\end{Section}

			\begin{Section}
				%[app:sec:simcampaigns:parallel]
				{Parallel \SIMcampaignProse+}
				%{parallel/content.tex}

				\label[appendix]{app:sec:simcampaigns:parallel}
				The answer to a \Ac{SLFV} problem $\SLFV = \SLFVdef$ (\ie the collection of the simulator outputs at the end of each scenario) can be computed by arbitrarily partitioning $\TraceSet$ into $\NbSlices \in \IntegersGZ$ subsets (\emph{slices}) $\TraceSet[_0], \ldots, \TraceSet[_{\NbSlices - 1}]$ (where $\NbSlices$ is the number of available computational nodes), and by computing and taking the union of the answers to the $\NbSlices$ smaller \Ac{SLFV} problems $\SLFV[_i] = \SLFVdef[_i]$, $i \in [0, \NbSlices - 1]$.
				In our simulation-based setting, this can be achieved using \NbSlices simulators for \SUV running as \NbSlices \emph{independent} processes (\eg in parallel in a \Ac{HPC} infrastructure) and \emph{independently} driven by \NbSlices \SIMcampaignProse+ \SIMcampaign[_1], \ldots, \SIMcampaign[_{\NbSlices}], 
				where, for all $i$, \SIMcampaign[_i] is a \SIMcampaignProse for \SLFV[_i]. 
				\Cref{def:simulation-campaign:slfv:parallel} formalises this concept.

				\begin{SupplMatStatement}{def:simulation-campaign:slfv:parallel}[\protect{\SIMcampaignParallelProse^()} for a \Ac{SLFV} problem]
					%\Input(absolute){40-simulation-based-slfv/parallel/def_sim_campaign_parallel.tex}
					%
					A \emph{\SIMcampaignParallelProse} for \Ac{SLFV} problem $\SLFV = \SLFVdef$ is a tuple 
						$\SIMcampaignParallel = \SIMcampaignParallelDef$
					such that there exists a partition of \TraceSet into sets
						$\TraceSet[_0], \ldots, \TraceSet[_{\NbSlices - 1}]$ 
					such that, 
					for all $i$, %
					\SIMcampaign[_i] is a \SIMcampaignProse for $\SLFV[_i] = \SLFVdef[_i]$.

					The \emph{length} of \SIMcampaign is $\SIMcampaignLen(\SIMcampaign) = \max_{i=0}^{\NbSlices - 1} \SIMcampaignLen(\SIMcampaign[_i])$.
					Given $\SIMmemorySize \in \IntegersGZ \cup \Set{\infty}$, \SIMcampaignParallel is a \emph{\SIMcampaignParallelMaxMemProse} if all \SIMcampaign[_i]{}s are \SIMcampaignMaxMemProse+.
				\end{SupplMatStatement}
			\end{Section}
		\end{Section}

		\begin{Section}
			%[app:sec:algo]
			{Parallel computation of \SIMcampaignParallelProse()+}
			%{50-algorithm/content.tex}

			\label[appendix]{app:sec:algo}
			\begin{Section}
				%[app:sec:algo:optimiser]
				{Computing a \SIMcampaignProse from each slice}
				%{30-optimiser/content.tex}

				\label[appendix]{app:sec:algo:optimiser}

				\let\tmpSIM\SIM%
				\def\SIM{\tmpSIM[_i]}%
				\let\tmpOPTstateID\OPTstateID%
				\def\OPTstateID{\tmpOPTstateID[_{i}]}%
				\let\tmpSIMcampaign\SIMcampaign%
				\def\SIMcampaign{\tmpSIMcampaign[_i]}%
				\let\tmpWholeTraceSet\TraceSet%
				\def\TraceSet{\tmpWholeTraceSet[_i]}%
				\let\tmpWholeTraceSetLabelled\TraceSetLabelled%
				\def\TraceSetLabelled{\tmpWholeTraceSetLabelled[_i]}%
				\let\tmpWholeTraceSetLabelledRnd\TraceSetLabelledRnd%
				\def\TraceSetLabelledRnd{\tmpWholeTraceSetLabelledRnd[i]}%

				\begin{Section}
					%[app:sec:algo:optimiser:bt]
					{Computing the \Acl{BT}}
					%{bt/content.tex}

					\label[appendix]{app:sec:algo:optimiser:bt}
					In the following, given two \TraceProse+ $\Trace[_a]$ and $\Trace[_b]$, we denote by
					$\Trace[_a] \IsPrefixOf \Trace[_b]$ (\respectively $\Trace[_a] \IsProperPrefixOf \Trace[_b]$) the fact that $\Trace[_a]$ (possibly the empty sequence) is a \emph{prefix} (\respectively, a \emph{proper prefix}) of $\Trace[_b]$.
					Also, by exploiting the fact that set $\InputSet$ is ordered, we denote by $\Trace[_a] \IsLexLess \Trace[_b]$ the fact that \Trace[_a] is \emph{lexicographically less} than \Trace[_b]. % 
						
					\Cref{def:lsp} defines the notions of \Acl{LSP} and that of \Acl{BT}.

					\begin{definition}[{\Acf[, ]{LSP}[]; \Acf[, ]{BT}[]}]
					\AcronymReset{LSP,BT}%
					\label{def:lsp}
					\label{def:bt}
					\label{def:lspt}
					Let \TraceSet be a finite collection of \TraceProse+\relax (\eg a slice of \tmpWholeTraceSet) with values in $\SUVinputSet$.

					A \Ac{LSP} for \TraceSet is a (possibly empty) sequence \Trace of inputs (\ie sequences of values of \SUVinputSet) such that there exist two traces $\Trace[_a]$ and $\Trace[_b]$ in \TraceSet such that: 
							$\Trace \IsPrefixOf \Trace[_a]$,
							$\Trace \IsPrefixOf \Trace[_b]$,
							and there exists no $\Trace[']$ in \TraceSet such that
								$\Trace \IsProperPrefixOf \Trace[']$,
								$\Trace['] \IsPrefixOf \Trace[_a]$, and
								$\Trace['] \IsPrefixOf \Trace[_b]$.

					A \Ac{BT} for \TraceSet is a tree $\BT = \BTdef$ such that:

					\begin{conditions}
						\item \label{item:bt:nodes}
						Nodes (set \BTvertexSet) denote \emph{distinct} \Acp{LSP} of \TraceSet. 

						\item \label{item:bt:edges}
						The parent node of \Trace (if one exists) is $\BTparent(\Trace) = \Trace[_p] \in \BTvertexSet$ such that
						$\Trace[_p] \IsProperPrefixOf \Trace$ and there exists no $\Trace['] \in \BTvertexSet$ such that $\Trace[_p] \IsProperPrefixOf \Trace['] \IsProperPrefixOf \Trace$.
					\end{conditions}

					\Cref{item:bt:edges} implies that a \Ac{BT} is a rooted tree.

					The following functions are defined over nodes of \BT (set $\BTvertexSet$):
					\begin{enumerate}[a)]
					\item 
					Function
					$\BTdepth : \BTvertexSet \to \IntegersGEZ$ associates to each node $\Trace = \TraceDef<d>$ of \BT its length $d$, which represents the time point $d \TimeQuantum$ reached by the simulator (starting from its initial state) after having injected input sequence $\Trace$. 
					The $\BTdepth$ of the node associated to the empty sequence is zero.

					\item 
					$\BTntraces : \BTvertexSet \to \IntegersGZ$, which associates to each node \Trace the number of traces in \TraceSet having \Trace as a (proper or non-proper) prefix, \ie
						$\BTntraces(\Trace) = 
						\SetCardinality{\Set{
							\Trace['] \in \TraceSet 
							~|~ 
							\Trace \IsPrefixOf \Trace[']
						}}$.
					\end{enumerate}

					A \Ac{BT} $\BT = \BTdef$ for \TraceSet is \emph{complete} if no \Ac{BT} $\BT['] = \BTdef[']$ exists for \TraceSet such that $\BTvertexSet \subset \BTvertexSet[']$.

					The \emph{size} of \Ac{BT} $\BT = \BTdef$ is $\BTsize(\BT) = \BTsizeDef$, \ie the number of its nodes.
					\end{definition}

					The goal of function \OPTbuildBT() is to build a complete \Ac{BT} for \TraceSet in central memory. 
					To this end, the algorithm scans \TraceSet in lexicographic order, since, under this ordering, deciding which trace prefixes are nodes of the tree is straightforward and memory-efficient. 

					To keep an as small as possible RAM footprint of the \Ac{BT}, the algorithm represents in central memory each of its nodes $\TraceDef<d>$ by a \emph{unique identifier} $\SIMstateID(\TraceDef*<d>)$.
					Unique identifiers for each trace prefix are available for free when traces are extracted from a scenario generator. 
					In case traces are taken from an input database, any efficiently computable \emph{injective} function of finite sequences of input values (or even a cryptographic hash function, when the probability of conflicts is small enough) can be used.

					Pseudocode of function \OPTbuildBT()  
					is reported in \cref{app:algo:optimiser:bt}.

					\begin{algorithm}
						%\Input{pseudocode/build-BT.tex}
						\AlgoFunction{\OPTbuildBT(\TraceSet)}{%
							\KwIn{$\TraceSet$, slice of traces}
							\KwOut{$\BT = \BTdef$, 
									\Acl{BT} of simulator states}
							
							\AlgoVarGets{$\BT$}{empty tree}\;
							\AlgoVarGets{\OPTlastTrace}{empty (will keep last trace)}\;

							\ForEach{$\Trace \in \TraceSet$ in lex order}{%
								\If{\Trace is \emph{not} the first trace in \TraceSet}{%
									\AlgoVarGets
										{\OPTlsp}{longest (possibly empty) prefix 
											shared by
											\Trace and \OPTlastTrace}%
									\; \label{line:algo:optimiser:bt:lsp}%
									
									\AlgoVarGets
										{\OPTpar}
										{longest node in \BTvertexSet s.t.\
											$\OPTpar \IsPrefixOf \OPTlsp$
											(possibly \OPTnone)					
										}%				
									\; \label{line:algo:optimiser:bt:par}%
								
									\If{$\OPTlsp \not\in \BTvertexSet$}{
										add \OPTlsp to \BTvertexSet\;
										\AlgoVarGets{$\BTparent(\OPTlsp)$}{\OPTpar}
									\; \label{line:algo:optimiser:bt:add}%
										
										\AlgoVarGets
											{\OPTchild}
											{shorter prefix of \OPTlastTrace s.t.\
												$\OPTlsp \IsProperPrefixOf \OPTchild \in \BTvertexSet$
												and $\BTparent(\OPTchild) = \OPTpar$ 
												(\OPTpar can be \OPTnone; at most one such node exists)%
											}
											\nllabel{line:algo:optimiser:bt:child}%
										\;				

										\If{$\OPTchild$ exists}{%
											\nllabel{line:algo:optimiser:bt:tree-rearrangement}%
											\AlgoVarGets{\BTparent(\OPTchild)}{\OPTlsp}\;

											\AlgoVarGets
												{$\BTntraces(\OPTlsp)$}
												{$\BTntraces(\OPTchild)$}%
												\nllabel{line:algo:optimiser:bt:weight-from-child}%
											\;
										}\lElse{%
											\AlgoVarGets
												{$\BTntraces(\OPTlsp)$}
												{1}%
												\nllabel{line:algo:optimiser:bt:weight-one}%
										}
									}% end \If new not in tree
									
									\AlgoVarGets{\AlgoVar{prefix}}{\OPTlsp}%
										\nllabel{line:algo:optimiser:bt:revise-weights:start}%
									\;
									\While{$\AlgoVar{prefix}$ is not \OPTnone}{%
										\AlgoVarIncr
											{$\BTntraces(\AlgoVar{prefix})$}
										\;
										\AlgoVarGets
											{\AlgoVar{prefix}}
											{$\BTparent(\AlgoVar{prefix})$}%
											\nllabel{line:algo:optimiser:bt:revise-weights:end}%
										\;
									}
								}
								\AlgoVarGets
									{\OPTlastTrace}{\Trace}
									\nllabel{line:algo:optimiser:bt:set-last}%
								\;
							}
							\Return{\BT}\;
						}
						
						\caption{Function \OPTbuildBT().}
						\label{app:algo:optimiser:bt}	
					\end{algorithm}

					%\Input{addnode.tex}
					\Cref{app:algo:optimiser:bt} aims at creating a new node of the \Ac{BT} for any simulator state associated to a prefix of the current trace \Trace that satisfies \cref{item:bt:nodes} of \cref{def:bt}.

					To recognise the trace prefixes to add as nodes, the function again exploits the fact that traces in \TraceSet can be accessed in lexicographic order. 
					This implies that (unique identifiers of) prefixes of \Trace that can be added as nodes of the \Ac{BT} must belong to the last-processed trace \OPTlastTrace. Also, at most one prefix of each \Trace can be added as a node of the \Ac{BT}.

					Thus, \cref{app:algo:optimiser:bt} proceeds as follows.

					\begin{steps}
					\item
					It selects the longest (possibly empty) prefix \OPTlsp shared by \Trace and \OPTlastTrace (line~\ref{line:algo:optimiser:bt:lsp}) and the longest prefix $\OPTpar \in \BTvertexSet$ such that $\OPTpar \IsPrefixOf \OPTlsp$ (possibly \OPTnone, line~\ref{line:algo:optimiser:bt:par}).

					\item
					It infers that the current trace \Trace and the previously processed trace \OPTlastTrace are identical up to \OPTlsp and differ at the next time point.

					\item
					If $\OPTlsp \in \BTvertexSet$, then \OPTlsp is already a node of the \Ac{BT}.
					Otherwise, \OPTlsp satisfies \cref{item:bt:nodes} of \cref{def:bt} and is added as a new node of the \Ac{BT}.
					In particular:
						\begin{steps}
							\item Node \OPTlsp is added to the \Ac{BT} as a \emph{child} node of \OPTpar (which can possibly be the empty prefix or \OPTnone; in the latter case, \OPTlsp becomes the root of the \Ac{BT}). 

							\item As node \OPTlsp could already have children in the \Ac{BT}, the tree may need to be rearranged to accommodate the new label \OPTlsp.
							Tree rearrangement is again very efficient thanks to the lexicographic order of the input traces. In fact, \OPTpar can have at most one child that needs to be moved and needs to become a child of \OPTlsp.

							This child, if exists, must be the shortest prefix \OPTchild of the previous trace \OPTlastTrace already in the \Ac{BT}, and such that $\OPTlsp \IsProperPrefixOf \OPTchild$ and $\BTparent(\OPTchild) = \OPTpar$ (where $\OPTpar$ can be \OPTnone, line~\ref{line:algo:optimiser:bt:child}).

							If such a child node exists, then it is moved as to become a child of \OPTlsp (line~\ref{line:algo:optimiser:bt:tree-rearrangement}) and value for $\BTntraces(\OPTlsp)$ is temporarily set to $\BTntraces(\OPTchild)$.

							Otherwise (node \OPTchild does not exist), value for $\BTntraces(\OPTlsp)$ is temporarily set to 1 (to account that it occurs in \OPTlastTrace).
						\end{steps}

						\item
						In both cases, the values of \BTntraces(\AlgoVar{prefix}) for each \Ac{BT} node $\AlgoVar{prefix} \IsPrefixOf \OPTlsp$ (including both \OPTlsp and the empty prefix, if the latter is in the \Ac{BT}) are incremented by 1, in order to take into account their occurrence in trace \Trace (lines~\ref{line:algo:optimiser:bt:revise-weights:start}--\ref{line:algo:optimiser:bt:revise-weights:end}).
					\end{steps}

					\begin{lemma}[Function \OPTbuildBT()]
						\label{thm:bt}
					 	%\Input{thm-bt_stmt.tex}
					 	Data structure \BT computed by function \OPTbuildBT() (\cref{app:algo:optimiser:bt}) is a complete \Ac{BT} of $\TraceSet$ according to \cref{def:bt}. 

						\begin{proof}
							%\Input{thm-bt_proof.tex}
							Let $\TraceSet = \TraceSetDef$.
							We prove the lemma by induction, by showing that, for each $j \in [1, \NbTraces]$, the tree $\BT$ computed by function \OPTbuildBT() after having processed the set $\TraceSet^j = \TraceSetDef<j>$ of the first $j$ traces of \TraceSet in lexicographic order is a complete \Ac{BT} for $\TraceSet^j$.

							\begin{description}[wide]
							\item[Base case] When $j=1$, $\TraceSet^j$ consists of a single disturbance trace $\Trace[_0] = \TraceDef$.
							In this case, function \OPTbuildBT() would just store $\Trace[_0]$ as $\OPTlastTrace$ and $\BT$ would be the empty tree, which is a complete \Ac{BT} when the set of traces is a singleton. 
							The thesis trivially follows.

							\item[Inductive case] Assume that, after having processed the first $j-1 \geq 1$ traces of \TraceSet (\ie set $\TraceSet^{j-1}$), data structure \BT computed by \cref{app:algo:optimiser:bt} is a complete \Ac{BT} for $\TraceSet^{j-1}$ according to \cref{def:bt}.
							We now show that, after processing trace \Trace[_{j-1}], the revised \BT is a complete \Ac{BT} for $\TraceSet^{j}$.

							As $\Trace[_{j-1}]$ is not the first processed trace, we have that: $\OPTlastTrace = \Trace[_{j-2}]$ (as set in the previous iteration, see line~\ref{line:algo:optimiser:bt:set-last} of \cref{app:algo:optimiser:bt}), \OPTlsp is set (line~\ref{line:algo:optimiser:bt:lsp}) to the longest (possibly empty) prefix shared by $\Trace[_{j-1}]$, and $\OPTlastTrace = \Trace[_{j-2}]$, and \OPTpar is set (line~\ref{line:algo:optimiser:bt:par}) to the longest prefix of \OPTlsp (hence, occurring in both $\Trace[_{j-1}]$ and $\Trace[_{j-2}]$) denoting a node already in $\BT$, and to `\OPTnone' if none exists.

							Clearly, $\OPTlsp$ is a \Ac{LSP} of $\TraceSet^{j}$ and, since traces in \TraceSet are lexicographically ordered, no additional \Acp{LSP} of $\TraceSet^{j}$ can exist.

							If $\OPTlsp$ is already a node in $\BT$, the algorithm ignores it and increments by 1 the value of $\BTntraces$ for all the \BT nodes occurring in the current trace $\Trace[_{j-1}]$ (lines~\ref{line:algo:optimiser:bt:revise-weights:start}--\ref{line:algo:optimiser:bt:revise-weights:end}), thus making such values satisfy again \cref{def:bt}.

							On the other hand, in case $\OPTlsp \not\in \BTvertexSet$, the algorithm adds it to the tree (line~\ref{line:algo:optimiser:bt:add}) making the tree complete \wrt $\TraceSet^{j}$. The new node is added to $\BT$ as a child of $\OPTpar$, which, by construction, is either the longest \emph{proper} prefix of $\OPTlsp$ already in $\BTvertexSet$ or `\OPTnone' (if no such prefix exists, in which case, the newly added $\OPTlsp$ becomes the new root of the tree), and thus satisfies \cref{item:bt:edges} of \cref{def:bt}.

							However, the introduction of a new node in $\BT$ ($\OPTlsp$) could make \cref{item:bt:edges} false for some of the pre-existing children nodes of $\OPTpar$ in $\BT$.

							For a child node $\OPTchild$ of $\OPTpar$ to violate \cref{item:bt:edges} in the current iteration, it must be that $\OPTpar \IsProperPrefixOf \Trace[_x] \IsProperPrefixOf \OPTchild$ for some $\Trace[_x] \in \BTvertexSet$.
							Since the only newly added node is \OPTlsp, it must be $\Trace[_x] = \OPTlsp$.
							Also, given that the input traces are in lexicographic order, at most one such child node exists.
							\Cref{app:algo:optimiser:bt}, by reassigning the parent of \OPTchild (if it exists) to \OPTlsp, makes \cref{item:bt:edges} of \cref{def:bt} true again.

							The only thing that remains to show is that the value of \BTntraces for all nodes of \BT is correctly revised.
							We show this in two steps:
							\begin{steps}
							\item \label{item:thm:bt:proof:revise-weights:past-traces} The value of \BTntraces for nodes of the tree as computed before line~\ref{line:algo:optimiser:bt:revise-weights:start}, is correct if we consider only traces processed up to iteration $j-1$.
							\item \label{item:thm:bt:proof:revise-weights:current-trace} The current trace $\Trace[_{j-1}]$ is correctly taken into account in the revision of \BTntraces values, in lines~\ref{line:algo:optimiser:bt:revise-weights:start}--\ref{line:algo:optimiser:bt:revise-weights:end}.
							\end{steps}

							As for \cref{item:thm:bt:proof:revise-weights:past-traces}, only the value of $\BTntraces$ for the newly inserted node $\OPTlsp$ must be set (in case such node is added to $\BT$). Node $\OPTlsp$ might have zero or one children (as seen above).

							If $\OPTlsp$ has a child node ($\OPTchild$), then the set of traces within $\TraceSet^{j-1}$ (hence, excluding the current trace $\Trace[_{j-1}]$) having prefix $\OPTlsp$ are exactly those having (the longer) prefix $\OPTchild$ (line~\ref{line:algo:optimiser:bt:weight-from-child}).

							Otherwise ($\OPTlsp$ has no children), trace $\Trace[_{j-1}]$ shares $\OPTlsp$ as a prefix only with (the previous) trace $\Trace[_{j-2}]$ among those seen so far.
							To see why, assume, for the sake of contradiction, that two traces $\Trace[_a]$ and $\Trace[_b]$ exist in $\TraceSet^{j-1}$ which both have $\OPTlsp$ as a prefix.
							Since $\TraceSet^{j-1}$ is lex-ordered, there must exists $\Trace[_p]$ such that
							$\OPTlsp \IsPrefixOf \Trace[_{p}] \IsPrefixOf \Trace[_a]$ and $\OPTlsp \IsPrefixOf \Trace[_{p}] \IsPrefixOf \Trace[_b]$, and such that \Trace[_a] and \Trace[_b] differ immediately after \Trace[_p].
							But this would mean that $\Trace[_p]$ would have been recognised as a \Ac{LSP} and added as a node of $\BT$ in a previous iteration of the algorithm, when the last trace between $\Trace[_a]$ and $\Trace[_b]$ was processed (contradiction). This proves the correctness of setting $\BTntraces(\OPTlsp)$ to 1 in line~\ref{line:algo:optimiser:bt:weight-one} (when the current trace $\Trace[_{j-1}]$ has not yet been considered in this computation).

							As for \cref{item:thm:bt:proof:revise-weights:current-trace}, lines~\ref{line:algo:optimiser:bt:revise-weights:start}--\ref{line:algo:optimiser:bt:revise-weights:end} increment by 1 the value of $\BTntraces$ for all the nodes of $\BT$ occurring as prefixes of the currently processed trace $\Trace[_{j-1}]$ (including $\OPTlsp$), thus taking into correct account the existence of the $\Trace[_{j-1}]$.
							\end{description}

							As a result of the above, $\BT$ is a complete \Ac{BT} for $\TraceSet^{j}$ and the thesis follows.
						\end{proof}
					\end{lemma}
				\end{Section}
			\end{Section}

			\begin{Section}
				%[app:sec:algo:correctness]
				{Algorithm correctness}
				%{40-correctness/content.tex}

				\label[appendix]{app:sec:algo:correctness}

				\begin{SupplMatStatement}{thm:correctness}[Correctness of \cref{algo:optimiser}]
					%\Input(absolute){50-algorithm/40-correctness/thm-correctness_stmt.tex}
					Let $\SLFV = \SLFVdef$ be a \Ac{SLFV} problem for \Ac{SUV} $\SUV = \SUVdef$,
						with $\TraceSet$ being defined as \TraceProse+ associated to time quantum $\TimeQuantum \in \TimeSet \setminus \Set{0}$, and let $\SIMmemorySize \in \IntegersGZ$.

					Given any partition $\Set{\TraceSet[_0], \ldots, \TraceSet[_{\NbSlices-1}]}$ ($\NbSlices \in \IntegersGZ$) of $\TraceSet$, let $\SIMcampaignParallel = \SIMcampaignParallelDef$ be the \SIMcampaignParallelProse such that $\SIMcampaign[_i]$ ($i \in [0, \NbSlices-1]$) is computed by \cref{algo:optimiser} on inputs $\TraceSet[_i]$ (under any user-defined order), $\TimeQuantum$, and $\SIMmemorySize$.

					We have that:
					\begin{points}
						\item \label{item:thm:correctness:sequence}
						For all $i \in [0, \NbSlices-1]$, the sequence $\SIMinputFunctionSeq(\SIMcampaign[_i])$ is $\TraceSet[_i]$;
						\item \label{item:thm:correctness:shortest}
						There exists $\SIMmemorySizeOpt \in \IntegersGZ$ such that, if $\SIMmemorySize \geq \SIMmemorySizeOpt$, all $\SIMcampaign[_i]$s ($i \in [0, \NbSlices-1]$) are \emph{shortest} \SIMcampaignMaxMemProse+.
					\end{points}

					\begin{proof}
						%\Input(absolute){50-algorithm/40-correctness/thm-correctness_proof.tex}
						\NewDocumentCommand{\THMdiscontinuityIdx}{m}{
							\ensuremath{j_{#1}}\xspace%
						}
						\NewDocumentCommand{\THMnbDiscontinuities}{O{i}}{\ensuremath{r_{#1}}\xspace}

						\NewDocumentCommand{\THMrunLength}{r[]}{%
							\ensuremath{
								\Fun{len}_{#1}%
							}\xspace%
						}
						\NewDocumentCommand{\THMrunLengthDef}{r[]}{%
							\ensuremath{
								%\TimeQuantum
								(\AlgoVar{end}_{#1} - \AlgoVar{start}_{#1} + 1)
							}\xspace%
						}
						\NewDocumentCommand{\THMrunDuration}{r[]}{%
							\ensuremath{
								\TimeQuantum \THMrunLength[#1]
							}\xspace%
						}

						We first make the following observations:

						\begin{enumerate}[a), wide]
						\item At any time during execution of \cref{algo:optimiser}, when sequence of simulator commands $\SIMcampaign$ (a prefix of the overall \SIMcampaignProse) has been computed, the set of simulator states $\SIMstateID$ such that $\OPTstored(\SIMstateID) = \True$ are \emph{exactly} those that would be available in the simulator memory after the actual execution of $\SIMcampaign$.
						In particular, $\OPTstored(\SIMstateID)$ is set to \True (respectively \False) immediately after appending to $\SIMcampaign$ command $\SIMstore(\SIMstateID)$ (respectively $\SIMfree(\SIMstateID)$).

						\item Each of the computed \SIMcampaignProse+ is \emph{executable}. 
						This is immediate, as $\SIMload(\SIMstateID)$/$\SIMfree(\SIMstateID)$ (respectively, $\SIMstore(\SIMstateID)$) commands are issued only for simulator state identifiers $\SIMstateID$ available (respectively, not available) in the simulator memory, as requested by \cref{def:simulator:transition}.
						Also, when $j > 0$, function $\OPTsimCmds()$ always finds state $\SIMstateID_{\OPTidxload}$ to load in \cref{algo:optimiser:sim_cmds:line:load}. This is because, if not, then the $j$-th trace would have no prefix in common with any of the previous traces, not even the \emph{empty} prefix, which is impossible.

						\end{enumerate}

						\medskip

						The simulation campaign \SIMcampaign generated by \cref{algo:optimiser} has the form:

						\begin{equation*}
							\SIMcampaign[_i] ~=~ \SIMcampaignStructureDef[i]
						\end{equation*}
						where $\NbTraces = \SetCardinality{\TraceSet[_i]}$ and, for each $j \in [0, \NbTraces-1]$, $\SIMcampaign[_{i,j}]$ is the sequence of simulator commands generated by function \OPTsimCmds() (\cref{algo:optimiser:sim_cmds}) on $\Trace[_j]$, the $j$-th trace of $\TraceSet[_i]$. 

						From \cref{algo:optimiser:sim_cmds}, we know that each $\SIMcampaign[_{i,j}]$ starts with a \SIMload command if $j > 0$ (omitted for $j=0$), and then continues with a sequence of \SIMrun commands (possibly interleaved with \SIMstore and/or \SIMfree commands). Each \SIMrun command covers a distinct constant portion of $\Trace[_j]$, and, together, they cover the entire postfix of $\Trace[_j]$ after the prefix loaded with \SIMload, if any. 
						Finally, $\SIMcampaign[_{i,j}]$ terminates with an \SIMout command.

						Executability of $\SIMcampaign[_{i}]$ guarantees that all \SIMload, \SIMstore and \SIMfree commands succeed, and \cref{def:simulator:commands,thm:simulation-campaign:input-functions} together ensure that the input time function associated to the \emph{final} state reached by the simulator when executing each $\SIMcampaign[_{i,j}]$ (for all $j$) is exactly $\InputFunction[_j]$.
						Thus, \cref{item:thm:correctness:sequence} immediately follows from \cref{def:simulation-campaign:input-functions}.

						As for the proof of \cref{item:thm:correctness:shortest}, if $\SIMmemorySize$ is at least the number of nodes of the \Ac{BT}, then each $\SIMcampaign[_i]$ is a shortest campaign. 
						This can be shown along the lines of the proof of \cref{thm:simulation-campaigns:shortest-any-order}, observing that, by \cref{thm:bt}, $\BT$ is a a complete \Ac{BT} for $\TraceSet[_i]$ (\cref{thm:bt}).

						However, we also observe that the number of nodes of $\BT$ is most often a \emph{very loose} upper bound for $\SIMmemorySizeOpt$. 
						In most cases, the simulation memory required to produce shortest campaigns is \emph{much} smaller than the number of nodes of the \Ac{BT}. 
						This is because function $\OPTsimCmds$ (\cref{algo:optimiser:sim_cmds}) greedily frees simulator memory (by injecting \SIMfree commands) \emph{has soon as} it discover that a state will not be needed to shorten the simulation of future traces (\ie as soon as its associated counter $\BTntraces$ becomes zero). 
						The actual value for $\SIMmemorySizeOpt$ of course depends on the (possibly random) simulation order chosen by the user.

					\end{proof}
				\end{SupplMatStatement}

			\end{Section}
			 
		\end{Section}

		\begin{Section}
			%[app:sec:expres:case-studies]
			{Case studies}
			%{60-expres/20-case-studies/content.tex}
			\label[appendix]{app:sec:expres:case-studies}
			Our scenario generators for our three case studies below are inspired from those in \cite{mancini-etal:2021:tse-supervisory}.

			In order to focus the \Ac{SLFV} activity on clearly selected portions of the space of inputs and to keep the overall number of traces under control, our scenario generators enforce various constraints on the entailed scenarios. 

			We chose such constraints from (slight variations of) those in \cite{mancini-etal:2021:tse-supervisory} with the final aim to have a number of entailed traces of around 50, 100, and 200 million for \Ac{BDC}, \Ac{ALMA}, and \Ac{FCS}, respectively.
			Multiple \SIMcampaignProse+, then, have been computed (see \cref{sec:expres}) for various random portions of such sets of traces (from 25\% to 100\%), different random seeds, and different optimisation settings.

			\begin{Section}
				{\Acf{BDC}}
				%{buck/content.tex}
				\label[appendix]{app:sec:expres:case-studies:buck}
				The \Ac{BDC} \Ac{SUV} model takes two inputs: the input voltage $V_i$ and the load $R$, which vary during time.

				Our scenario generator enforces the following constraints on the time course of its inputs:

				\begin{enumerate}
				\item Both $V_i$ and $R$ may vary during time up to at most $\pm 30\%$ of their nominal values, in steps of $\pm 5\%$ and $\pm 10\%$ of their initial values.

				\item Values for $V_i$ and $R$ are stable for 5 and 6 \Acp{tu}, \respectively.

				\item To have a proper set-up, $V_i$ and $R$ are assumed stable to their nominal values for the first 2 \Acp{tu}

				\item $V_i$ and $R$ do not change simultaneously.

				\item Whenever $V_i$ changes, $R$ will change after 7 \Acp{tu}
				\end{enumerate}

				By enforcing the constraints above and a time horizon of \DataGet<experiments/scenarios/buck>{horizon} \Acp{tu}, our scenario generator entails \DataGet<experiments/scenarios/buck>{number} \TraceProse+, each one defining a piecewise constant function (time quantum $\TimeQuantum = 1$ \Ac{tu}) over an input space of \DataGet<experiments/scenarios/buck>{input space/size} different values.
			\end{Section}

			\begin{Section}
				{\Acf{ALMA}}
				%{apollo/content.tex}

				\label[appendix]{app:sec:expres:case-studies:apollo}
				The \Ac{ALMA} \Ac{SUV} model takes as inputs attitude change requests along each of the three axes (``Yaw'', ``Pitch'', ``Roll'') as well as events signalling temporary failures of actuators (reaction jets).

				Our scenario generator enforces the following constraints on the time course of its inputs:

				\begin{enumerate}
					\item Attitude requests arrive at each \Ac{tu}, and each request may ask for a unitary positive or negative change of the attitude along \emph{at most} one axis. 

					\item Attitude requests do not ask the autopilot to immediately undo the rotation requested along any axis in the preceding \Ac{tu}

					\item Two consecutive requests for attitude changes along the same axis are 10 or 11 \Acp{tu} apart.

					\item Only two given reaction jets (number 14 and 15) can be subject to temporary unavailability, which are always recovered within 6 to 7 \Acp{tu}. Jet unavailability events occur every 12 to 13 \Acp{tu}

					\item To obey each received attitude request, the autopilot decides which reaction jets must be used at each \Ac{tu}. We further constrain attitude requests so that no jet is used consecutively for more than 3 \Acp{tu}. Also, when a jet (among number 14 and 15) is used for 2 \Acp{tu} in a row, it will certainly become unavailable within 3 to 4 \Acp{tu}
				\end{enumerate}

				The following additional constraints were enforced to further limit the focus of the verification activity and the overall number of traces.

				\begin{enumerate}[resume]
					\item  
					Jet number 14 always becomes unavailable (respectively, becomes available) immediately after the reception of a request for a negative (respectively, positive) change in the Yaw attitude. 

					\item
					Jet number 15 always becomes unavailable (respectively, becomes available) immediately after the reception of a request for a negative (respectively, positive) change in the Roll or Pitch attitude.

					\item
					Whenever a positive (respectively, negative) attitude Roll request is received, the current attitudes along Yaw and Pitch are at least (respectively, less then) a given threshold value.

					\item
					Whenever a positive (respectively, negative or null) attitude Pitch request is received, the current attitudes along Yaw and Roll are at least (respectively, less then) a given threshold value.
				\end{enumerate}	

				By enforcing the constraints above and a time horizon of \DataGet<experiments/scenarios/apollo>{horizon} \Acp{tu}, our scenario generator entails \DataGet<experiments/scenarios/apollo>{number} \TraceProse+, each one defining a piecewise constant function (time quantum $\TimeQuantum = 1$ \Ac{tu}) over an input space of \DataGet<experiments/scenarios/apollo>{input space/size} different values.
			\end{Section}

			\begin{Section}
				%[app:sec:expres:case-studies:fcs]
				{Fault Tolerant \Acf{FCS}}
				%{fcs/content.tex}

				\label[appendix]{app:sec:expres:case-studies:fcs}
				The \Ac{FCS} \Ac{SUV} model takes as inputs failure events on its four sensors (``throttle'', ``speed'', ``ECO'', ``MAP'').

				Our scenario generator enforces the following constraints on the time course of its inputs:

				\begin{enumerate}
				\item At time zero, all sensors are functioning properly.

				\item Each faulty sensor recovers within the following time bounds (in \Acp[, ]{tu}[]):
					3--5 (throttle), 5--7 (speed), 10--15 (EGO), 13--17 (MAP).

				\item At most one sensor is faulty at any given time.

				\item A sensor fault occurs every 15--20 \Acp{tu}

				\item Whenever a fault on the throttle sensor occurs, a fault on the speed sensor will occur within 18 or 19 \Acp{tu}

				\item Whenever a fault on the EGO sensor occurs, a fault on the MAP sensor will occur within 20 or 21 \Ac{tu}
				\end{enumerate}

				By enforcing the constraints above and a time horizon of \DataGet<experiments/scenarios/fcs>{horizon} \Acp{tu}, our scenario generator entails \DataGet<experiments/scenarios/fcs>{number} \TraceProse+, each one defining a piecewise constant function (time quantum $\TimeQuantum = 1$ \Ac{tu}) over an input space of \DataGet<experiments/scenarios/fcs>{input space/size} different values.
			\end{Section}
		\end{Section}

	%\Bibliography[style=plain, verbosity=XS]
	\providecommand{\MCLabProceedingsOf}[1]{#1}
	
\end{BibliographyUnit}

\BoxArticleDisclaimer

\begin{thebibliography}{10}

		\bibitem{abbas-etal:2013:tecs}
		H.~Abbas, G.~Fainekos, S.~Sankaranarayanan, F.~Ivan\v{c}i\'{c}, and A.~Gupta.
		\newblock Probabilistic temporal logic falsification of cyber-physical systems.
		\newblock {\em ACM TECS}, 12(2s), 2013.

		\bibitem{adimoolam-etal:2017:cav}
		A.~Adimoolam, T.~Dang, A.~Donz\'{e}, J.~Kapinski, and X.~Jin.
		\newblock Classification and coverage-based falsification for embedded control
		  systems.
		\newblock In {\em \MCLabProceedingsOf{CAV~2017}}, volume 10426 of {\em LNCS}.
		  Springer, 2017.

		\bibitem{agha-etal:2018:surveysmc}
		G.~Agha and K.~Palmskog.
		\newblock A survey of statistical model checking.
		\newblock {\em {ACM} Trans.\ Model.\ Comput.\ Simul.}, 28(1), 2018.

		\bibitem{alturki-etal:2011:calco}
		M.~AlTurki and J.~Meseguer.
		\newblock {PVeStA}: A parallel statistical model checking and quantitative
		  analysis tool.
		\newblock In {\em \MCLabProceedingsOf{CALCO~2011}}, volume 6859 of {\em LNCS}.
		  Springer, 2011.

		\bibitem{annpureddy-etal:2011:tacas}
		Y.S.R. Annpureddy, C.~Liu, G.~E. Fainekos, and S.~Sankaranarayanan.
		\newblock S-{T}a{L}i{R}o: A tool for temporal logic falsification for hybrid
		  systems.
		\newblock In {\em \MCLabProceedingsOf{TACAS~2011}}, volume 6605 of {\em LNCS}.
		  Springer, 2011.

		\bibitem{bak-etal:2017:cav}
		S.~Bak and P.S. Duggirala.
		\newblock Simulation-equivalent reachability of large linear systems with
		  inputs.
		\newblock In {\em \MCLabProceedingsOf{CAV~2017}}, volume 10426 of {\em LNCS}.
		  Springer, 2017.

		\bibitem{basu-etal:2010:bookChapter}
		A.~Basu, S.~Bensalem, M.~Bozga, B.~Caillaud, B.~Delahaye, and A.~Legay.
		\newblock Statistical abstraction and model-checking of large heterogeneous
		  systems.
		\newblock In {\em Formal Techniques for Distributed Systems}. Springer, 2010.

		\bibitem{bogdoll-etal:2011:bookChapter}
		J.~Bogdoll, L.M.F. Fioriti, A.~Hartmanns, and H.~Hermanns.
		\newblock Partial order methods for statistical model checking and simulation.
		\newblock In {\em Formal Techniques for Distributed Systems}. Springer, 2011.

		\bibitem{bostrom:2016:contract}
		P.~Bostr\"{o}m and J.~Wiik.
		\newblock Contract-based verification of discrete-time multi-rate {S}imulink
		  models.
		\newblock {\em SoSyM}, 15(4), 2016.

		\bibitem{boyer-etal:2013:qest}
		B.~Boyer, K.~Corre, A.~Legay, and S.~Sedwards.
		\newblock {PLASMA}-lab: A flexible, distributable statistical model checking
		  library.
		\newblock In {\em \MCLabProceedingsOf{QEST~2013}}. Springer, 2013.

		\bibitem{clarke-etal:2010:probabilistic}
		E.M. Clarke, A.~Donz\'{e}, and A.~Legay.
		\newblock On simulation-based probabilistic model checking of mixed-analog
		  circuits.
		\newblock {\em Form.\ Meth.\ Sys.\ Des.}, 36(2), 2010.

		\bibitem{clarke-etal:2011:atva}
		E.M. Clarke and P.~Zuliani.
		\newblock Statistical model checking for cyber-physical systems.
		\newblock In {\em \MCLabProceedingsOf{ATVA~2011}}, volume~11. Springer, 2011.

		\bibitem{deshmukh-etal:2015:stochastic}
		J.~Deshmukh, X.~Jin, J.~Kapinski, and O.~Maler.
		\newblock Stochastic local search for falsification of hybrid systems.
		\newblock In {\em \MCLabProceedingsOf{ATVA~2015}}. Springer, 2015.

		\bibitem{donze:2010:breach}
		A.~Donz\'{e}.
		\newblock Breach, a toolbox for verification and parameter synthesis of hybrid
		  systems.
		\newblock In {\em \MCLabProceedingsOf{CAV~2010}}, volume 6174 of {\em LNCS}.
		  Springer, 2010.

		\bibitem{dreossi-etal:2015:efficient}
		T.~Dreossi, T.~Dang, A.~Donz\'{e}, J.~Kapinski, X.~Jin, and J.V. Deshmukh.
		\newblock Efficient guiding strategies for testing of temporal properties of
		  hybrid systems.
		\newblock In {\em \MCLabProceedingsOf{NFM~2015}}. Springer, 2015.

		\bibitem{fainekos-etal:2009:robustness}
		G.E. Fainekos and G.J. Pappas.
		\newblock Robustness of temporal logic specifications for continuous-time
		  signals.
		\newblock {\em TCS}, 410(42), 2009.

		\bibitem{fan-etal:2017:cav}
		C.~Fan, B.~Qi, S.~Mitra, and M.~Viswanathan.
		\newblock Dry{VR}: Data-driven verification and compositional reasoning for
		  automotive systems.
		\newblock In {\em \MCLabProceedingsOf{CAV~2017}}, volume 10426 of {\em LNCS}.
		  Springer, 2017.

		\bibitem{gonschorek-etal:2017:memocode}
		T.~Gonschorek, B.~Rabeler, F.~Ortmeier, and D.~Schomburg.
		\newblock On improving rare event simulation for probabilistic safety analysis.
		\newblock In {\em \MCLabProceedingsOf{MEMOCODE~2017}}. ACM, 2017.

		\bibitem{grosu-etal:2004:isola}
		R.~Grosu and S.A. Smolka.
		\newblock Quantitative model checking.
		\newblock In {\em \MCLabProceedingsOf{ISoLA~2004}}, 2004.

		\bibitem{grosu-etal:2005:tacas}
		R.~Grosu and S.A. Smolka.
		\newblock {M}onte {C}arlo model checking.
		\newblock In {\em \MCLabProceedingsOf{TACAS~2005}}, volume 3440 of {\em LNCS}.
		  Springer, 2005.

		\bibitem{hayes-etal:2016:tsg}
		B.P. Hayes, I.~Melatti, T.~Mancini, M.~Prodanovic, and E.~Tronci.
		\newblock Residential demand management using individualised demand aware price
		  policies.
		\newblock {\em {IEEE} Trans.\ Smart Grid}, 8(3), 2017.

		\bibitem{hoxha-etal:2017:sttt}
		B.~Hoxha, A.~Dokhanchi, and G.~Fainekos.
		\newblock Mining parametric temporal logic properties in model based design for
		  cyber-physical systems.
		\newblock {\em STTT}, 2017.

		\bibitem{jegourel-etal:2012:tacas}
		C.~Jegourel, A.~Legay, and S.~Sedwards.
		\newblock A platform for high performance statistical model checking--plasma.
		\newblock In {\em \MCLabProceedingsOf{TACAS~2012}}, volume 7214 of {\em LNCS}.
		  Springer, 2012.

		\bibitem{jegourel-etal:2013:cav}
		C.~Jegourel, A.~Legay, and S.~Sedwards.
		\newblock Importance splitting for statistical model checking rare properties.
		\newblock In {\em \MCLabProceedingsOf{CAV~2013}}, volume 8044 of {\em LNCS}.
		  Springer, 2013.

		\bibitem{kim-etal:2013:validating}
		Y.J. Kim, O.~Choi, M.~Kim, J.~Baik, and T.{-}H. Kim.
		\newblock Validating software reliability early through statistical model
		  checking.
		\newblock {\em {IEEE} Softw.}, 30(3), 2013.

		\bibitem{kim-kim:2012:hybrid}
		Y.J. Kim and M.~Kim.
		\newblock Hybrid statistical model checking technique for reliable safety
		  critical systems.
		\newblock In {\em \MCLabProceedingsOf{ISSRE~2012}}. IEEE, 2012.

		\bibitem{maler-etal:2004:monitoring}
		O.~Maler and D.~Nickovic.
		\newblock Monitoring temporal properties of continuous signals.
		\newblock In {\em \MCLabProceedingsOf{FORMATS/FTRTFT~2004}}, volume 3253 of
		  {\em LNCS}. Springer, 2004.

		\bibitem{mancini-etal:2013:cav}
		T.~Mancini, F.~Mari, A.~Massini, I.~Melatti, F.~Merli, and E.~Tronci.
		\newblock System level formal verification via model checking driven
		  simulation.
		\newblock In {\em \MCLabProceedingsOf{CAV~2013}}, volume 8044 of {\em LNCS}.
		  Springer, 2013.

		\bibitem{mancini-etal:2017:ipl}
		T.~Mancini, F.~Mari, A.~Massini, I.~Melatti, I.~Salvo, and E.~Tronci.
		\newblock On minimising the maximum expected verification time.
		\newblock {\em Inf.\ Proc.\ Lett.}, 122, 2017.

		\bibitem{mancini-etal:2014:dsd-anytime}
		T.~Mancini, F.~Mari, A.~Massini, I.~Melatti, and E.~Tronci.
		\newblock Anytime system level verification via random exhaustive hardware in
		  the loop simulation.
		\newblock In {\em \MCLabProceedingsOf{DSD~2014}}. IEEE, 2014.

		\bibitem{mancini-etal:2015:gandalf}
		T.~Mancini, F.~Mari, A.~Massini, I.~Melatti, and E.~Tronci.
		\newblock Simulator semantics for system level formal verification.
		\newblock {\em EPTCS}, 193, 2015.

		\bibitem{mancini-etal:2016:micpro}
		T.~Mancini, F.~Mari, A.~Massini, I.~Melatti, and E.~Tronci.
		\newblock Anytime system level verification via parallel random exhaustive
		  hardware in the loop simulation.
		\newblock {\em Microprocessors and Microsystems}, 41, 2016.

		\bibitem{mancini-etal:2016:fundam}
		T.~Mancini, F.~Mari, A.~Massini, I.~Melatti, and E.~Tronci.
		\newblock {SyLVaaS}: System level formal verification as a service.
		\newblock {\em Fundam.\ Inform.}, 149(1--2), 2016.

		\bibitem{mancini-etal:2021:jlamp}
		T.~Mancini, F.~Mari, A.~Massini, I.~Melatti, and E.~Tronci.
		\newblock On checking equivalence of simulation scripts.
		\newblock {\em J.\ Log.\ Algebr.\ Meth.\ Program.}, 120, 2021.

		\bibitem{mancini-etal:2014:smartgridcomm}
		T.~Mancini, F.~Mari, I.~Melatti, I.~Salvo, E.~Tronci, J.~Gruber, B.~Hayes,
		  M.~Prodanovic, and L.~Elmegaard.
		\newblock Demand-aware price policy synthesis and verification services for
		  smart grids.
		\newblock In {\em \MCLabProceedingsOf{SmartGridComm~2014}}. IEEE, 2014.

		\bibitem{mancini-etal:2018:smartgridcomm}
		T.~Mancini, F.~Mari, I.~Melatti, I.~Salvo, E.~Tronci, J.K. Gruber, B.~Hayes,
		  and L.~Elmegaard.
		\newblock Parallel statistical model checking for safety verification in smart
		  grids.
		\newblock In {\em \MCLabProceedingsOf{SmartGridComm~2018}}. IEEE, 2018.

		\bibitem{mancini-etal:2015:dsd}
		T.~Mancini, F.~Mari, I.~Melatti, I.~Salvo, E.~Tronci, J.K. Gruber, B.~Hayes,
		  M.~Prodanovic, and L.~Elmegaard.
		\newblock User flexibility aware price policy synthesis for smart grids.
		\newblock In {\em \MCLabProceedingsOf{DSD~2015}}. IEEE, 2015.

		\bibitem{mancini-etal:2021:tse-supervisory}
		T.~Mancini, I.~Melatti, and E.~Tronci.
		\newblock Any-horizon uniform random sampling and enumeration of constrained
		  scenarios for simulation-based formal verification.
		\newblock {\em IEEE TSE}, 2021.

		\bibitem{mancini-etal:2015:iwbbio}
		T.~Mancini, E.~Tronci, I.~Salvo, F.~Mari, A.~Massini, and I.~Melatti.
		\newblock Computing biological model parameters by parallel statistical model
		  checking.
		\newblock In {\em \MCLabProceedingsOf{IWBBIO~2015}}, volume 9044 of {\em LNCS}.
		  Springer, 2015.

		\bibitem{mari-etal:2014:tosem}
		F.~Mari, I.~Melatti, I.~Salvo, and E.~Tronci.
		\newblock Model based synthesis of control software from system level formal
		  specifications.
		\newblock {\em {ACM} {TOSEM}}, 23(1), 2014.

		\bibitem{meenakshi-etal:2006:simulink}
		B.~Meenakshi, A.~Bhatnagar, and S.~Roy.
		\newblock Tool for translating {S}imulink models into input language of a model
		  checker.
		\newblock In {\em \MCLabProceedingsOf{ICFEM~2006}}. Springer, 2006.

		\bibitem{miskov-etal:2013:bcb}
		N.~Miskov-Zivanov, P.~Zuliani, E.M. Clarke, and J.R. Faeder.
		\newblock Studies of biological networks with statistical model checking:
		  Application to immune system cells.
		\newblock In {\em \MCLabProceedingsOf{ACM-BCB~2013}}. ACM, 2013.

		\bibitem{rajhans-2021:specification}
		A~Rajhans, A.~Mavrommati, P.J. Mosterman, and R.G. Valenti.
		\newblock Specification and runtime verification of temporal assessments in
		  simulink.
		\newblock In {\em \MCLabProceedingsOf{RV~2021}}. Springer, 2021.

		\bibitem{sankaranarayanan-etal:2017:SIGBED}
		S.~Sankaranarayanan, S.A. Kumar, F.~Cameron, B.W. Bequette, G.~Fainekos, and
		  D.M. Maahs.
		\newblock Model-based falsification of an artificial pancreas control system.
		\newblock {\em {ACM SIGBED} Review}, 14(2), 2017.

		\bibitem{sinisi-etal:2021:fmu}
		S.~Sinisi, V.~Alimguzhin, T.~Mancini, and E.~Tronci.
		\newblock Reconciling interoperability with efficient verification and
		  validation within open source simulation environments.
		\newblock {\em Simul.\ Model.\ Pract.\ Theory}, 109, 2021.

		\bibitem{sinisi-etal:2020:bioinf}
		S.~Sinisi, V.~Alimguzhin, T.~Mancini, E.~Tronci, and B.~Leeners.
		\newblock Complete populations of virtual patients for in silico clinical
		  trials.
		\newblock {\em Bioinformatics}, 36(22--23), 2020.

		\bibitem{so-etal:1996:buck}
		W.-C. So, C.K. Tse, and Y.-S. Lee.
		\newblock Development of a fuzzy logic controller for {DC/DC} converters:
		  Design, computer simulation, and experimental evaluation.
		\newblock {\em {IEEE} Trans.\ Pow.\ Electr.}, 11(1), 1996.

		\bibitem{sontag:1998:book}
		E.D. Sontag.
		\newblock {\em Mathematical Control Theory: Deterministic Finite Dimensional
		  Systems (2nd Ed.)}.
		\newblock Springer, 1998.

		\bibitem{tripakis-etal:2005:translating}
		S.~Tripakis, C.~Sofronis, P.~Caspi, and A.~Curic.
		\newblock Translating discrete-time {S}imulink to {L}ustre.
		\newblock {\em ACM TECS}, 4(4), 2005.

		\bibitem{mancini-etal:2014:fmcad}
		E.~Tronci, T.~Mancini, I.~Salvo, S.~Sinisi, F.~Mari, I.~Melatti, A.~Massini,
		  F.~Dav{i'}, T.~Dierkes, R.~Ehrig, S.~R\"{o}blitz, B.~Leeners, T.H.C.
		  Kr\"{u}ger, M.~Egli, and F.~Ille.
		\newblock Patient-specific models from inter-patient biological models and
		  clinical records.
		\newblock In {\em \MCLabProceedingsOf{FMCAD 2014}}. IEEE, 2014.

		\bibitem{tuncali-etal:2019:rapidly}
		C.E. Tuncali and G.~Fainekos.
		\newblock Rapidly-exploring random trees for testing automated vehicles.
		\newblock In {\em \MCLabProceedingsOf{ITSC~2019}}. IEEE, 2019.

		\bibitem{whalen-etal:2007:integration}
		M.W. Whalen, D.D. Cofer, S.P. Miller, B.H. Krogh, and W.~Storm.
		\newblock Integration of formal analysis into a model-based software
		  development process.
		\newblock In {\em \MCLabProceedingsOf{FMICS~2007}}, volume 4916 of {\em LNCS}.
		  Springer, 2007.

		\bibitem{younes-etal:2006:sttt}
		H.L.S. Younes, M.Z. Kwiatkowska, G.~Norman, and D.~Parker.
		\newblock Numerical vs.\ statistical probabilistic model checking.
		\newblock {\em STTT}, 8(3), 2006.

		\bibitem{younes-etal:2002:probabilistic}
		H.L.S. Younes and R.G. Simmons.
		\newblock Probabilistic verification of discrete event systems using acceptance
		  sampling.
		\newblock In {\em \MCLabProceedingsOf{CAV~2002}}, volume 2404 of {\em LNCS}.
		  Springer, 2002.

		\bibitem{zuliani-etal:2013:fmsd}
		P.~Zuliani, A.~Platzer, and E.M. Clarke.
		\newblock {B}ayesian statistical model checking with application to
		  {S}tateflow/{S}imulink verification.
		\newblock {\em Form.\ Meth.\ Sys.\ Des.}, 43(2), 2013.

	\end{thebibliography}

\begin{thebibliography}{1}
		\bibitem{mancini-etal:2017:ipl}
		T.~Mancini, F.~Mari, A.~Massini, I.~Melatti, I.~Salvo, and E.~Tronci.
		\newblock On minimising the maximum expected verification time.
		\newblock {\em Inf.\ Proc.\ Lett.}, 122, 2017.

		\bibitem{mancini-etal:2015:gandalf}
		T.~Mancini, F.~Mari, A.~Massini, I.~Melatti, and E.~Tronci.
		\newblock Simulator semantics for system level formal verification.
		\newblock {\em EPTCS}, 193, 2015.

		\bibitem{mancini-etal:2021:jlamp}
		T.~Mancini, F.~Mari, A.~Massini, I.~Melatti, and E.~Tronci.
		\newblock On checking equivalence of simulation scripts.
		\newblock {\em J.\ Log.\ Algebr.\ Meth.\ Program.}, 120, 2021.

		\bibitem{mancini-etal:2021:tse-supervisory}
		T.~Mancini, I.~Melatti, and E.~Tronci.
		\newblock Any-horizon uniform random sampling and enumeration of constrained
		  scenarios for simulation-based formal verification.
		\newblock {\em IEEE TSE}, 2021.

		\bibitem{sontag:1998:book}
		E.D. Sontag.
		\newblock {\em Mathematical Control Theory: Deterministic Finite Dimensional
		  Systems (2nd Ed.)}.
		\newblock Springer, 1998.
	\end{thebibliography}
\end{document}